\documentclass[12pt,twoside,a4paper]{article}
\usepackage{oldgerm}
\usepackage[margin=1in,nofoot,bottom=0.8in,right=1.4in,left=1.0in]{geometry}
\usepackage{epsfig}
\usepackage{fancyhdr}
\usepackage{cite}
\usepackage{makeidx}
\normalfont
\makeindex
\begin{document}
\pagestyle{fancy}
\rhead[K. P. N. Murthy]{\thepage}
\lhead[\thepage]{Monte Carlo for Statistical Physics}
\chead{}
\lfoot{}\cfoot{}\rfoot{}
\thispagestyle{empty}
\renewcommand{\thepage}{\roman{page}}
\vskip 20mm
\centerline{\Large\bf{\cal AN INTRODUCTION TO} }
\vskip 4mm
\centerline{\Large\bf {\cal  MONTE CARLO SIMULATIONS IN} }
\vskip 4mm
\centerline{\Large\bf{\cal STATISTICAL PHYSICS
  }}
\vskip 50mm
\centerline{by}
\vskip 8mm
\centerline{\Large \sffamily {\bf K. P. N. Murthy} }
\vskip 5mm
\centerline{\large Theoretical Studies Section,}
\centerline{\large Materials Science Division.}
\vfill
\centerline{\large Indira Gandhi Centre for Atomic Research,}
\centerline{\large Kalpakkam 603 102, Tamilnadu.}
\centerline{\large INDIA}
\vskip 5mm
\centerline{\large 16 December   2003}

\newpage
\thispagestyle{empty}
\ \ \ \ \ \ \ \ \ \ \ \ 
\newpage
\newpage
\normalfont
\thispagestyle{empty}
\addcontentsline{toc}{section}{\numberline{}Preface}

\hfill {\Huge\sffamily\bfseries Preface}
\vskip 20mm
\setlength{\fboxrule}{.8pt}
\setlength{\fboxsep}{5mm}
\begin{center}
\fbox{
{\parbox{.83\linewidth}{
    Science is what we understand well enough
to explain to a computer.
\vskip 5mm
\hfill  Donald E. Knuth \ 
}
}
}
\end{center}
\smallskip
\vskip 10mm
Computing facilities - like work stations and PCs, with
high speed and large memory are 
becoming increasingly and easily available to researchers and students 
since recent times. This is particularly true 
in  our country;  the growth has been rapid and is most 
likely to become more rapid in the coming years.
As a result, computer
simulation methods are gaining in importance and popularity
as viable and useful tools for research and education. 

Monte Carlo is an early and an important computer simulation 
technique. Named after the city (in the province of Monoco, 
south of France) famous for 
its (gambling) casinos, the Monte Carlo method makes extensive use of  
 random numbers. It is employed for simulation
of a large variety of phenomena in very many different disciplines. 
However, in this monograph, I shall confine myself to elucidating the 
technique for simulation of statistical physics systems. 

We begin with a quick look at what do we mean by 
ensembles in general and Gibbs ensemble
in particular. We discuss briefly  microcanonical ensemble 
that models isolated system, canonical ensemble that describes closed system
and  grand canonical ensemble, useful in the study of open system. 
In the following  few sections  I  present 
briefly the basic ingredients of the  method of Monte Carlo. These 
include discussions on random numbers, generation and testing of a 
sequence of pseudo random numbers, random sampling from different distributions, 
importance sampling and statistical error bars from uncorrelated and
correlated data. I have tried to keep the presentation as simple as
possible. 

These preliminary discussions are followed by a reasonably 
detailed description of the 
Monte Carlo methods for simulating a canonical ensemble of microstates
of a classical statistical mechanics system. The topics covered are:
Markov chains,  Metropolis algorithm, Ising model, 
continuous phase transition,
critical exponents, finite size scaling, n-fold way,  critical slowing down,
cluster algorithms, cluster counting, percolation, histogram
techniques, supercritical slowing down,
multicanonical /entropic sampling, 
Wang-Landau algorithm and Jarzynski's equality relating nonequilibrium
work done to equilibrium free energies. I have tried to present 
these topics the way I have seen them, the way I have learnt them and 
the way I am teaching them. 

My first word of thanks goes to Prof. A. K. Bhatnagar, the Central University 
of Hyderabad and  presently the 
vice chancelor of the Pondichery University. 
Prof. Bhatnagar met me during a conference in Kurukshetra and spoke  of 
his plans to start a post-graduate course on computations in the school of physics
at the Hyderabad University.  
He asked of me if I would  be willing to handle the 
portions on Monte Carlo and Molecular dynamics. I readily agreed.
On his invitation  
I went to Hyderabad during November and December 1998 and 
gave lectures on Monte Carlo and Molecular dynamics to  the first batch of 
students of the  M. Tech. Computational physics programme. On my request 
 Prof. V. S. S. Sastri, Prof. K. Venu,
Prof. Bansal and Prof. Srivatsav attended all the lectures and 
gave me valuable criticisms and suggestions. I take this opportunity 
to say thanks to them and the students of the course. 
Prof. Bhatnagar invited me again during January-February
2000 for giving the same course to the second batch of students.
While writing this monograph I have drawn heavily from  
the notes I prepared for the Hyderabad courses. I thank Prof. 
Bhatnagar for giving me this wonderful opportunity to teach and to learn.
I requested Prof. Bhatnagar to give a foreword to this monograph and he readily
agreed.  

Subsequently I got opportunities to speak on Monte Carlo theory and practice in several places 
some of which
are listed below.
\begin{enumerate}
\item[]
Guru Ghasidas University, Bilaspur ({\small tutorial lecture at the
Solid State Physics Symposium:  26-30, December  2000}),
\item[]
Madurai Kamaraj University, Madurai ({\small Seminar on Computer
Simulations in Physics, 4-5, February 2002}),
\item[]
 Calicut University, Calicut (Refresher course for the college teachers,
25-30 November 2002),
\item[]
  Bharathidasan University,
Tiruchirappalli ({\small UGC refresher course in Physics,
10-30 December 2002}),
\item[]
 Annamalai University,
Annamalai Nagar ({\small UGC sponsored refresher course,
19 December 2002 - 8 January 2003}),
\item[]
and
\item[]
the  Materials Science Division, Indira Gandhi Centre for Atomic Research,
Kalpakkam (to several doctoral students, post doctoral students and
colleagues) for more than a decade. 
\end{enumerate}

I thank
M.~Ramanadham (Mumbai),
K.~Ramachandran (Madurai),
P.~Remesan (Calicut),
K.~Ramamurthi (Tiruchirappalli) and
A.~N.~Kannappan (Annamalai Nagar) for the invitations.

I am thankful to 
V.~Sridhar (Kalpakkam),
R.~Indira (Kalpakkam),
V.~S.~S.~Sastri (Hyderabad),
K.~Venu (Hyderabad),
M.~C.~Valsakumar (Kalpakkam),
S.~L.~Narasimhan (Mumbai),
K.~Satyavathi (Hyderabad),
Karl-Heinz~Herrmann (J\"ulich, Kalpakkam),
Giancarlo~Franzes (Italy),
S.~Sivakumar (Kalpakkam),
R.~Harish (Kalpakkam),
Jean-Feng~Gwan (Taiwan, J\"ulich)
and 
Vishal~Mehra (New Delhi, J\"ulich)
for
criticisms and suggestions.

I owe a special word of thanks to my friend Dr. A. Natarajan (Kalpakkam); 
Dr. Natarajan is a man with very big heart and is genuinely conccerned 
with lives and careers of his collegues and juniors; he always finds time
to actively involve himself with new ventures of his colegues and encourage them
in their work. Dr. Natarajan  read through the entire 
manuscript carefully and critically; he marked several corrections in the manuscript and 
suggested revisions in several places to improve the readability.  
I am extremely thankful to Dr. A. Natarajan.

I have intended this monograph only  as an introduction to this 
vast and rapidly growing field. The aim is to help you obtain
a feel for the method  and an appreciation for the 
nuances in the different techniques  including the recent ones. 
I have tried to be
simple in the  presentation, up-to-date in the coverage of several and 
important topics, and exhaustive in the bibliography. I hope these
notes shall become a useful addition to the bookshelves of researchers 
and students.
\vskip 20mm
\noindent
Kalpakkam\\
9 November 2003 \hfill K. P. N.

\newpage
\ \ \ \ \ \ \ \ \ \
\newpage
\tableofcontents
\listoffigures
\newpage
\listoftables
\newpage
\ \ \ \ \ \ \ \ \ 
\newpage
 
\pagenumbering{arabic}
\setcounter{page}{1}
\setcounter{table}{0}
\addtolength{\headheight}{3pt}
\pagestyle{fancy}
\rhead[K. P. N. Murthy]{\thepage}
\lhead[\thepage]{Monte Carlo for Statistical Physics}
\chead{}
\lfoot{}\cfoot{}\rfoot{}
\thispagestyle{empty}
\normalfont
\mathversion{normal}
\section{Prologue}\label{prologue}

A macroscopic system  consists of a very large 
number of  microscopic constituents. 
A gas in a container is a simple example.
The microscopic constituents are the molecules of the gas. The number of 
molecules $ N$  is typically of the order of $10^{23}$. 
Each molecule 
requires three position and three momentum coordinates for complete 
specification in classical mechanics. The entire macroscopic 
system at any given 
instant of time can thus be  specified by a string of $6 N$ numbers,
which defines, in a sense, a microstate. In a $6N$ dimensional phase 
space\index{phase space}, the system is represented by a point; or more precisely,  
by a
phase space element of volume $h^{3N}$ (owing its origin to our {\it
quantum mechanical}  inability to specify precisely the position and 
its conjugate momentum simultaneously).  Here, $h$ is  
the Planck's constant.
\footnote{The Planck's constant\index{Planck's constant} 
$h=6.626\times 10^{-34}$  Joules\index{Joule}.second or 
$4.136\times 10^{-15}$ electron volts.second\index{electron volt}.} 
It is clear that 
the number of microstates associated with the system is extremely large. 
The (equilibrium) system is  switching  from one microstate to the other 
all the time. In a classical picture, the  point representing the system 
traces a trajectory in the $6N$ dimensional phase space\index{phase space}
 as time proceeds.  
Keeping track of $6N$ numbers  as a function of time is neither feasible 
nor practical. A statistical approach would be  helpful. 

Any experimentally measured  equilibrium macroscopic property is  a time 
averaged\index{time average} quantity : averaged over the phase space 
\index{phase space} trajectory traced by the macroscopic system during the observation time. 
Let us assume that  during the experimental observation
time, the system visits all the (typical?)  microstates, consistent 
with the constraints 
on volume, number of particles, energy {\it etc.}
This is called ergodicity\index{ergodicity} assumption.  
This assumption is fine since  
the observation time scale is very large compared to the time scale over which
the system switches from one microstate to the other. 
We can then 
equate the \ \lq experimental\rq\ time average\index{time average} 
to an average over a 
suitable static Gibbs  ensemble\index{ensemble!Gibbs} of microstates.  
For such a scheme to be 
successful we must, at the outset, recognize that a  macroscopic property  
is basically statistical in nature. For example, 
pressure\index{pressure} is the average 
momentum transferred by the molecules colliding with the walls of the 
container;  entropy\index{entropy}  is the logarithm of the number of microstates 
accessible to the system;  temperature is average energy; 
specific heat\index{specific heat} is a 
manifestation of energy fluctuations\index{fluctuations}; {\it etc.} 

When in equilibrium, the macroscopic properties of a system
do not change 
with time; hence a macroscopic property can  be defined as a time  average\index{time average}  
over an underlying (stationary) stochastic process\index{stochastic process} or equivalently  an 
average over  a suitably defined Gibbs  ensemble\index{ensemble!Gibbs}; the associated 
fluctuations\index{fluctuations} 
are inversely proportional to the  square root of the system\index{system size} size and 
hence are usually negligibly small.  This is  directly a consequence of 
the  Central Limit Theorem\index{Central Limit Theorem}: {\it the distribution of the sum of $N$ 
independent and  finite variance\index{variance} random variables converges to a 
Gaussian\index{Gaussian} in the 
limit $N\to\infty$, and has a variance\index{variance} proportional to $N$}~\cite{cltref}. 
\footnote{For an account of the rate of convergence of the sum to a Gaussian\index{Gaussian}
random variable see~\cite{pap}. For an elegant demonstration  of the Central
Limit Theorem\index{Central Limit Theorem}  employing renormalization and scaling 
ideas, see~\cite{sornette}.}
Thus the standard deviation of the arithmetic mean\index{arithmetic mean}
 of these $N$ random
variables is inversely proportional to $\sqrt{N}$. 
Physically this implies that the fluctuations
of a macroscopic property from its average is extremely small and inversely proportional to the
size (number of microscopic constituents) of the system. Clearly it is
the largeness of the number of  microscopic constituents that gives rise
to a certain robustness to the macroscopic behaviour of the system.

\subsection{What is an ensemble? What is a Gibbs ensemble?\label{ensemble}\index{ensemble!Gibbs}}

To understand the  notion of an ensemble\index{ensemble} consider an experiment  
which has more than one outcome. Let us collect all the possible outcomes
of the experiment in a set called the sample space\index{sample space} 
$\Omega = \{\omega_i : i=1,\ 2,\ \cdots\ ,\ {\hat \Omega}\}$, where $\omega_i$ denotes  
an outcome  and ${\hat \Omega}$ the number of outcomes of the 
experiment. Let $P(\omega_i) \ge 0$ be the probability of the outcome $\omega_i$.
We have $\sum_{i=1}^{ {\hat \Omega} } P(\omega_i) =1$. Thus the sample space\index{sample space}
 $\Omega$ and 
the probability $P$ define the experiment. Physicists however use a single notion 
of an  ensemble\index{ensemble}, introduced by Maxwell and denoted by $\Omega_{{\rm E}}$. 
The members of the ensemble\index{ensemble} are  drawn from the 
sample space\index{sample space} $\Omega$. However an out come $\omega_i$ appears in the ensemble
$n(\omega_i)$ times so that $ n(\omega_i)/{\hat\Omega}_E \equiv P(\omega_i)$, where
$ {\hat \Omega}_{{\rm E}}$ is the size of the ensemble\index{ensemble}; ${\hat\Omega}_{{\rm E}}$ 
is taken adequately
large  so that 
all outcomes of the experiment appear in the ensemble\index{ensemble} in correct proportions. 
We do not specify any rule for the  construction of an ensemble\index{ensemble};
we can employ Monte 
Carlo (subject of this monograph) or Molecular Dynamics or for that matter any other 
technique to construct an ensemble\index{ensemble}.
 
A Gibbs ensemble\index{ensemble!Gibbs} contains  microstates of an
equilibrium macroscopic system under the macroscopic constraints  operating on the
system. Depending on the nature of the constraints we have different Gibbs 
ensembles\index{ensemble!Gibbs}. 

\subsubsection
{Isolated system: Microcanonical ensemble}\label{microcanonical_ensemble}
Consider an isolated\index{isolated system} macroscopic equilibrium system. The constraints
are on  : Energy (E), Volume (V) and Number of particles (N). 
Collect all possible
microstates of the system into a set $\Omega_{{\rm IS}}$,  called the sample
space\index{sample space}. 
All microstates of an  isolated system are 
equally probable.
\footnote{This is an axiom; the entire edifice of statistical 
mechanics is built on this axiom.} 
Also, because of the constraints,  all the microstates are of the same energy and volume and have 
the same number of particles.
Let $O({\cal C})$ be a macroscopic property of the system when in its 
microstate ${\cal C}$. Formally the average of $O$ is given by
\begin{eqnarray}
\left\langle O \right\rangle &= &\ {{1}\over{ {\hat \Omega}_{{\rm IS}} }}\ \ 
\sum_{ {\cal C}\ \in\ \Omega_{{\rm IS} }} O({\cal C})
\end{eqnarray}
where ${\hat \Omega}_{{\rm IS} }$ is the total number of microstates 
of the system. In fact the entropy{\index{entropy}  of the system is given by
$S=k_{{\rm B}} \ln {\hat \Omega}_{{\rm IS} }$, where $k_{{\rm B}}$ is the Boltzmann constant.
\footnote{The Boltzmann constant 
$k_{{\rm B}}$  equals
$1.381\times 10^{-23}$Joules\index{Joule} per degree Kelvin\index{Kelvin} or
$8.671\times 10^{-5}$ electron volts per degree Kelvin\index{Kelvin}. }
Thus the probability $P({\cal C})$ for an isolated system to be 
in a microstate ${\cal C}$ is  
$1/{\hat\Omega}_{{\rm IS} }$  or $\exp (-S/k_{{\rm B}})$, and is the same for all the 
microstates.  

Equivalently, we can construct a microcanonical\index{ensemble!microcanonical} 
ensemble $\Omega_{\mu {\rm CE}}$
of microstates; a microstate ${\cal C} \in \Omega_{{\rm IS} }$ 
occurs in $\Omega_{\mu {\rm CE} }$
as often as to reflect its probability $P({\cal C})$.
In other words, the number of times the microstate 
${\cal C}$ occurs in the microcanonical ensemble\index{ensemble!microcanonical} 
$\Omega_{\mu {\rm CE}}$ divided by the 
size of the ensemble (denoted by ${\hat\Omega}_{\mu {\rm CE} }$) 
equals $P({\cal C})$.  For an isolated system all microstates 
belonging to $\Omega_{{\rm IS}}$ occur in equal proportions 
in $\Omega_{\mu {\rm CE}}$. 

\subsubsection{Closed system: Canonical ensemble\index{ensemble!canonical}}\label{canonical_ensemble}

Consider now an equilibrium macroscopic system in thermal contact 
with a heat bath\index{heat bath} at temperature 
say $T$. Only the volume (V) and the number of particles (N) are 
fixed. The energy is not fixed. Instead the temperature of the system
equals that of the heat bath when in equilibrium. This is a 
closed system\index{closed system}: 
the system exchanges energy with the 
heat bath but not particles. Let  $\Omega_{{\rm CS}}$ 
denote the set of all possible microstates 
of the closed system at temperature $T$. $\Omega_{{\rm CS}}$ is 
called the sample space\index{sample space} 
of 
the closed system; since the system 
can exchange energy with 
the heat bath, different microstates belonging to $\Omega_{{\rm CS}}$ 
can have different energies.
\footnote{There can be many microstates 
having the same energy $E$. Logarithm of this number multiplied 
by $k_{{\rm B}}$ shall be called suggestively as 
microcanonical entropy $S(E)$. We shall find this terminology
useful when we consider entropic/multicanonical sampling in 
sections (\ref{entropic_sampling}) and (\ref{entropic_multi_canonical}).}
\index{microcanonical entropy}
Also the probability with which a closed system 
can be found in its microstate 
${\cal C}$ is proportional to $\exp [-\beta E({\cal C})]$ 
where $\beta = 1/(k_{{\rm B}} T)$. 
The probability of occurrence of a microstate ${\cal C}$ 
in a closed system depends on the 
energy of the microstate and the temperature of the system. 
Thus to each ${\cal C} \in \Omega_{{\rm CS} }$
we attach a Boltzmann weight $\exp[-\beta E({\cal C})]$. 
The sum of the Boltzmann weights 
taken over all the microstates belonging to $\Omega_{{\rm CS}}$ 
is called the canonical partition 
function\index{partition function!canonical} $Z(\beta)$, given by,
\begin{eqnarray}
Z(\beta ,V,N) & = & \sum_{ {\cal C}\  \in\  \Omega_{{\rm CS} } }
\exp \left[ -\beta E( {\cal C})\right].
\end{eqnarray}
Let $O({\cal C})$ be a macroscopic property of the closed system 
when in its microstate
${\cal C}$. The average of $O$ is formally given by,
\begin{eqnarray}
\left\langle O(\beta)\right\rangle &=&
{{1}\over{Z(\beta,\ V,\ N)}}\  
\sum_{ {\cal C}\  \in\  \Omega_{{\rm CS} } }  O ({\cal C})
\exp\left[ -\beta E ({\cal C} )\right]
\end{eqnarray}

To calculate $\langle O(\beta)\rangle$, we can take an alternate approach.
Let $\Omega_{{\rm CE} }$ denote a canonical ensemble\index{ensemble!canonical} of microstates 
belonging to the 
closed system\index{closed system}. A microstate 
${\cal C}\ \in\ \Omega_{{\rm CS} }$ occurs in the 
ensemble\index{ensemble!canonical} $\Omega_{{\rm CE}}$ as 
often as to reflect its probability 
$P({\cal C})=\exp[-\beta E( {\cal C})]/Z(\beta,\ V,\ N)$.
Thus the number of times a microstate 
${\cal C}\in\Omega_{{\rm CS} }$ occurs in the  canonical ensemble\index{ensemble!canonical}
$\Omega_{{\rm CE} }$ divided by the  size,  
${\hat \Omega}_{{\rm CE}}$ of the ensemble\index{ensemble!canonical} is given by $P( {\cal C})$. 
The advantage of constructing
a canonical ensemble is that a macroscopic property can be 
calculated as a simple arithmetic 
mean\index{arithmetic mean},
\begin{eqnarray}
\left\langle O(\beta)\right\rangle & = & 
{{1}\over{ {\hat \Omega}_{{\rm CE} } }}\ \  \ 
\sum_{ {\rm C}\  \in\  \Omega_{{\rm CE}} }\ \ 
O( {\rm C} )\ .
\end{eqnarray}  
The size ${\hat\Omega}_{{\rm CE} }$ of the canonical 
ensemble\index{ensemble!canonical} is usually taken to be  large  
so that even those microstates 
with very low probability are present in the ensemble in right proportions. 
We shall follow the convention of denoting an element of 
a sample space\index{sample space}
by script letters, {\it e.g.} ${\cal C}$ and that of an 
ensemble\index{ensemble} by
capital letters {\it e.g.} C. 

\subsubsection{Open system: 
Grand canonical ensemble\index{ensemble!grand canonical}}\label{GCE}

We can proceed further and consider an open system in 
which there is an exchange of energy as well as
particles with the surrounding bath. The constraints are on the 
temperature (T), volume (V) and chemical potential\index{chemical potential}
 ($\mu$). 
Let $\Omega_{{\rm OS} }$ denote  the sample space.\index{sample space}
The open system\index{open system}
can be found in a microstate ${\cal C}$ 
with a probability proportional to
the Gibbs factor given by 
$\exp[ -\beta E({\cal C}) + \mu\beta N({\cal C})]$, where  
$N({\cal C})$ is the number of particles in the open system 
when it is in its  microstate
${\cal C}$. Thus to each microstate ${\cal C}$ we attach  a Gibbs weight 
$\exp[ -\beta E({\cal C}) + \mu\beta N({\cal C})]$. The 
sum of the Gibbs weights taken over all the microstates of 
the open system is called the grand canonical 
partition function\index{partition function!grand canonical},
\begin{eqnarray}
{\cal Q} (\beta, V, \mu) & = & \sum_{ {\cal C}\  \in\  \Omega_{{\rm OS}}} 
\exp\left[ -\beta E( {\cal C}) + \mu\beta N({\cal C})\right]\ .
\end{eqnarray}
The average of a macroscopic property  $O$ is formally given by,
\begin{eqnarray}
\left\langle O(\beta, V, \mu)\right\rangle  & = & {{1}\over{ {\cal Q}}}
\sum_{ {\cal C}\  \in\  \Omega_{{\rm OS} } } O( {\cal C} ) 
\exp\left[ -\beta E( {\cal C}) + \mu\beta N({\cal C})\right].
\end{eqnarray}
We can calculate the average by constructing 
a grand canonical ensemble\index{ensemble!grand canonical} 
$\Omega_{{\rm GCE}}$ of microstates; the number of times 
a microstate ${\cal C}$, belonging to $\Omega_{{\rm OS}}$
occurs in $\Omega_{{\rm GCE}}$ divided by the size of 
$\Omega_{{\rm GCE} }$ (denoted by ${\hat \Omega}_{{\rm GCE}}$)
 is given by $\exp[ -\beta E({\cal C}) 
+ \mu\beta N({\cal C})]/{\cal Q}$. Thus  $\langle O\rangle $ 
is given by a 
simple arithmetic average\index{arithmetic mean},
\begin{eqnarray}
\left\langle O (\beta,V, \mu) \right\rangle & = & 
{{1}\over{ {\hat \Omega}_{{\rm GCE} } }}\ \  \sum_{ {\rm C}\  \in\  \Omega_{{\rm GCE}} } 
O( {\rm C})
\end{eqnarray}

Besides the microcanonical\index{ensemble!microcanonical}, 
canonical\index{ensemble!canonical} and grand canonical\index{ensemble!grand canonical} 
ensembles, we can construct other physical ensembles
like isobaric-isothermal ensembles {\it etc.}, depending on the problem 
under considerations. The choice of the 
ensemble is purely a matter of convenience dictated by 
the nature of the physical system under investigation and the 
nature of the macroscopic properties we want to calculate. 
Indeed we shall see later that even construction of   
 unphysical ensembles like multicanonical ensemble\index{ensemble!multicanonical} 
proves to have  certain distinct advantages. 

However, in these notes  
I shall concentrate on the calculation  of a  macroscopic property as
an average over a canonical ensemble\index{ensemble!canonical} 
of microstates generated by
the technique of Monte Carlo\index{Monte Carlo}. 

\section{What is a Monte Carlo method?}\label{MC} 

Monte Carlo\index{Monte Carlo}  is a numerical technique that makes 
use of random numbers to solve a problem.  Historically, the first 
large scale Monte Carlo work carried out dates back to the middle 
of the twentieth century.  This work pertained to studies of neutron\index{neutron!multiplication} 
multiplication,  scattering\index{neutron!scattering}, propagation\index{neutron!propagation}
 and eventual absorption\index{neutron!absorption} in 
a medium or leakage\index{neutron!leakage} from it.  Stanislav Ulam\index{Ulam}, 
John von Neumann\index{von Neumann} and Enrico Fermi\index{Fermi} were 
the first to propose and employ the  Monte Carlo  method as a viable 
numerical technique for solving practical problems. The earliest published
work on Monte Carlo is perhaps the paper of Metropolis and Ulam\cite{mu} in the 
year 1949.
\footnote{There were of course several isolated and 
perhaps not fully developed instances 
earlier, when Monte Carlo has been used in 
some form or the other. An example is the 
experiment performed in the middle of the 
nineteenth century, 
consisting of throwing a needle
randomly on a  board notched with parallel 
lines and inferring the numerical value
of $\pi$ from the number times the needle 
intersects a line; this is known as 
Buffon's needle problem,\index{Buffon's needle} 
an early description of which can be found in \cite{Hall}.
The Quincunx\index{Quincunx} constructed by 
Galton \cite{Galton}\index{Galton} in the second 
half of the nineteenth 
century, consisted of balls rolling down 
an array of pins 
and getting  collected 
in the vertical compartments placed at the bottom. 
A pin deflects the rolling ball randomly and with equal probability
to its left or right. 
The heights of the balls in the 
compartments approximate a binomial 
distribution\index{binomial distribution}. 
Perhaps this is an early technique 
of physically  sampling  random numbers 
from a binomial distribution. In 
nineteen twenties, Karl 
Pearson\index{Pearson} perceived 
the use of random numbers for solving complex
problems in probability theory and statistics 
that  were not amenable to exact analytical solutions. Pearson
encouraged L. H. C. Tippet\index{Tippet} to 
produce a table of random numbers to help in such studies, and a book
of random sampling numbers \cite{tippet} was published 
in the year 1927. 

In India, P. C. Mahalanbois\index{Mahalanbois} 
\cite{pcm} exploited \lq\ random sampling \rq\ 
technique to solve
a variety problems like the choice of 
optimum sampling plans in survey work, 
choice of optimum size
and shape of plots in experimental work 
{\it etc.}, see \cite{crrao}.

Indeed,  descriptions of several modern Monte Carlo 
techniques appear in a paper by
Kelvin \cite{KELVIN}\index{Kelvin}, written nearly 
more than hundred  years ago, in the context of a discussion on the
Boltzmann equation. But Kelvin was more interested in the 
results  than  in the technique,
which to him was {\it obvious}!}
 
Thus, for carrying out a Monte Carlo simulation, we  require a 
sequence of numbers which are random, independent, real  and  
uniformly distributed in the range zero to one. Strictly, we can call 
a sequence of numbers random, if and only if it is generated by a 
random physical process like radioactive decay\index{radioactive decay}, 
thermal noise\index{thermal noise} in 
electronic devices, cosmic ray arrival times\index{cosmic ray arrival times}, tossing of a coin
 {\it etc.}  These phenomena are known to be truly random at least
according to the present day theories. Generate 
once and for all,  a fairly long sequence of random numbers from a 
random physical process. Employ this sequence in your Monte Carlo\index{Monte Carlo} 
calculations.  This is always a safe procedure. Indeed, tables of
random numbers were generated and employed in early Monte Carlo  
practice. Census reports~\cite{tippet}, telephone 
directories~\cite{kbs-1}, spinning wheel~\cite{kbs-1,kbs-2}, 
automatic electronic roulette~\cite{rand}, radioactive 
decay\index{radioactive decay}~\cite{ikyim}, 
thermal noise\index{thermal noise} in a semi-conductor 
device\footnote{M. Richter: millions of physical random numbers were generated by measuring thermal noise 
of a semiconductor device; these are available upon an anonymous ftp call to site 
at dfv.rwth-aachen.de} {\it etc.}, were employed to generate large 
tables of random digits.  

A table of random numbers is  useful  
if the Monte Carlo\index{Monte Carlo} calculation is  carried
out manually. However, for computer calculations, 
use of these tables is impractical. 
A computer, having a relatively small internal memory,  
 cannot hold a large table of random numbers. 
One can keep the  random numbers in an 
external storage device like a disk or a tape; 
but then, constant retrieval from these 
peripherals would considerably slow down the 
computations. Hence it is often   
desirable to generate a random number as and 
when  required, employing simple arithmetic operations 
that do not take much computer time and the algorithm itself 
does not require much memory for storage.
These numbers, generated
 by deterministic algorithms, 
are therefore predictable and reproducible. 
Hence by no stretch of 
imagination can they be called random. We shall  call them 
 pseudo random\index{pseudo random numbers}.  We shall be content  with pseudo random 
numbers\index{pseudo random numbers} if we find  they go through tests of 
randomness satisfactorily. In fact, it is always
a good practice to check if the sequence of random numbers 
you are going to use in your simulation,  goes through 
several standard tests for randomness. 
It is also desirable to devise special tests 
of randomness,  depending on the particular 
application you have in mind.  
I shall come to tests for randomness a bit later.

\section{How do we generate  a sequence of 
pseudo random numbers\index{pseudo random numbers}?}\label{prn}

\subsection{Congruential Generators}\label{cg}

In a linear   congruential generator\index{congruential generator}, 
for example, we start with an integer
$R_1$ and generate successive integers by the recursion\index{recursion},
\begin{equation}
R_{i+1}=a\times R_i + b\  ({\rm mod}\   m),
\end{equation}
where $a$, $b$ and $m$ are integers.  $a$ is called the generator 
or multiplier\index{multiplier}, $b$ the increment\index{increment}
 and $m$ the modulus\index{modulus}. If the increment\index{increment} 
$b$ is zero, we call the generator  
multiplicative\index{multiplicative congruential} 
congruential  and if $b > 0$, 
it is called the mixed congruential\index{mixed congruential}.  
The above equation means the following. Start with an integer $R_1$, between 
zero and $m-1$. This is your choice. Calculate $a\times R_1 +b$.
Divide the result by $m$ and find the remainder. Call it $R_2$. Calculate
$a\times R_2 +b$; divide the result by $m$; find the remainder and call it 
$R_3$. Proceed in the same fashion and generate a sequence of integers 
$\{ R_1 , R_2 , \cdots \} $,  initiated by the seed $R_1$. 
Thus $\{ 0\le R_i \le m-1 : i=1,2,\ \cdots\  \}$ is 
a sequence of pseudo random integers\index{pseudo random numbers}. 
This sequence can be converted to floating point\index{floating point} 
numbers by dividing each 
by $m$. 

A  congruential generator\index{congruential generator} is robust,  
simple and  fast; the theory is reasonably well 
understood. It  
gives a fairly long sequence of \ \lq good\rq\  pseudo random 
numbers\index{pseudo random numbers}. The values of $a$, $b$ and $m$ must be chosen 
carefully.  Clearly, whatever be the choice of $a$, $b$ and $m$, 
the sequence  shall repeat itself after 
utmost $m$ steps. The sequence is therefore
periodic. Hence in applications, we must ensure that the number of random
numbers required for any single simulation must be much less than the
period.  Usually $m$ is taken to be very large to permit this.
We can always get a sequence with full period, if  
$m$, $b$ and $a$ are properly chosen.
\footnote{We must ensure that
(i) $m$ and $b$ are relatively prime to each other; 
{\it i.e.}\ $\gcd(m,b)=1 $; (ii)  $a \equiv\  1\  (\rm{mod}\  p)$ 
for every prime factor $p$ of $m$; and (iii) $a\equiv\  1\  (\rm{mod}\ 4)$, 
if $m\equiv\  0\  \rm{(mod}\  4)$.}
  The modulus\index{modulus} $m$ is usually 
taken as $2^{t-1}-1$, where $t$ is the number of bits 
used to store an integer and hence is machine specific. One of the $t$ bits 
is used up for storing the sign of the integer. The choice $a=7^{5}=16807$, 
$b=0$ and $m=2^{31}-1$, for a $32$ bit machine has been shown to yield good 
results~\cite{PTVF}. The period of this generator is $2^{31}-1=2.15\times 10^9$, which
is really not long for several applications. Another popular choice of 
parameters~\cite{Marsaglia_1} 
consists of $a=69069$, $b=1$, and $m=2^{32}$, which also has a short period 
of $4.29\times 10^{9}$.  The choice of $a=13^{13}$, $b=0$ and 
$m=2^{59}$ in the generator G05FAF of the NAG library~\cite{NAG}  has a much longer period 
of nearly $5.76\times 10^{17}$. The   
congruential generators\index{congruential generator} have been successful 
and invariably most of the present day pseudo random number
\index{pseudo random numbers} 
generators are of this class.  

\subsection{Marsaglia lattice structure}\label{marsaglia_defect}

Let $\xi_1 , \xi_2 , \xi_3 , \cdots $ denote a sequence of random numbers 
from a linear congruential generator.
\index{congruential generator} Form from these numbers, a sequence 
of $d$-tuples: 
$v_1=(\xi_1 , \xi_2 , \cdots , \xi_d )$, $\ v_2 = (\xi_2 , \xi_3 ,
 \cdots , \xi_{d+1})$,
$v_3 = (\xi_3 , \xi_4 , \cdots , \xi_{d+2}) \cdots .$  View  the 
d-tuples $\{ v_1\ , v_2 , \ v_3 , \cdots \}$ as  points in the unit hyper
cube of $d$  dimensions. You would expect these points to be 
more or less uniformly distributed in the hyper cube. But invariably you will 
find that these points are confined to a relatively smaller number 
of parallel planes, called the Marsaglia\index{Marsaglia} lattice structure named after its 
discoverer~\cite{Marsaglia_2}. The points on several of the Marsaglia
planes  form regular patterns. Such a hidden Marsaglia\index{Marsaglia} order is an 
inherent feature of all linear congruential generators\index{congruential generator}. The randomness is 
truly {\it pseudo}  indeed! Hence you must check very carefully if the 
inevitable presence of Marasaglia defect in the pseudo 
random number\index{pseudo random numbers} sequence  
affects significantly the results of your Monte Carlo\index{Monte Carlo} application. 
Hopefully it doesn't. But then one never knows.  Somehow the Monte Carlo 
community has learnt to live with the Marsaglia\index{Marsaglia} defect over the last three 
decades!

You can increase the number of Marsaglia\index{Marsaglia} 
planes by a second order recursion\index{recursion}, 
\begin{eqnarray}
R_{i+2} & = & a\times R_{i+1} + b\times R_i \ ({\rm mod}\   m),
\end{eqnarray}
and 
choosing properly the recursion\index{recursion} parameters $a$ and $b$.
Note that the above recursion requires two seeds.

\subsection{Shift register generators}\label{shift_register}\index{shift register}

An alternative that does not suffer from the correlations invariably present 
in the congruential generators is based on the generalized feedback shift 
register algorithms \cite{tglwhp}. These are also called the Tausworthe generators 
\cite{tausworthe}.\index{Tausworthe generator}
 Let $R_1$, $R_2$ $\cdots$ denote a sequence of random binary integers. 
We define a recursion as follows,
\begin{eqnarray}
R_i & = & R_{i-p} + R_{i-1}\  ({\rm mod}\ \  2)\ \ \ q\ <\ p\ <\  i\ .
\end{eqnarray}
It is easily seen that the above addition of binary digits is the same as the 
exclusive-OR  operation (\ .XOR.\ ),\index{exclusive-OR} which can be carried out in a computer rather fast:
\begin{eqnarray}
R_i &=& R_{i-p} \ \ .XOR.\ \  R_{i-q}\ .
\end{eqnarray} 
For proper choice of values of $p$ and $q$ one can 
get a maximal period of $2^{p}-1$. A standard choice proposed by Kirkpatrick and 
Stoll \cite{skeps} consists of $p=250$ and $q=103$.\index{Kirkpatrick}\index{Stoll} This is called 
the R250 generator;\index{R250 generator}
it has a long period
of $2^{250}-1 \approx 1.81\times 10^{75}$. We need to provide 250
seeds  for which we can employ another random number generator {\it e.g.} the linear 
congruential generator with $a=16807$, $b=1$ and $m=2^{31}-1$ as recommended by 
Kirkpatrick and Stoll \cite{skeps}. We shall see more about the R250 generator later on 
in section \ref{r250_generator}.  

There are also several other alternatives that  
have been proposed. These include 
for example the inversive~\cite{IGC} and explicit inversive~\cite{EIGC} generators 
which employ nonlinear recursions\index{recursion}.
I do  not want to go into the details of random number generation and 
instead refer you to~\cite{rngenlit} for some additional  literature on this important topic.  

\section{How do we know that a given sequence of numbers is 
random?}\label{randomness_test}

This is not an easy question to answer, especially when 
we know that the sequence has been generated by a deterministic 
algorithm and hence is predictable and reproducible. 
We take a practical attitude and say that it suffices if we 
establish that  the pseudo random number\index{pseudo random numbers} sequence is indistinguishable
statistically from a sequence of real random numbers. This brings us to tests
for randomness. 

A randomness  test, in a general sense, consists of constructing a function\\ 
$\psi$ $(r_1,$ $r_2,$ $\cdots )$, where $r_1 ,$ $r_2 ,$ $\cdots $, 
are independent variables. 
Calculate the value of this  function for a sequence of  
pseudo random numbers\index{pseudo random numbers} by setting 
$r_i = \xi_i\ \forall \ i=1,\ 2,\ \cdots$.  
Compare this value with the value that $\psi$ 
is expected to have if $\{ r_i : i=1,2,\cdots \}$ 
were truly random and  distributed independently and 
uniformly in the range zero to unity. 

\subsection{Test based on calculation of the mean}\label{test_mean}

For example, the simplest test one can think of, is to set
\begin{equation}
\psi (r_1, r_2, \cdots  )={{1}\over{N}}\sum_{i=1}^{N} r_i ,
\end{equation}
which defines the average of $N$ numbers. 
For a sequence of $N$ truly random numbers,
we expect $\psi$ to lie between $.5-\epsilon$ \& $.5+\epsilon$ 
with a certain probability $p(\epsilon)$. 
Notice that for $N$ large, from the 
Central Limit Theorem\index{Central Limit Theorem},
 $\psi$ is Gaussian\index{Gaussian} with mean $0.5$ and variance\index{variance} $\sigma^{2}_{\psi}
 = \sigma ^{2}_{r}/N$, where $\sigma^2 _r = 1/12$ is the variance\index{variance} of
the random variable $r$, uniformly distributed in the interval $(0,1)$.
If we take $\epsilon = 2\sigma_{\psi}=1/\sqrt{3N}$, 
then $p(\epsilon)$ is the area under the 
Gaussian\index{Gaussian} between $.5-\epsilon$ and $.5+\epsilon$ 
and is equal to $0.9545$. 
$\epsilon$ is called two-sigma 
confidence\index{statistical error!two-sigma confidence} interval. I shall discuss in details the issue of 
confidence interval / statistical error   very soon.  
Thus for a sequence of $N$ truly random numbers, 
we expect its mean to be within $\pm\epsilon$ around 
$0.5$ with $.95$ probability, for large $N$. 
If a sequence of $N$ pseudo random numbers\index{pseudo random numbers} 
has an average that falls outside the interval 
$(0.5-\epsilon ,\  0.5+\epsilon)$ then we say that 
it fails the test at $5\%$ level. 

\subsection{Run test}\label{run_test}

In practice, one employs more sophisticated tests, 
by making $\psi$ a complicated function of its arguments.
An illustrative  example is the  run test\index{run test} which is  
sensitive to the correlations\index{correlations}.  The idea 
is to calculate the length of a run of 
increasing (run-up) or decreasing (run-down) size. 
We say a run-down   length is $l$ if we have a 
sequence of random numbers such that 
$\xi_{l+1} > \xi_l < \xi _{l-1}  < \xi_{l-2} < \cdots < \xi_2 < \xi_1$. 
We can similarly define the  run-up  length.

Let me illustrate the meaning of run-down length 
by considering a sequence of integers between  
$0$ and $9$, given below.
\begin{eqnarray}
\begin{array}{ccccccccccccccccccccccccccc}
 &6  &  &4  &  & 1 &  &3  &  &4 &  &5  &  &3  &  & 2 &  &7  &  &4  & &8 &
\\ &  &  &  &  &  &  &  &  & &  &  &  &  &  &  &  &  &  &  &  &  & \\ | &6
&  &4  &  & 1 &  &3  & | &4 &  &5  & | &3  &  & 2 &  &7  &|  &4  & &8 &|
\\ &  &  &  &  &  &  &  &  & &  &  &  &  &  &  &  &  &  &  &  &  & \\ | &
&  &  &  &  &  &3  & | & &  &1  & | &  &  &  &    &2  &|  &  & & 1 &|
\end{array}\nonumber
\end{eqnarray}
The first row displays the sequence; 
the second  depicts the same sequence with numbers 
separated into groups by vertical lines. 
Each group contains $n+1$ numbers, the first $n$ 
of which are in descending order.  {\it i.e.} 
these numbers are running down. The descent is 
broken by the $(n+1)^{{\rm th}}$ number 
which is greater than the $n^{{\rm th}}$ number. 
The third row  gives the run-down length, for each group, which is $n$.

We can calculate the probability distribution 
of the run-down  length as follows.  
Let $P(L\ge l)$ be the probability that 
the run-down length $L$  is greater than or 
equal to $l$. To calculate $P(L\ge l)$ 
we consider a sequence of $l$ distinct 
random numbers. There are $l!$ ways of 
arranging these numbers. For a sequence of truly 
random numbers, each of these arrangements is 
equally likely. Of these, there is only one 
arrangement which has all the $l$ numbers in 
the descending order.  Hence $P(L\ge l)=1/l!$. 
Since $P(L=l)=P(L\ge l)-P(L\ge l+1)$, we get
\begin{equation}\label{rundown_eq}
P(l) = {{1}\over{l!}} - {{1}\over{(l+1)!}}\quad.
\end{equation}
Alternately,  for the probability of run-down length  we have,
\begin{eqnarray}
P(L\ge l) & = & \int_{0}^{1}d\xi_1\int_{0}^{\xi_1}d\xi_2 \cdots
\int_{0}^{\xi_{l-1}}d\xi_{l}\nonumber\\
 & = & \int_0 ^1 {{\xi_1 ^{l-1} }\over{(l-1)!}} d\xi_1\nonumber\\ & = &
{{1}\over{l!}}
\end{eqnarray}
In the test, we determine  numerically, the 
distribution of the run length on the sequence 
of pseudo random numbers\index{pseudo random numbers} and compare it with the 
exact distribution given by Eq.~(\ref{rundown_eq}).  
Figure~(\ref{RUNDOWN_PS}) depicts the results of a run-down  test\index{run test}. 
\begin{figure}[htp]
\centerline{\psfig{figure=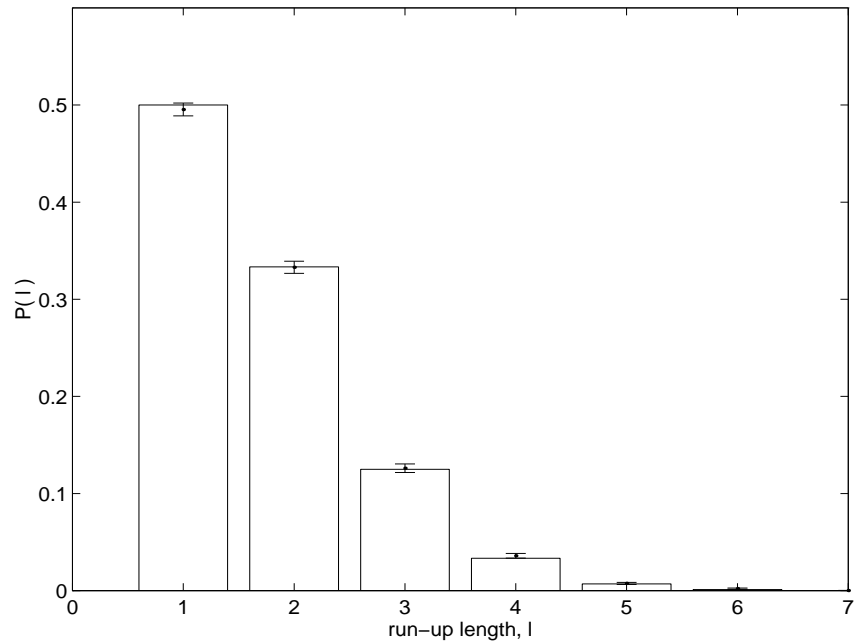,height=08.5cm,width=12.5cm}}
\caption{\protect\small
Results of a run-down test\index{run test}. The histogram\index{histogram} plot 
corresponds to what is expected from a 
sequence of truly random numbers. 
See Eq.~(\ref{rundown_eq}). The points 
with error bars correspond to the ones obtained 
with random number generator routine RAND of MATLAB  version 4.1. 
The run-down lengths were generated from a sequence 
of $10,000$ numbers. The statistical errors\index{statistical error!one-sigma confidence} 
in the calculated values correspond to one-sigma confidence interval.}
\label{RUNDOWN_PS}
\rule{1.0\columnwidth}{0.4pt}
\end{figure}
\subsection{The rise and fall of R250\index{R250 generator}}\label{r250_generator} 

Generation and testing of random numbers is a specialized  area of research. 
Several important pieces of work  have been  carried out in these areas.  
The field remains active even today. The reason for this  is simple: 
There exists a large number of questions that remain unanswered.    
In fact, even the very basic issue  of what is meant by randomness 
of a sequence of numbers, is not clearly understood. 
I shall not discuss  these  and related issues here and  
instead,  refer you  to~\cite{rnlit} for some interesting literature on 
this intriguing subject. For some literature  
on testing of randomness of a pseudo random number\index{pseudo random numbers} sequence  
see~\cite{rntestlit}. The use of pseudo random numbers\index{pseudo random numbers} 
in a Monte Carlo\index{Monte Carlo} 
simulation is perhaps all right for  most of the application; in any case 
we have no other choice; it is always prudent to be watchful;  
the subtle correlations\index{correlations} and patterns present in a sequence of 
pseudo random  numbers\index{pseudo random numbers} may have non 
trivial effects on the results of your 
simulation. 

The (hi)story of the rise and fall of R250 generator 
\index{R250 generator}
should warn us of  how delicate the 
issue of pseudo random number generation is. I have already 
talked about the generalized feedback shift register algorithm  in section \ref{shift_register}.
R250 is an early implementation of the shift register algorithm \cite{skeps}.
It  was introduced in the early nineteen eighties and became an
instant hit amongst the Monte Carlo practitioners. Troubles started in the early
nineteen nineties. In the year 1992 
Ferrenberg,\index{Ferrenberg} 
Landau\index{Landau} and Wong\index{Wong} 
\cite{Ferrenberg_Landau_Wong} reported 
severe problems with the generator in the Monte Carlo simulation of a two dimensional
Ising\index{Ising} model employing Wolff cluster algorithm.\index{Wolff algorithm}
\footnote{We shall discuss the Ising model and 
the Wolff cluster algorithm sometime later; see sections \ref{Ising_model} and \ref{wolff_algorithm} }  The calculated values of energy 
and specific heat\index{specific heat} were erroneous; however, the same algorithm gave correct results when 
other random number generators were employed. These startling findings  were confirmed by
Coddington \cite{Coddington}\index{Coddington} in the year 1994. The culprit was eventually 
traced to the triplet correlations\index{triplet correlations} in R250\index{R250 generator} in particular\cite{ahbdamf} and 
in the shift register\index{shift register} algorithms for certain specific choices of 
the parameters, in general\cite{lnshwjp}.
See also \cite{Others} for more literature on how the correlations present 
in the random numbers would affect the results of the simulations. 
Perhaps one is tempted to agree with  
Compagner\cite{Compagner}\index{Compagner} when, 
tongue in cheek,
 he says that 
\lq\lq\  random sequences are easy to generate in a computer except when needed for 
reliable Monte Carlo simulations\ \rq\rq\  ! But as I said earlier, we have no other choice;
we have to depend on deterministic algorithms for generating random numbers.
The lesson from the above story of the R250 generator\index{R250 generator} is that do not 
use your  \lq  pet\rq\  recursions or algorithms for generating random numbers 
 for your applications.
Follow the advise of Janke \cite{Janke}:\index{Janke} Use only well known generators which have been tested 
and which are being  employed by  a large number of  Monte Carlo practitioners. If such a 
generator poses problems in your specific application, then you can at least 
be assured that the Monte Carlo community would work hard to track down the 
problem and perhaps resolve it, like it did in the case of R250 generator. 
\index{R250 generator}

\section{How do we sample from a given distribution?}\label{random_sampling}

\subsection{Analytical inversion and sampling from exponential}\label{analytical_inversion}

Let $\left\{\xi _i :i=1,\ 2,\ \cdots\ \right\}$  denotes the 
sequence of pseudo random numbers\index{pseudo random numbers}.  
 These are independent,
real and uniformly distributed in the range $(0,1)$.
\index{uniform distribution}   
From $\left\{ \xi_i\right\}$,  we can construct a sequence 
of independent random numbers, $\left\{ x_i \right\}$ 
having the desired distribution, say $f(x)$. Let $F(x)$ denote the
cumulative distribution\index{cumulative distribution} defined as,
\begin{equation}
F(x) = \int _{-\infty}^{x}f(x')dx'\ .
\end{equation}
Then $\{ x_i = F^{-1}(\xi_i):\ i=1,\ 2,\ \cdots \}$ constitute 
an ensemble\index{ensemble} of real numbers,  
whose distribution is the desired $f(x)$. 
For example,  
$$\left\{ x_i = -\lambda \ln (1-\xi _i ) :\  i=1,\ 2,\ \cdots  \ \right\}$$ 
are independent random numbers 
distributed as per the exponential\index{exponential} distribution,
\begin{eqnarray}
f(x)  =\cases {  \lambda^{-1} \exp (-x/\lambda) \   &  for\    $x \ge  0$\ ,  \cr
                                &                     \cr
                   0            & for          $x < 0$\ .\cr}
\end{eqnarray}  

\subsection{ Fern\'andez-Rivero\index{Fern\'andez}\index{Rivero} 
technique for sampling from exponential}\label{f_r_technique}

To sample from an exponential\index{exponential} distribution, Fern\'andez
and Rivero~\cite{FRexp}\index{Fern\'andez}\index{Rivero} proposed a simple algorithm inspired by 
statistical physics.  Start with $N$ particles indexed by integers $i=1,\
2,\ \cdots\ N$.  Initialize $x(k) = \lambda \ \forall\ k$.  
The particles \ \lq interact\rq\  in the following 
fashion. Select independently and randomly 
two particles, say $i$ and $j$ with $i\ne j$.
Let $S=x(i) + x(j)$; split $S$ randomly
into two parts, assign them to the particles $i$ and $j$ and return. 
Repeat the 
above several times until the system {\it equilibrates}. Some $4N$ 
iterations are recommended for {\it equilibration}, see~\cite{FRexp}.  
After the system has {\it equilibrated}, every time you 
select two particles, say $l$ and $k$ and $l\ne k$,   for interaction, you
have $x(k)$ and $x(l)$ as two independent random numbers distributed 
exponentially\index{exponential}.  For more  details see~\cite{FRexp}.

\subsection{Sampling from Gaussian}\label{sampling_gaussian}

\subsubsection{Fern\'andez-Criado technique}\index{Fern\'andez}\index{Criado}\label{f_c}

In the same spirit as in section (\ref{f_r_technique}),
set $x^{(0)}(i)=1\ \forall\  i$ and let the 
randomly chosen particles $i$ and $j$ with $i\ne j$ interact
 in the following fashion, 
\begin{eqnarray}
x^{(k+1)} (i) & =& \eta ^{(k)}\  {{x^{(k)} (i)+x^{(k)} (j)}\over{\sqrt{2} }}\ ,\nonumber\\
x^{(k+1)} (j) & = & \eta ^{(k)}\  {{ x^{(k)} (i)-x^{(k)} (j)}\over{ \sqrt{2} }} \ ,
\end{eqnarray}
where the superscript is the iteration index and 
$\{ \eta ^{(k)} : k=0,1,\cdots\}$ are identically distributed independent
random variables taking values $\pm 1$ with equal probabilities.  
 The \ \lq equilibrium\rq\  distribution of $x$ 
is Gaussian\index{Gaussian} with mean zero and 
variance\index{variance} unity, see~\cite{FCgaussian}.

\subsubsection{Box-M\"uller algorithm}\label{Box_Muller}   
There is another ingenious way of  generating 
Gaussian\index{Gaussian} random numbers
called the Box-M\"uller\index{Box-M\"uller algorithm}  
method~\cite{Box_Muller}. 
The Gaussian distribution\index{Gaussian} of zero 
mean and unit variance\index{variance} is 
given by,
\begin{eqnarray}
f(x) & = & {{1}\over{\sqrt{2\pi}}}\exp\left( - {{x^2}\over{2}}\right).
\end{eqnarray}
The corresponding  cumulative distribution\index{cumulative distribution}
 is given by
\begin{eqnarray}
F(x) & = & {{1}\over{\sqrt{2\pi}}}\int_{-\infty}^{x} dy 
                           \exp \left( - {{y^2}\over{2}}\right) \nonumber\\
     &   & \nonumber\\
     & = & {{1}\over{2}}\left[ 1+ {\rm erf} \left( {{x}\over{\sqrt{2}}}\right)\right]
\end{eqnarray}
where ${\rm erf} (\cdot)$ denotes the error function\index{error function}, 
which is not analytically invertible. Hence 
we consider a two dimensional  distribution obtained as  the product of 
two identical Gaussians\index{Gaussian} of zero mean and 
unit variance\index{variance} given by,
\begin{eqnarray}
f_2 (x,y) & = & f_1(x)f_1 (y) = {{1}\over{2\pi}}\exp \left( -{{x^2 + y^2}\over{2}}\right)
\end{eqnarray}
Let us express the above in terms of polar coordinates $r$ 
and $\theta$ with the transformations
defined by: $x=r\cos (\theta)$ and $y=r\sin (\theta)$. We get
\begin{eqnarray}
f_2 (x,y)dx dy & = \exp\left( - {{r^2}\over{2}}\right) r dr 
{{d\theta}\over{2\pi}}\ ,
\end{eqnarray}
which suggests that the angle $\theta$ is distributed 
uniformly between $0$ and $2\pi$ and 
$r$ can be sampled by analytical  
inversion\index{analytical inversion}  since,
\begin{eqnarray}
F(r) &=& \int_0 ^r dr_1 r_1 \exp\left( -{{r_1 ^2}\over{2}}\right)\nonumber\\ 
& = & 1-\exp\left( -r^2 /2\right)
\end{eqnarray}
Thus we get the Box-M\"uller\index{Box-M\"uller algorithm}  
sampling: from two  
random numbers $\xi_1$ and $\xi_2$, independently and 
uniformly distributed in the 
unit interval $(0,1)$ we get two independent 
Gaussian\index{Gaussian} random numbers,
\begin{eqnarray}
q_1 & =& \sqrt{-2\ln \xi_1} \cos (2\pi\xi_2)\nonumber\\
q_2 & = & \sqrt{-2\ln \xi_1}\sin (2\pi\xi_2).
\end{eqnarray}
For Gaussian  there is an algorithm directly based on the 
Central Limit Theorem. 

\subsubsection{Algorithm based on the Central Limit Theorem}\label{algorithm_clt}

According to the Central Limit Theorem\index{Central Limit Theorem}, 
the sum of  $N$ independent 
random variables
each with finite variance, tends to a Gaussian\index{Gaussian}  in the limit
$N\to\infty$. Let $y$ be the sum of $N$ random numbers (independently sampled 
from  uniform distribution\index{uniform distribution}
 in the interval $0$ to $1$). The distribution
of $y$ tends to a Gaussian\index{Gaussian}
 with mean $N/2$ and variance\index{variance} $N/12$ as $N\to\infty$.
Define
\begin{eqnarray}
x & = & \left( y-{{N}\over{2}}\right) \sqrt{ {{12}\over{N}}}.
\end{eqnarray}
 The random variable $x$ has  mean zero and unit variance. The value of $x$ is confined 
between $-\sqrt{3N}$ and $+\sqrt{3N}$ and hence $x$ tends to a 
Gaussian strictly in the limit  
of $N\to\infty$.  But as shown in \cite{abramowitz}, even a  convenient 
choice of $N=12$ gives good Gaussian random numbers 
with errors not exceeding one percent or so inside a two sigma 
interval $(-2,+2)$.

A very large number of transformations, tricks and algorithms have been 
discovered for random sampling from non uniform distributions. In fact this 
activity continues to be a favorite pastime  of
the  Monte Carlo\index{Monte Carlo} practitioners. 
Standard texts on Monte Carlo\index{Monte Carlo} 
methods~\cite{mcref1,mcref2,mcref3,mcref4} 
contain detailed descriptions of several random sampling techniques. 

\section{How do we evaluate an integral by Monte Carlo\index{Monte Carlo}\\ method?}\label{MC_integral}

For purpose of illustrating the  technique of Monte Carlo\index{Monte Carlo}, 
let us consider a simple problem of evaluating the following integral, 
\begin{eqnarray}\label{integral}
I&=&\int_{a}^{b} \Phi (x) dx.
\end{eqnarray}
In a finite difference\index{finite difference} approximation, we divide the range $(a,b)$ 
of $x$ into $N$ equal intervals.  Let $x_i$ denote the mid-point 
of the $i^{ {\rm th}}$ interval. The integral is approximated by the sum,
\begin{eqnarray}\label{int} 
I_N & \approx & {{b-a}\over{N}}\sum_{i=1}^{N}\Phi (x_i),\\
  &         &                                     \nonumber\\
 x_i & = & a+ \left( i - {{1}\over{2}}\right) {{b-a}\over{N}} .  
\end{eqnarray}
In the above, instead of choosing $x_i$ at regular intervals, we can
 choose them randomly and independently from a uniform distribution\index{uniform distribution} in the 
interval $a$ to $b$. In other words, we set,
\begin{eqnarray}
x_i &=&a+(b-a)\ \xi_i \ : i=1,\ 2,\ \cdots\ ,\ N
\end{eqnarray}
where $\{ \xi_i \}$ are independent
random numbers uniformly distributed in the range $(0,1)$,
and carry out the sum in Eq.~(\ref{int}). 
Let ${\overline I}_N$ denote the 
Monte Carlo\index{Monte Carlo!estimate} estimate of the integral, obtained from a sample of 
size $N$. It is intuitively clear that  
as $N\to\infty$  the estimate ${\overline I}_N \to I$. 
How large should $N$ be so 
that ${\overline I}_N$ is a reliable estimate
of $I$ ?  Before we answer this question let us consider 
the evaluation of the mean of a function $h(x)$ of a random variable $X$.

Formally we have,
\begin{equation}\label{analogue_score}
\left\langle h\right\rangle _f =\int_{-\infty}^{+\infty} h(x)f(x)dx,
\end{equation}
where $f(x)$ is the probability density function of the random
variable $X$. $h(x)$ is usually called the score function.\index{score function}
How do we estimate $\left\langle h\right\rangle _f$~?  
Let us first consider the so called 
 analogue simulation.\index{analogue Monte Carlo}
We sample randomly a sequence
of $\{ x_i\  :i=1,2,\cdots ,N\}$, from the density $f(x)$ and write,
\begin{equation}
\overline{h}_N = {{1}\over{N}}\sum_{i=1}^{N} h(x_i) .
\end{equation}
In the limit $N\to\infty$, $\overline{h}_N \to\langle f\rangle _h $. Also by the
Central Limit Theorem\index{Central Limit Theorem}, in the limit $N\to\infty$,
the probability density of the random variable $\overline{h}_N$ tends to a
Gaussian\index{Gaussian} with mean $\langle h\rangle _f$ and variance\index{variance} $\sigma^2 /N$, where
\begin{equation}
\sigma^2 = \int_{-\infty}^{+\infty}
         \left[ h(x)-\left\langle h \right\rangle _f\right]^2 f(x)dx .
\end{equation}
Thus we say that  analogue Monte Carlo\index{Monte Carlo!estimate}\index{analogue Monte Carlo} estimate of $\langle h\rangle _f$
is given by $\overline{h}_N \pm \sigma/\sqrt{N}$,
where $\pm\sigma/\sqrt{N}$ defines the
   one-sigma
confidence\index{statistical error!one-sigma confidence} interval. This means that with a probability $p$ given by,
\begin{eqnarray}
 p & = & {{\sqrt{N} }\over{\sigma\sqrt{2\pi} } }
\int_{ \left\langle h \right\rangle _f 
 -(\sigma/\sqrt{N})}^{\left\langle h \right\rangle _f 
 +(\sigma/\sqrt{N})} \exp\left[
-{{N(x-\left\langle h \right\rangle _f)^2}\over{2\sigma^2}}\right] dx\nonumber\\
\  & \  \nonumber\\
& = & {{1}\over{\sqrt{2\pi}}} \int_{-1}^{+1}\exp
           \left[ -{{y^2}\over{2}}\right] dy\nonumber\\
\ & \  \nonumber \\
& = & 0.68268
\end{eqnarray}
we expect $\overline{h}_N$ to lie within $\pm \sigma/\sqrt{N}$
around $\langle h\rangle _f$, if $N$ is sufficiently large.
First we notice that we do not know $\sigma$.
Hence we approximate it by its Monte Carlo\index{Monte Carlo!estimate} 
estimate $S_N$, given by,
\begin{equation}
S^2 _N = {{1}\over{N}}\sum_{i=1}^{N}h^2 (x_i ) -\left[ {{1}\over{N}}
\sum_{i=1}^{N} h(x_i )\right] ^2 .
\end{equation}
The quantity $\pm S_N /\sqrt{N}$ is called the  statistical error\index{statistical error!one-sigma confidence}.
Notice that the sample size $N$ must be large for the above estimate
of the error (and of course of the mean) to be reliable.    
 
Now, returning to the question of the reliability of 
${\overline I}_N$ as an estimate of the integral $I$ given by  Eq.~(\ref{integral}),
we immediately see that the associated statistical error\index{statistical error!one-sigma confidence} is
$\pm S_N /\sqrt{N}$, where $S^{2}_{N} = J_N - [ {\overline I}_{N}]^2$, and

\begin{equation}
J_N = {{b-a}\over{N}}\sum_{i=1}^{N} \left[ \Phi (x_i)\right] ^2 \ .
\end{equation}
The statistical error\index{statistical error} is independent 
of the dimensionality of the integral.  The error in a 
deterministic algorithm, on the other hand,  
is  proportional  
to $N^{-1/d}$, where $d$ is the dimensionality 
of the integral. 
Thus, Monte Carlo\index{Monte Carlo} method 
becomes numerically advantageous, for $d \ge 3$. 

It is clear from the above that the 
statistical error\index{statistical error} is directly proportional
to $\sigma$ and inversely proportional to $\sqrt{N}$. The computer time 
however is directly proportional to $N$. If the problem on hand  has 
inherently large $\sigma$, then for calculating  averages within desired (small) statistical
error bars, we shall need a very large sample of microstates; generating a large sample 
is often not possible within meaningful computer time. Then analogue simulation\index{analogue Monte Carlo}
is going to be extremely difficult and most  often impossible. 
We need to resort to 
techniques that reduce the variance\index{variance}
without in any way changing the averages of the desired quantities. These are called
 variance reduction\index{variance reduction} techniques and in what follows I shall
describe one of them  called  
 importance sampling\index{importance sampling}.

\section{What is the basic principle of importance sampling\index{importance sampling}?}
\label{importance_sampling}
Importance sampling helps us sample from the important
regions of the sample space\index{sample space}.  Consider the problem
described in section \ref{MC_integral}; see Eq. (\ref{analogue_score}).
 Let $f(x)$ be high where the 
score $h(x)$ is low and $f(x)$ be low where $h(x)$ is high. 
Then, high score regions of $x$ shall get very poorly sampled. 
A finite sample Monte Carlo\index{Monte Carlo!estimate} estimate of $\langle h\rangle _f$ would 
suffer from poor statistics and would  often be erroneous. This is a 
typical situation whence importance sampling\index{importance sampling} becomes imperative. 

Let $g(x)$ denote an importance density\index{importance density}. 
We decide to sample randomly and independently from
the importance density\index{importance density} instead of $f(x)$. 
Define a modified score function\index{score function}
$H(x)$ as
\begin{equation}
H(x)=h(x){{f(x)}\over{g(x)}}.
\end{equation}
The expectation of $H(x)$ over the importance density\index{importance density}  $g(x)$ is identically
equal to the expectation of $h(x)$ over $f(x)$:
\begin{eqnarray}
\left\langle H\right\rangle _g &=&\int_{-\infty}^{+\infty} H(x)g(x) dx\nonumber\\
\ & &\ \nonumber\\
\      &=&\int_{-\infty}^{+\infty} h(x) {{f(x)}\over{g(x)}} g(x) dx\nonumber\\
\ & &\ \nonumber\\
\      &=&\int_{-\infty}^{+\infty} h(x)f(x)dx\nonumber\\
\ & &\ \nonumber\\
\      &=&\left\langle h\right\rangle _f .
\end{eqnarray}
 
Thus we sample $\Omega_g = \{ x_i :i=1,2,\cdots ,N\}$ randomly and
 independently
from the
 importance density\index{importance density}  $g(x)$.
For each $x_i \in \Omega_g$,
calculate the unweighting - reweighting\index{reweighting} factor $f(x_i)/g(x_i)$. Monte 
Carlo estimate of $\langle H\rangle _g$ is given by,
\begin{equation}
\overline{H}_N={{1}\over{N}}\sum_{i=1}^{N} H(x_i) = 
    {{1}\over{N}}\sum_{i=1}^{N}h(x_i ){{f(x_i )}\over{g(x_i )}}\ \ \  
    {}^{\ \ \ \sim}_{N\to\infty}\ \  \left\langle H\right\rangle _g
\end{equation}
Notice that as $N\to \infty$, 
$\overline{H}_N \to \langle H\rangle _g \equiv \langle h\rangle _f$.

Let us now calculate
the statistical error\index{statistical error} associated with $\overline{H}_N$. It is adequate
if we consider the second moment,\index{second moment} since we have formally shown
that the average value  of $h$ over $f$- ensemble is identically 
equal to the average of $H$ over $g$-ensemble\index{ensemble}.  In other words the mean 
is preserved under importance sampling\index{importance sampling}. The 
second moment\index{second moment} of the $H$, evaluated over the $g$-ensemble\index{ensemble}  
is given by, 
\begin{eqnarray}\label{m2bh}
M_{2}^{B} (H) &=& \int_{-\infty}^{+\infty} H^2 (x)g(x) dx\ ,\ \nonumber\\
\ & &\ \nonumber\\
\             &=& \int_{-\infty}^{+\infty}
                   {{h(x) f(x)}\over{g(x)}} {{h(x)f(x)}\over{g(x)}} g(x)
dx\ ,\ \nonumber\\
\ & &\ \nonumber\\
\             &=&\int_{-\infty}^{+\infty}
                    \left[ {{f(x)}\over{g(x)}}\right]  h^2 (x) f(x) dx\  .
\end{eqnarray}
On the other hand the second moment\index{second moment} of $h$ over the $f$-ensemble\index{ensemble}  is given by,
\begin{equation}
M_{2}^{A} (h)=\int_{-\infty}^{+\infty} h^2 (x)f(x) dx .
\end{equation}
We find $M^B _2 (H) \ne M^A _2 (h)$. If we choose 
the importance density\index{importance density} function $g(x)$ properly,
we can make
$M_{2}^{B} (H)$ to be much less than $M_{2}^{A}(h)$. In other words 
we can get substantial variance reduction\index{variance reduction} under importance sampling\index{importance sampling}.
The averages estimated employing importance sampling\index{importance sampling} will be 
statistically more reliable than the ones calculated by sampling from
$f(x)$.   This in essence 
is the principle of importance sampling\index{importance sampling}.

\subsection{An illustration of importance sampling\index{importance sampling}}
\label{importance_sampling_illustration}
A simple problem would drive home this important  point.   
Let $f(x) = \exp(-x)$
defined for $x\ge 0$. Consider a score function\index{score function} defined as,
\begin{eqnarray}
h(x)=\cases{   0\   {\rm for} \   & $x < T$   \cr
                           &   \        \cr  
              1 \ {\rm }{\rm for} \ & $ x \ge T$\  .    \cr }
\end{eqnarray}
These two functions are depicted in Fig.~(\ref{IMP_PS}a) for $T=5$. Notice that 
the score function\index{score function} $h(x)$ is zero for $x < T$ - a region of $x$ where 
the probability is very high.  Hence most of the values of $x$ sampled 
from the density $f(x)$ will all be in this region and the corresponding 
scores are 
zero. On the other hand, the score function\index{score function}  $h(x)$ is unity for $x > T$ -
a region of $x$ where the probability is negligibly small. This high score 
region is going to be scarcely sampled; often this region is almost never represented 
in a finite sample. 
Let us consider an importance density\index{importance density} $g(b,x)=b\exp(-bx)$ 
defined for $x \ge 0$ and $0 <  b \le 1$ where $b$ is
a parameter to be optimized for minimum variance\index{variance}.   
Fig.~(\ref{IMP_PS}b)  depicts the importance density\index{importance density}  and 
the modified score function\index{score function} 
$H(x)=h(x)f(x)/g(b,x)$ for $T=5$ and $b=\hat{b}=0.18$.
\begin{figure}[bhp]
\centerline{\psfig{figure=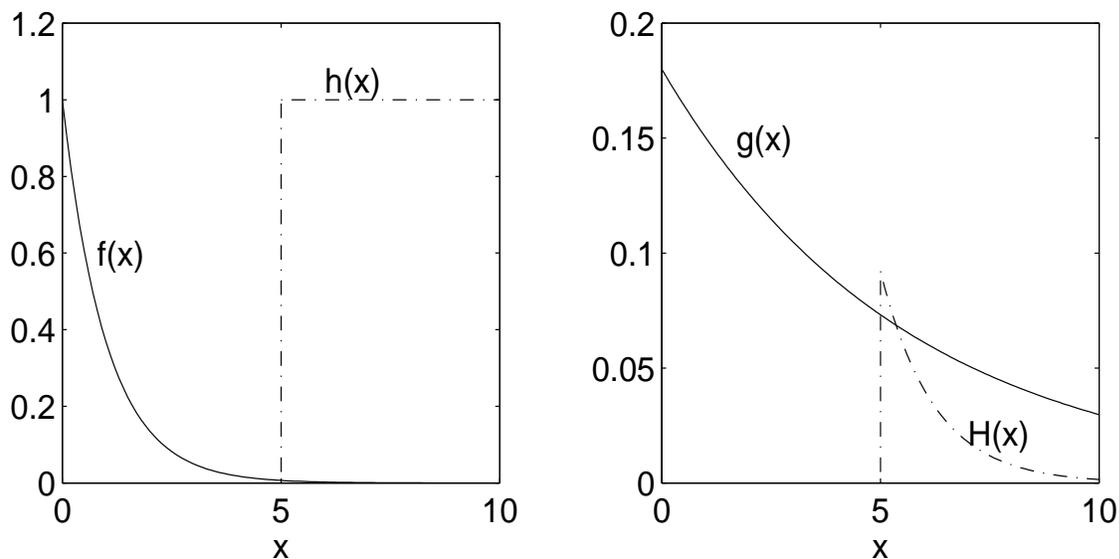,height=7.6cm,width=15.0cm}}
\caption{\protect\small  (a) The probability density $f(x)$ and the score
function $h(x)$ for $T=5$. The score function\index{score function} $h(x)$ is zero for $x < 5$, where 
the probability is high; $h(x) = 1$ for $x>5$ where the probability is negligibly 
small.  (b) The importance density\index{importance density} function
$g(\hat{b}, x)$ 
and the modified score function\index{score function} $H(x)=
h(x)f(x)/g({\hat b}, x)$,
where $\hat{b} = 0.18$ . 
 }
\label{IMP_PS}
\vskip 3mm
\hrule 
\end{figure}
 For this problem all
the statistics can be calculated  analytically. The mean and variance\index{variance}
of $h$ over the $f$-ensemble\index{ensemble} are given by,
\begin{eqnarray}
\left\langle h\right\rangle _f&=& \exp (-T)\nonumber\\
\sigma ^2 (h)               &=& \left\langle h\right\rangle _f\left(
                            1 - \left\langle h\right\rangle _f\right) 
\end{eqnarray}
Table (1)
gives the mean and the relative standard deviation  of $h$ over the 
$f$-ensemble\index{ensemble} for several values of $T$. 
The last column gives the size of the sample required
 in a Monte Carlo\index{Monte Carlo} simulation, to estimate the mean within $\pm10\%$ statistical
 error\index{statistical error!one-sigma confidence} by random sampling from the 
$f$-ensemble\index{ensemble}.
\begin{table}[htp]\label{anasimexact_tab}
\caption{\protect\small  $f$-ensemble\index{ensemble}:  exact analytical results}
\bigskip
\begin{center}
\begin{tabular}{|c|c|c|c|}
\hline
     &                      &                   &                     \\
$T$ & $\left\langle h\right\rangle _f$  & $\sigma (h) /\left\langle h\right\rangle _f$ &  
$N=100\sigma^2 (h)/\left\langle h\right\rangle _f ^2$\\
& &                   &                     \\
\hline
   &                        &                      &  \\ 
$3$  &
$4.98\times 10^{-2}$ & $4.37$            & $1.91\times 10^3   $\\[2mm] 
$5$  &
$6.74\times 10^{-3}$ & $1.21\times 10^1$ & $1.47\times 10^4   $\\[2mm] 
$10$ &
$4.54\times 10^{-5}$ & $1.48\times 10^2$ & $2.20\times 10^6   $\\[2mm] 
$20$ &
$2.06\times 10^{-9}$ & $2.20\times 10^4$ & $4.85\times 10^{10}$\\
 &
&                   &                     \\
\hline
\end{tabular}
\end{center}
\end{table}
We see from 
Table (1)
that as $T$ increases
$\left\langle h\right\rangle _f$ decreases
and the relative fluctuation\index{fluctuations} increases. We find that the sample size  
 required to predict the mean within $\pm 10\%$ statistical
error\index{statistical error!one-sigma confidence} is over $48$ billion for $T=20$, a task which is nearly  
impossible on any computer.

Let us see how the use of importance sampling\index{importance sampling} renders possible
the impossible. The mean of $H$ over the $g$-ensemble\index{ensemble} is identically
equal to the mean of $h$ over the $f$-ensemble\index{ensemble}. However the variance\index{variance} 
of $H$ over $g$-ensemble\index{ensemble} is different and is given by,
\begin{equation}
\sigma ^2 (H,b)= {{ e^{-T(2-b)} }\over{b(2-b)}}-e^{-2T}.
\end{equation}
It is now quite straight forward to calculate the value of $b=\hat{b}$ 
for which
the variance\index{variance} is minimum. We get,
\begin{equation}
\hat{b} = 1+ {{1}\over{T}}-\sqrt{1+{{1}\over{T^2}} }.
\end{equation}
Table (2) presents 
$T, \hat{b}, \sigma (H,b=\hat{b})/\left\langle H\right\rangle _g$ and the 
sample size required to estimate the mean within $\pm 10\%$ statistical error 
\index{statistical error!one-sigma confidence} 
under importance sampling\index{importance sampling} Monte Carlo\index{Monte Carlo}.
We find from 
Table (2)
that  use of importance sampling\index{importance sampling} would lead to a
considerable reduction of variance\index{variance}
especially for large $T$. Take the case with $T=5$.
Use of importance
sampling\index{importance sampling} would reduce the statistical
error\index{statistical error} by a factor five or so.  
As a consequence
a sample of size $650$ is adequate to estimate $\left\langle H\right\rangle _g
= \left\langle h\right\rangle _f$ 
whereas from the $f$-ensemble\index{ensemble} 
we would require to  sample  $14,700$ realizations  to
estimate the average within $\pm 10\%$ statistical error\index{statistical error!one-sigma confidence}. 
The results for $T=20$  are more dramatic
and bring home the need and power of importance sampling\index{importance sampling}. The statistical
error\index{statistical error} gets reduced by a factor of
$4247$ when we employ importance sampling\index{importance sampling}.  
We need only a sample of size $2687$ from the $g$-ensemble\index{ensemble} 
to estimate the mean
within $\pm 10\%$ of the exact, as compared to the sample of size $48$ billion 
from the $f$-ensemble\index{ensemble}. 
\begin{table}[h]\label{bse_tab}
\caption{\protect\small  Importance sampling\index{importance sampling}: exact analytical results}
\bigskip
\begin{center}
\begin{tabular}{|c|c|c|c|}
\hline
  &         &               &  \\
\ \ $T$\ \  & \ \ $\hat{b}$\ \  &\ \  $\sigma (H,\hat{b})/\left\langle H\right\rangle _g$ \ \  &\ \  
$N= 100\sigma^2 (H, \hat{b}) /\left\langle H\right\rangle _g ^2$\ \ \\
  &         &               &  \\
\hline
    &           &                 &      \\ 
$\ 3$ & $.28$     & $1.95$
& $\ 381$\\[2mm] 
$\ 5$ & $.18$ & $2.55$ & $\ 651$\\[2mm] 
$10$ & $.09$ & $3.65$ &
$1329$\\[2mm] $20$ & $.048$& $5.18$ & $2687$ \\ 
&        &        &       \\
\hline
\end{tabular}
\end{center}
\end{table}
 
Let us see below if the above conclusions based on analytical 
theories are borne out by Monte 
Carlo simulations. We have sampled explicitly $10,000$ realizations 
from the importance density\index{importance density} 
$\hat{b}\exp (-\hat{b}x)$, 
employing analytical inversion\index{analytical inversion}  technique, described earlier, and
calculate $\left\langle H\right\rangle _g$ as,
\begin{equation}
\bar{H}_N = {{1}\over{N}}\sum_{i=1}^N H(x_i) \approx \left\langle H\right\rangle_g {\rm and} N=10,000,
\end{equation}
where,
\begin{equation}
H(x_i)=\left\{ \begin{array}{lll} {{1}\over{\hat{b}}}\exp \left[ -\left(
1-\hat{b}\right) x_i\right] , \ & {\rm if }& x_i\ \ge \ T,\\ &
\\ 0,                    & {\rm if} & x_i\  < \  T.  \end{array}\right.
\end{equation}
The statistical error\index{statistical error!one-sigma confidence} is calculated as,
\begin{equation}
\Delta \bar{H}_N = \pm {{1}\over{\sqrt{N}}}\sqrt{
                      {{1}\over{N}}\sum_{i=1}^{N}H^2 (x_i)-\bar{H}_N ^2 }.
\end{equation}
Table (3)
gives the estimated mean
$\bar{H}_N$, the relative statistical error\index{statistical error!one-sigma confidence}, and the actual
deviation of the Monte Carlo\index{Monte Carlo!estimate} estimate from the exact value
$\left\langle h\right\rangle _f$ .
We observe from 
Table (3)
that we are able to make a very good estimate of the desired average 
employing importance sampling\index{importance sampling}.
The corresponding results obtained by random sampling from the density $f(x)$ 
\begin{table}[h]\label{tenthousand_tab}
\caption{\protect\small  Results of Monte Carlo\index{Monte Carlo} sampling  of $10,000$ realizations
from the $g$-ensemble}\index{ensemble}  
\bigskip
\begin{center}
\begin{tabular}{|c|c|c|c|}
\hline
   &        &     & \\ $T$&$\bar{H}_N$ & ${{\Delta
\bar{H}_N}\over{\bar{H}_N}}\times 100$ & $ {{\bar{H}_N
-\left\langle h\right\rangle _f }\over{\left\langle h\right\rangle _f }}\times 100$\\ &                             &
&       \\
\hline
   &                                  &                & \\[2mm] $3$   &
$4.94\times 10^{-2}$         &   $\pm 2.0\%$  &$ -0.8\%$\\[2mm] $5$   &
$6.76\times 10^{-3}$         &   $\pm 2.6\%$  &$ +0.3\%$ \\[2mm] $10$  &
$4.52\times 10^{-5}$         &   $\pm 3.7\%$  & $-0.4\%$\\[2mm] $20$  &
$1.96\times 10^{-9}$         &   $\pm 5.4\%$  & $-4.9\%$\\ &
&                & \\
\hline
\end{tabular}
\end{center}
\end{table}
\begin{table}[ht]\label{fiftythousand_tab}
\caption{\protect\small  Results of Monte 
Carlo  sampling of  $50,000 $ realizations from the $f$ ensemble}
\bigskip
\begin{center}
\begin{tabular}{|c|c|c|c|}
\hline
    &              &                    &\\ $T$ & $\bar{h}_N$ & ${{ \Delta
\bar{h}_N}\over{\left\langle h\right\rangle _f}}\times 100$ & ${{\bar{h}_N-\langle h\rangle _f }\over{\left\langle h\right\rangle _f}}\times
100$\\ &    &                                           &         \\
\hline
    &                        &                 &          \\[2mm] $3$  &
$5.03\times 10^{-2}$   &   $\pm 1.9\%$   &$+ 1.8\% $ \\ $5$  &$6.68\times
10^{-3}$   &   $\pm 5.5\%$   &$-0.9\%$  \\[2mm] $10$ &$6.16\times 10^{-4}$   &
$\pm 57.7\%$  &$+35.7\%$  \\[2mm] $20$ &$\cdot$                 &   $\cdot$
&$\cdot$  \\ &                       &                  &\\
\hline
\end{tabular}
\end{center}
\end{table}
of $50, 000$ realizations (five times more than what we have
considered for sampling from the $g$-ensemble\index{ensemble})
are given in 
Table (4).
We find from 
Table (4)
that Monte carlo simulation based on sampling from the 
$f$-ensemble\index{ensemble} to estimate $\left\langle h\right\rangle _f$ with $T=20$  is impossible.
On the average, we can expect one in $49$ million realizations  sampled from
the exponential\index{exponential} density, to have a value greater than $20$.
The chance of getting a score in a simulation of $50,000$ histories is practically nil.

The key point is that importance sampling\index{importance sampling} helps you pick up
rare  events\index{rare events} - events that have very  low probability of occurrence. 
We need these rare events\index{rare events} in our Monte Carlo\index{Monte Carlo} sample because 
they have high scores; {\it i.e.} we are interested in investigating
a phenomenon that concerns these  rare events\index{rare events}. For example we would like
to have in our Monte Carlo\index{Monte Carlo} sample the rare neutrons from a source that 
manage to penetrate a very thick shield and enter the detector
kept on the other side. We shall employ, for example, an importance sampling\index{importance sampling}
scheme based on  
exponential\index{exponential} biasing~\cite{kahn,kpnm}. In the problem of Monte Carlo\index{Monte Carlo} 
calculation of averages over canonical ensemble\index{ensemble!canonical},  we would like to  
sample microstates with 
high Boltzmann\index{Boltzmann} weight. 
We shall employ the Metropolis\index{Metropolis!sampling} 
importance sampling\index{importance sampling}~\cite{metropolis}.
In the study
of first order phase transition\index{phase transition!first order}, we would be interested in sampling the rare
microstates
that describe interfaces\index{interface}
 between the ordered and disordered phases.
We would resort to  
multicanonical\index{ensemble!multicanonical} sampling~\cite{multicanonicalsampling}. 
In fact we can say that without importance 
sampling\index{importance sampling} techniques, Monte Carlo\index{Monte Carlo} method  
is of rather very little practical use for
simulation of a large variety of problems, be it in statistical physics or
in radiation transport.   

With these preliminaries let us now 
turn our attention to 
sampling from a
canonical ensemble\index{ensemble!canonical} of microstates of a macroscopic system,
employing Monte Carlo\index{Monte Carlo}. I 
shall illustrate this taking ferromagnetism\index{ferromagnetism} as an example.
We shall consider  Ising\index{Ising} spin model of magnetism.

\section{What is an Ising\index{Ising} spin model?}
\label{Ising_model}
Unpaired electron spins 
couple and align. The sum of such tiny magnetic moments  results 
in macroscopic magnetism. 
In the year 1923, Prof. Lenz\index{Lenz} proposed a very 
simple model of magnetism to his student, 
Ising\index{Ising}, who solved it analytically in one dimension. 
Ever since, this model is called the  Ising\index{Ising} model. 
For an historical introduction to  
Ising\index{Ising} model, see~\cite{sgb}. In the Ising\index{Ising}  model, 
a spin has only two states: an {\it up} state and
a {\it down} state. Let us denote the spin variable by the symbol $S_i$, 
and in the model, we say that $S_i $ can take only values of $+1$,
 denoting the {\it up} state ($\uparrow$)
or $-1$, denoting the {\it down} state ($\downarrow$). We organize the spins on  a lattice 
and the index $i$ refers to the lattice site. The lattice can be   
in general $d$ dimensional.
Let $i$ and $j$ refer to two nearest neighbour sites on the lattice and
let $S_i$ and $S_j$ be the spins on these lattice sites.
Energy associated with the pair of nearest neighbour spins is 
given by $\epsilon _{i,j} = -J S_i S_j$.  
Thus when the two spins  are aligned (up or down),
the energy associated with them is $-J$; when not, 
the energy associated with the pair is $+J$.  
The value of  $J$ measures the strength of the 
spin-spin interaction. If $J$ is positive, 
the interaction is ferromagnetic\index{ferromagnetism}. 
If $J$ is negative, the interaction is 
anti-ferromagnetic. We consider $J\ >\ 0$. Figure~(\ref{ising1_eps}) 
depicts two-spins configurations. There are a total of four
possible configurations. The two  configurations marked 
(a) are each
of energy $-J$, since the spins are aligned. 
The two configurations marked (b) are each 
of energy $+J$  since 
the spins are not aligned. 
Consider now Ising\index{Ising} spins on a lattice. Energy associated with a 
spin configuration ${\cal C}$ is given by, 
\begin{eqnarray}\label{isingh}
 E({\cal C} ) & = & 
-J \sum_{\left\langle i,j\right\rangle}
S_i ({\cal C}) S_j ({\cal C})\ , 
\end{eqnarray}
where, the symbol $\left\langle i,j\right\rangle$ 
denotes that the sites $i$ and 
$j$ are nearest neighbours. The sum is taken over all   pairs
of nearest neighbour sites in the lattice.  
We shall concentrate
on ferromagnetism\index{ferromagnetism} and hence $J$ is positive. The energy is lowest when 
all the spins are aligned, either up or down.
The Ising\index{Ising} Hamiltonian\index{Hamiltonian},
see Eq.~(\ref{isingh}), remains the same if all the spins are flipped.
An external magnetic field that couples to all the 
spins, if present, would break this symmetry\index{symmetry} 
of the Hamiltonian\index{Hamiltonian}. Let us consider an external 
field in the {\it up}($\uparrow$) direction. The external field
couples to each of the Ising spin in the system with a strength
$B$. The energy of the system when in the configuration ${\cal C}$
is given by,
\begin{eqnarray}\label{isingh_external_field}
 E({\cal C} ) & = &
-J \sum_{\left\langle i,j\right\rangle}
S_i ({\cal C}) S_j ({\cal C}) - B\sum_i S_i ({\cal C})\ ,
\end{eqnarray}
When the temperature is lowered below a critical value,
even in the absence of an external field ($B=0$), the symmetry is broken.
This is called  spontaneous symmetry breaking. 
\begin{figure}[tp]
\centerline{\psfig{figure=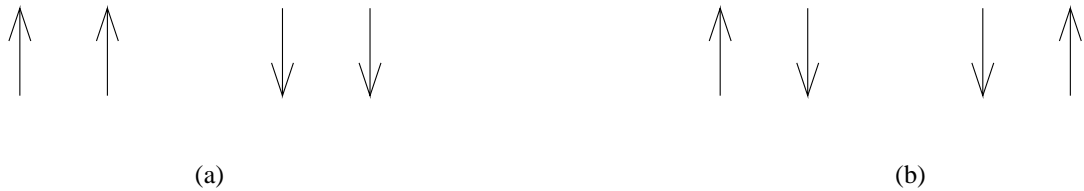,width=150mm}}
\caption{\small\protect{Two-spins configurations: 
                  (a) There are two aligned spin configurations.
                       The energy of each configuration is 
                       $\epsilon_{i,j}  = -J\  S_i\  S_j  =  - J $. 
     (b) There are two configurations in each of which the spins 
           are not aligned.   The energy of each 
                      configuration is  $\epsilon_{i,j} =-J\  S_i\  S_j
                       =  
                       +J$ . }}  
\label{ising1_eps}
\vskip 3mm
\hrule
\end{figure}

It is  clear that  Ising\index{Ising} spin system at
very low temperature, $k_{{\rm B}} T << J$, will have low energy, aligned spins and  
large magnetization. On the other hand, at very high temperature, 
the spin system 
will have high energy, randomly oriented spins and hence no net macroscopic 
magnetization. The system transforms from a disordered  
paramagnetic\index{paramagnetism} to 
an ordered ferromagnetic\index{ferromagnetism} phase, when 
the temperature is lowered below a critical value  called the Curie 
temperature\index{Curie temperature}.

Ising\index{Ising} solved the model in one dimension analytically and showed that there is 
no phase transition\index{phase transition}: Magnetization $M(T)$ decreases 
slowly and continuously as $T$ 
increases. The susceptibility\index{susceptibility} 
$\chi = \partial M /\partial T $ is
finite at all temperature. There is no divergence, either of the 
specific heat\index{specific heat}
or of the susceptibility\index{susceptibility}.  But, under a mean field\index{mean field} approximation, the 
Ising\index{Ising} model exhibits phase transition\index{phase transition} at $T = T_C = \eta J/k_{{\rm B}}$,
where $\eta$ is the coordination number of the lattice. 
\footnote{
The coordination number\index{coordination number} $\eta$ is $2$ for one 
dimensional lattice; $4$ for two dimensional square lattice;
$6$ for two dimensional hexagonal lattice; 
$6$ for three dimensional cubic lattice; {\it etc.} 
Under mean field\index{mean field}, 
the critical temperature does not depend on the dimensionality; it depends
only on $\eta$ - the number of nearest neighbour interactions 
and $J$ - the strength of 
interaction.} 
This raised a serious
doubt whether the statistical mechanics machinery can at all describe   
the phenomenon of phase transition\index{phase transition}. 
After all, one can argue that the phase transition\index{phase transition} 
predicted is an artifact of the 
mean field\index{mean field} approximation. This serious dilemma was settled,
 once and for all, 
 by Onsager\index{Onsager}~\cite{onsager}.
He  solved the two dimensional Ising\index{Ising} model  
exactly; it shows phase transition just like real magnets,
 at finite temperature. 
Onsager\index{Onsager} showed that the  transition 
temperature for two dimensional 
Ising\index{Ising} system is given by,
$k_{{\rm B}} T_c / J = 2/\ln ( 1+\sqrt{2})$. 

\section{How do we simulate an Ising\index{Ising} spin system?}
\label{Ising_model_simulation}

Let $P({\cal C})$ denote 
the probability with which  a microstate  ${\cal C}$ (of an Ising\index{Ising} spin system)
occurs in a closed system\index{closed system} at temperature $T=1/(k_{{\rm B}}\beta)$. 
It is given by, 
\begin{equation}
P({\cal C}) = {{1}\over{Z(\beta)}}  \exp \left[ -\beta E({\cal C})\right]\ ,
\end{equation} 
where the normalization, known as   the   canonical partition 
function\index{partition function!canonical}, is given by
\begin{equation}
Z(\beta)= \sum_{\cal C} \exp \left[-\beta E({\cal C})\right]\ ,
\end{equation}
and  $E({\cal C})$ is the energy of the microstate ${\cal C}$. 
We say that a  Boltzmann\index{Boltzmann} weight given by $\exp [- \beta E( {\cal C})]$ is
associated with the microstate ${\cal C}$.  
Let  
$V$ denote the total number of spins in the system. It is readily seen
that the number of possible spin configurations (microstates) is  
$2^V$. Let us consider a  simple minded 
 approach  to simulate the system; we shall call this analogue Monte Carlo.\index{analogue Monte Carlo}
It consists of sampling 
microstates randomly, independently and with 
equal probabilities. Assemble an ensemble of $N$ equi-probable microstates;
carry out (Boltzmann\index{Boltzmann}) weighted average of 
the macroscopic quantity, say magnetization $M$, over the ensemble as 
given below.   
\begin{eqnarray}
\left\langle M\right\rangle = {}^{ \ Lim.}_{N\to\infty} \overline{M}_N 
&=&
{{ \sum_{i=1} ^N M({\rm C}_i) \exp [-\beta E({\rm C}_i)]}\over{
  \sum_{i=1}^{N} \exp[-
\beta E({\rm C}_i)]}} \ .
\end{eqnarray}
A real magnetic system would  contain an extremely 
large number of spins; {\it i.e.}
$V$ shall be of the order of  
$10^{23}$. Let us consider a modest model system 
with some hundred spins
on a two dimensional $10\times 10$ square lattice. 
The number of spin configurations 
 is $2^{100} \approx 10^{30}$. 
Even if we assume optimistically that it takes a 
nano second  to generate a spin configuration,  
the total time required to
sample all the spin configurations is nearly of the order of
thirty thousand billion years! 
In a Monte Carlo\index{Monte Carlo} simulation, we sample 
only a very small number of spin configurations, 
say  ten thousand or so; with modern high speed computers it may be possible
to generate say a few billion microstates.  Unfortunately, because
of the Boltzmann\index{Boltzmann} weight  $\exp [-\beta E(C)]$, most of the 
spin configurations randomly generated would contribute very negligibly
to the sum. The associated fluctuations shall be very large. 
  Hence we resort to  importance sampling\index{importance sampling}. 

In importance sampling\index{importance sampling}, the idea is to select 
spin configurations randomly
and independently from the distribution $P({\cal C})=
\exp [-\beta E({\cal C})]/Z$. These spin  configurations, when sampled 
adequately
large in  number, constitute, to a very good approximation, a canonical
ensemble. Therefore average of a macroscopic property can be 
calculated as a simple arithmetic mean over the sampled microstates. 
But to carry out this task
we need to know the partition function\index{partition function!canonical},  $Z(T)$. 
But then $Z$ is precisely what we want to calculate in the first place. 
Thus there is a catch.   
But then fortunately there is a way out, as shown by Metropolis\index{Metropolis} 
and his co-workers~\cite{metropolis}.

Construct a  Markov\index{Markov!chain} chain of configurations, starting from an arbitrary
initial configuration ${\rm C}_0$. Let ${\rm C}_0\to {\rm C}_1 
\to \cdots\to {\rm C}_n \to
{\rm C}_{n+1} \to \cdots $ represent a Markov\index{Markov!chain} chain of spin configurations. 
The index $n$ can be viewed as denoting time.  
If we  establish that, for large $n$, the set  
$\{ {\rm C}_{n+1}, {\rm C}_{n+2},\cdots \} $ 
constitutes  an  equilibrium canonical ensemble,
we have done the job.

\section{What is a Markov\index{Markov!chain} chain?}
\label{Markov_chain}

A Markov\index{Markov!chain} chain is   a discrete time 
stochastic process\index{stochastic process}.
Consider the Ising spin system which can be described at any discrete time $n$ as being in 
any one of the microstates (spin configurations) belonging to the closed 
system at given temperature; the microstates are denoted by the script alphabet:
$\{ {\cal C}_1 ,\ {\cal C}_2 , \cdots {\cal C}_{\hat{\Omega}_{{\rm CS} }}\}$. The set of all
microstates are denoted by $\Omega_{{\rm CS}}$ and the number of microstates belonging 
to the closed system\index{closed system} is denoted by $\hat{\Omega}_{{\rm CS} }$. Let $C_n$ be the microstate of the 
system at the discrete time $n$. $C_n$ is  random  and can be any one of the 
microstates belonging to 
$\Omega_{{\rm CS}} = \{ {\cal C}_1 ,{\cal C}_2 , \cdots {\cal C}_{ \hat{\Omega}_{{\rm CS}}}\}$
with certain probabilities. If the system is in equilibrium the probability 
for ${\rm C}_n$ to be ${\cal C}_i$ is given by $P({\rm C}_n = {\cal C}_i) = exp[-\beta E( {\cal C}_i)]/Z$, 
and is independent of the time index $n$. 

The system present in microstate ${\rm C}_n\in \Omega_{{\rm CS}}$ at time 
$n$ makes a transition to microstate ${\rm C}_{n+1}\in\Omega_{{\rm CS}}$ at time $n+1$, according 
to certain probabilistic rules. The discrete time stochastic evolution of the system 
starting from an initial state ${\rm C}_0\in\Omega_{{\rm CS}}$ at time zero until time $n$ 
is completely specified by the joint probability $P({\rm C}_0, {\rm C}_1, \cdots {\rm C}_n)$; this joint 
probability can be expressed in terms of conditional probabilities
\footnote{Let $B$ be an event with non-zero probability, $P(B)\ >\ 0$. Consider an event $A$. We define
the conditional probability of $A$ assuming $B$ has occurred as $P(A\vert B)=P(A\cap B)/P(B)$. In words, 
$P(A\vert B)$ equals the probability of that part of $A$ included in $B$ divided by the probability of $B$},  
see page 33 and 34  of A. Papoulis\index{Papoulis} listed in \cite{cltref}  as, 
\begin{eqnarray}
P({\rm C}_n,{\rm C}_{n-1},\cdots {\rm C}_1, {\rm C}_0)  = 
                                            &    &  P({\rm C}_n\vert {\rm C}_{n-1},{\rm C}_{n-2},\cdots {\rm C}_1, {\rm C}_0)\times\nonumber\\
                                            &    &  P({\rm C}_{n-1}\vert{\rm C}_{n-2},{\rm C}_{n-3},\cdots,{\rm C}_1, {\rm C}_0)\times\nonumber\\
                                             &    &  P({\rm C}_{n-2}\vert{\rm C}_{n-3},{\rm C}_{n-4},\cdots,{\rm C}_1, {\rm C}_0)\times\nonumber\\
                                          &   &   \cdots\cdots\cdots\cdots\cdots\cdots\times \nonumber\\ 
                                            &   &  P({\rm C}_2 \vert {\rm C}_1, {\rm C}_0)\times\nonumber\\
                                             &   &  P({\rm C}_1\vert {\rm C}_0)\times \nonumber\\ 
                                             & \ &  P({\rm C}_0 )
\end{eqnarray}
The sequence of random variables ${\rm C}_0, {\rm C}_1, \cdots {\rm C}_n$ constitutes 
a Markov chain, if 
\begin{eqnarray}
P({\rm C}_k\vert {\rm C}_{k-1},{\rm C}_{k-2}\cdots {\rm C}_1, {\rm C}_0)=P({\rm C}_k\vert {\rm C}_{k-1})\ \forall\  k=1,n\ .
\end{eqnarray}
As a consequence we have
\begin{eqnarray}
P({\rm C}_n, {\rm C}_{n-1}, \cdots {\rm C}_1, {\rm C}_0)&=& 
P({\rm C}_0)P({\rm{\rm C}_1\vert {\rm C}_0)P({\rm C}_2\vert {\rm C}_1)\cdots P({\rm C}_n \vert {\rm C}_{n-1}) }\ .
\end{eqnarray}
Physically the Markovian assumption implies that the microstate the system is going to visit 
at time $n+1$ depends only on the state it is  present now at time $n$ and not where 
it was at all the previous times. 
\footnote{The Markov process\index{Markov!process} is the stochastic 
equivalence of the familiar Newtonian dynamics\index{Newtonian dynamics} }
{\it The past has no influence over the 
future once the present is specified}.
We restrict ourselves to a time invariant or also called a 
time homogeneous Markov chain\index{Markov!chain} 
for which, regardless of the time index $k$, we have,
\begin{eqnarray}
 P({\rm C}_{k+1} = {\cal C}_i \vert {\rm C}_k = {\cal C}_j) & = & W_{i,j}\ \ \ \ \  \ \forall\ k
\end{eqnarray}
where $W_{i,j}$ is the probability for the transition from microstate ${\cal C}_j$ to
microstate ${\cal C}_i$ in a single step. Thus $W_{i,j}$ is the $i,j-$th element of 
the $\hat{\Omega}_{{\rm CS} }\times \hat{\Omega}_{{\rm CS}}$ square matrix. The stochastic
matrix $W$ completely specifies the stochastic dynamical evolution of the system given the 
initial state ${\rm C}_0 $ at time $n=0$ 

Let $P({\cal C}_i ,n)$ denote the 
probability that the Ising\index{Ising} spin system is in a 
configuration ${\cal C}_i$ at time
$n$. In other words $P({\cal C}_i ,n)$ is the probability that 
${\rm C}_n = {\cal C}_i$. Formally we have the discrete time Master equation,\index{Master equation}
\begin{eqnarray}\label{master_equation}
P( {\cal C}_i, n+1) & = & \sum_{j\ne i} W_{i,j} P({\cal C}_j , n)
               + \left( 1-\sum_{j\ne i} W_{j,i}\right) P({\cal C}_i , n)\nonumber\\
                    & \ & \nonumber\\ 
                    & = & \sum_j \bigg[ W_{i,j}P({\cal C}_j , n)
                                        - W_{j,i}P({\cal C}_i , n)\bigg]
                           + P({\cal C}_i , n)
\end{eqnarray}
We need 
\begin{eqnarray}
P( {\cal C}_i, n+1) &=& P( {\cal C}_i, n)=\pi ({\cal C}_i) = \pi_i\ \  \forall\ \ i\ ,
\end{eqnarray}
when  ($n\to\infty$),  for asymptotic equilibrium. In the above $\pi (C_i) = \pi _i$ is the 
equilibrium probability of the microstate $C _i$. {\it i.e.} $\pi _i = \exp [-\beta E(C_i)]/Z$. 
We can ensure evolution to asymptotic equilibrium by 
demanding that each term in the sum over $j$ in the RHS of
Eq.~(\ref{master_equation}) be zero. Notice that this condition is 
sufficient but not necessary for equilibration. This is called the detailed 
balance condition:
\begin{eqnarray}
W_{i,j}\pi _j & = & W_{j,i}\pi _i\ , 
\end{eqnarray}
which implies that 
\begin{eqnarray}\label{detailed-balance2}
{{W_{i,j}}\over{W_{j,i}}}  & = &
{{\pi _i}\over{\pi _j }}
=\exp\bigg[ -\beta \big\{ E({\cal C}_i)-E({\cal C}_j )\big\}\bigg]
= \exp\left[ - \beta \Delta E\right]  ,
\end{eqnarray}
The important point is that only the ratios of the equilibrium
probabilities appear in the above. These ratios are known; they are just the ratios of the 
Boltzmann weights; we do not need to know the normalization (partition function) for constructing
a  transition matrix, see discussions toward the end of section (\ref{Ising_model_simulation}).  

A task now is to construct an algorithm that takes the system from one microstate 
to the next as per a transition matrix whose elements obey Eq. (\ref{detailed-balance2}).
We have already seen that the number of microstates of a closed system is
$\hat{\Omega}_{{\rm CS}} = 10^{30}$ even for a small system of Ising spins on a 
$10\times 10$ square lattice; the $W$ matrix for this problem  will thus contain some $10^{60}$ elements!
Constructing explicitly such a matrix, storing it, and carrying out operations with it are 
neither feasible nor required. What we need is an algorithm that 
takes the system from one microstate to another which in effect mimics the transition 
induced by the matrix $W$. The Metropolis\index{Metropolis!algorithm} algorithm does this. 

\section{What is the Metropolis Algorithm?}\label{metropolis_algorithm}
The Metropolis algorithm defines a stochastic 
dynamics which generates, 
starting from an arbitrary initial microstate ${\rm C}_0$,
a Markov chain\index{Markov!chain} of microstates, given by,
\begin{eqnarray}\label{markov_chain}
{\rm C}_0 \to   {\rm C}_1 
\to {\rm C}_{2}\to \cdots\to 
{\rm C}_k 
\to  {\rm C}_{k+1}\to\cdots
\end{eqnarray}
Let us see how does the transition: 
${\rm C}_{k} \to {\rm C}_{k+1}$, take place as
per Metropolis algorithm.\index{Metropolis!algorithm}
We select one of the spins 
randomly and with equal probability~($= 1/V$); let $S_i$ denote the 
selected spin which is at lattice site $i$.  
We flip the selected spin ($S_i \to -S_i$) and get a 
trial configuration denoted by ${\rm C}_t$. 
This step  constitutes  the selection part
of the transition matrix $W$.  
Let $\Delta E = E({\rm C}_t) - E({\rm C}_{k})$.
We accept ${\rm C}_t$ as the next microstate 
in the Markov chain\index{Markov!chain} with a probability
given by,
\begin{eqnarray}\label{pmetro}
p & = & {\rm min}\bigg( 1, {{\pi ({\rm C}_t)}
\over{\pi ({\rm C}_{k})}}\bigg)\ ,\nonumber\\
  &   & \nonumber\\
  & = & {\rm min} \bigg( 1,\exp \left[ -\beta \Delta E \right] \bigg)\ .
\end{eqnarray}
In other words,
\begin{eqnarray}
{\rm C}_{k+1} =\left\{ \begin{array}{lll} {\rm C}_t\ &{\rm with\  probability}& p\\
                                                   &                          &   \\
                                       {\rm C}_{i} &  {\rm with\  probability} &  1-p\\
                         \end{array}\right.
\end{eqnarray}
This step constitutes the acceptance / rejection part of the transition $W$. 
Executing these two steps  constitutes 
one Monte Carlo Step (MCS)\index{Monte Carlo!step (MCS)}. It is easily verified
that the Metropolis transition\index{Metropolis!transition} described above obeys the detailed balance 
\index{detailed balance} condition. 

Let me recapitulate the different steps involved in the practical implementation of the 
Metropolis algorithm.\index{Metropolis!algorithm}:
\begin{enumerate}
\item[$\bullet$]
Start with an arbitrary initial microstate ${\rm C}_0 \in \Omega_{{\rm CS}}$ and evolve it :
${\rm C}_0\to {\rm C}_1\to {\rm C}_2 \cdots$.
\item[$\bullet$]
select randomly and with equal probability one of  the spins in the current configuration,
${\rm C}_{k}$. Let $S_i$ denote the selected spin. Flip the spin ($S_i \to -S_i$) and get 
the trial configuration ${\rm C}_t$. Calculate $\Delta E = E({\rm C}_t) - 
E({\rm C}_{k})$.
\item[$\bullet$]
If $\Delta E \le 0$, accept the trial state; ${\rm C}_{k+1} = {\rm C}_t$
\item[$\bullet$]
if $\Delta E > 0$ draw a random number $\xi$
\begin{itemize}
\item[$\circ$]
if $\xi \le \exp (-\beta \Delta E)$, accept the trial state; ${\rm C}_{k+1} = {\rm C}_t$
\item[$\circ$]
if $\xi >  \exp (-\beta \Delta E)$ reject the trial state; ${\rm C}_{k+1} = {\rm C}_{k}$
\footnote{The rejection part of the algorithm is important for realizing  
          equilibrium ensemble; there are however rejection free algorithms;
in these, the information on the dynamics is transferred to the  time axis; 
these are called event 
driven algorithms\index{Monte Carlo!event driven}; the Metropolis algorithm\index{Metropolis!algorithm}  is 
 time driven\index{Monte Carlo!time driven}. We shall  discuss 
an event-driven algorithm called the n-fold way, later. See section (\ref{n-fold_way})}
\end{itemize}
\end{enumerate}
Iterating the above we get a Markov chain of microstates.
In a single Monte Carlo Step (MCS)\index{Monte Carlo!step (MCS)} the system switches from the 
current state ${\rm C}_k$ to the next state ${\rm C}_{k+1}$. A consecutive 
set of $V$ number of   
MCS constitutes 
one Monte Carlo Step per Spin (MCSS).\index{Monte Carlo!step per spin (MCSS)}
The 
time scale for the Metropolis dynamics is set by an MCSS. 
We construct an ensemble of microstates 
from the Markov Chain by picking up microstates at the end of 
every MCSS: $\{ {\rm C}_0 , {\rm C}_V, {\rm C}_{2V}, \cdots \}$. 
Discard configurations at the beginning of the chain;
Then the configurations in the asymptotic part of the chain
constitute  a  canonical ensemble\index{ensemble!canonical}, which can be employed for 
calculating the desired macroscopic properties of the system.

Fig.~(\ref{metro_ps}) depicts the probability $p$ of accepting a 
trial configuration ${\cal C}_t$ produced from the current configuration
${\rm C}_k$, as a function of $x=\beta [ E({\rm C}_t) - E({\rm C}_k)]$. 
It should be remarked that we can devise several other transition matrices 
consistent with detailed balance. For example, in the Metropolis algorithm\index{Metropolis!algorithm},
the derivative of the acceptance probability $p(x)$ is not continuous at
$x= 0$, see the solid  line in Fig.~(\ref{metro_ps}). 
 Hence Glauber\cite{Glauber}\index{Glauber} proposed a dynamics\index{Glauber dynamics} 
where the probability $p$ of accepting a trial state ${\rm C}_t$,  constructed 
from the current state ${\rm C}_k$, is given by
\begin{eqnarray}
p &= & {{\exp\left[ -\beta E({\rm C}_t) \right] }
\over{ \exp\left[ -\beta E({\rm C}_{k})\right] + 
     \exp\left[ -\beta E({\rm C}_t)\right]}}\nonumber\\
  & & \nonumber\\
  & =&  
 {{1}\over{1+\exp\left[\beta\left\{  E( {\rm C}_t)  -E({\rm C}_{k})\right\}\right] }}\nonumber\\
  & & \nonumber\\
  & = & {{1}\over{1+\exp(x)}} \ .
\end{eqnarray}
The above  choice of acceptance probability is also
consistent with the detailed balance\index{detailed balance}. 

\begin{figure}[tp]
\centerline{\psfig{figure=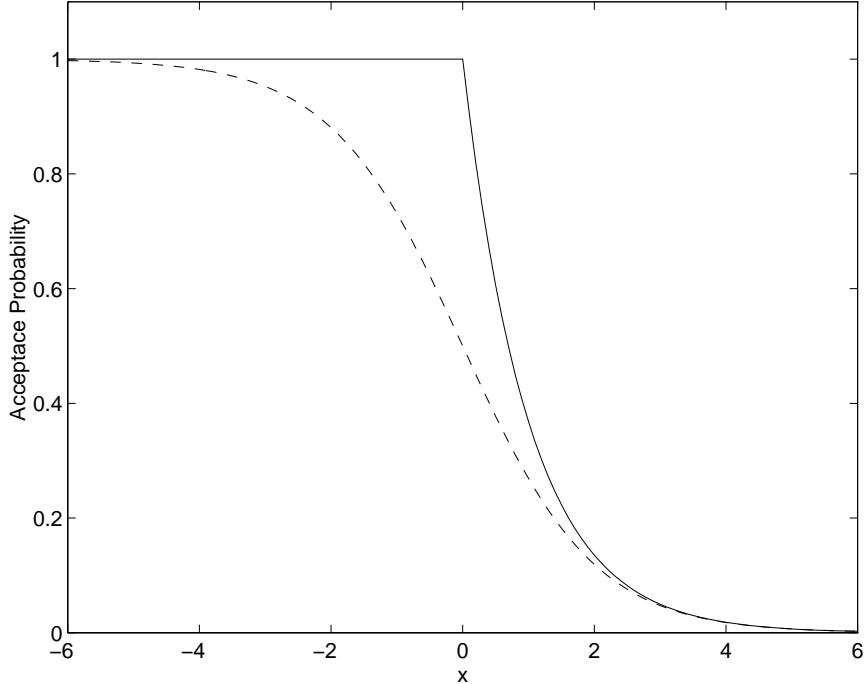,height=09.20cm,width=11.50cm}}
\caption{\small\protect{Acceptance probability $p$.
The solid  line refers to
Metropolis\index{Metropolis!sampling} sampling: $p = min\{ 1,\exp(-x)\}$. The dashed curve refers
 to Glauber dynamics:\index{Glauber dynamics} $p=1/[1+\exp(+x)]$.  In the above
$ x=\beta \Delta E$.}}
\label{metro_ps}
\vskip 3mm
\hrule
\end{figure}

There is another dynamics in vogue called the heat-bath algorithm.  
\index{heat-bath algorithm}
We define $h_i = \sum_j S_j$, where the index $j$ runs over all the nearest 
neighbour spins of $S_i$. We define a probability 
\begin{eqnarray}
p_i = {{ \exp (\beta J h_i)}\over{ \exp (\beta J h_i) + \exp(-\beta J h_i) }}\ .
\end{eqnarray}
The heat-bath algorithm \cite{HHED} consists of updating the spin $S_i (k)$ to $S_i (k+1)$ where 
$S_i (k+1) = \pm 1$ with probabilities $p_i$ and $1-p_i$ respectively. This is implemented
by drawing a random number $\xi$ (uniformly distributed in the interval $0$ to $1$) and 
setting $S_i (k+1) = SIGN (p_i -\xi)$, where  the function 
$SIGN (\eta)$ equals $+1$ if $\eta > 0$ and $-1$ if $\eta < 0$. Thus the spin at site $i$
behaves as if it is in its private heat bath constructed by its nearest  neigbbours!
Let us now cast the Glauber algorithm\index{Glauber dynamics}
 in the language of the heat-bath algorithm employing $p_i$. 

In the Glauber algorithm, if $S_i (k) = +1$, then the probability $S_i (k+1)=-1$ is 
$p=1-p_i$; therefore, $S_i (k+1) = - SIGN (1-p_i -\xi)$. On the other hand 
if $S_i (k) = -1$, then the probability that $S_i (k+1) = +1$ equals 
$p=p_i$ and hence $S_i (k+1) = SIGN (p_i -\xi)$. It is in this sense that 
the Glauber dynamics\index{Glauber dynamics} differs from the heat-bath algorithm.
\index{heat-bath algorithm}  

However, it readily seen that, in the Glauber algorithm,
\begin{eqnarray}
P\bigg( S_i (k+1) = +1 \bigg\vert S_i (k) = +1\bigg) = P\bigg( S_{i} (k+1) = +1\bigg\vert S_i (k) = -1\bigg) = p_i\nonumber\\
P\bigg( S_i (k+1) = -1 \bigg\vert S_i (k) = +1\bigg) = P\bigg( S_{i} (k+1) = -1\bigg\vert S_i (k) = -1\bigg) = 1-p_i\ ,
\end{eqnarray}
and these probabilities are identical to those of the heat-bath algorithm; 
\index{heat-bath algorithm} hence the Glauber 
algorithm \index{Glauber dynamics} differs from the heat-bath algorithm only  in a minor and irrelevant technical 
detail.  
\index{heat-bath algorithm}
 
I think we understand why do the Metropolis transitions take the system to equilibrium
eventually. A system has a natural tendency to lower its energy; hence if  a trial microstate
has  lower energy it is accepted with unit probability. But there are fluctuations present all the 
time; fluctuations are a part and parcel of an equilibrium system. It is precisely 
because of these fluctuations that an equilibrium system manages to remain in equilibrium.
The fluctuations also correspond to measurable physical properties of the macroscopic system,
{\it e.g.} the specific heat corresponds to energy fluctuations.
\footnote{Fluctuation dissipation theorems\index{fluctuation dissipation theorem}
 relate the equilibrium fluctuations 
to  nonequilibrium or more precisely near equilibrium responses.}
Hence when the trial microstate is of higher energy, we still accept it but with a probability
less than unity; larger the energy increase, lower is the acceptance probability, since in equilibrium
larger fluctuations are rarer. 

That the Metropolis algorithm assures asymptotic equilibrium can also be shown rigorously
mathematically.    
The transition matrix $W$,\index{transition matrix} whether based  on the Metropolis dynamics or the 
Glauber/heat-bath  dynamics\index{Glauber dynamics}, has the following properties:
$W_{i,j} \ge 0\ \ \forall\  i,j$ and $\sum_j W_{j,i} =1 \ \forall\ i$. This second property
is just a normalization condition: the system has to be in  any one of the microstates
including the present one, after a transition. 
Using the normalization in Eq.~(\ref{master_equation}), we get,
\begin{eqnarray}
P( {\cal C}_i , n+1) & = & \sum_{j} W_{i,j} P({\cal C}_j , n)\ \ \forall\ i.
\end{eqnarray}
The above can be written in a convenient vector notation
\begin{eqnarray}
\vert P(n+1)\rangle & = & W\vert P(n)\rangle 
\end{eqnarray}
where the components of the column vector\index{vector} $\vert P(n)\rangle$ are the probabilities of the 
microstates of the closed system\index{closed system} at time $n$ and the dimension of the 
vector\index{vector} is equal to
the size of the sample space\index{sample space}. 
Specifying the sample space\index{sample space} and the probabilities of the 
microstates of the sample space uniquely defines an  ensemble\index{ensemble}. Hence we shall call
$\vert P(n)\rangle$ an ensemble\index{ensemble} at time $n$. Thus the transition matrix acts on the 
ensemble $\vert P(n)\rangle$ and transforms it to the ensemble $\vert P(n+1)\rangle$.
Let us start with an arbitrary ensemble $\vert P_0\rangle$, such that 
$\langle \pi \vert P_0\rangle \ne 0$; in other words the initial ensemble 
chosen has a non-zero overlap with the equilibrium ensemble.  
Let $W$ act on $\vert P_0\rangle$ successively some $n$ times and transform it to
$\vert P(n)\rangle$. 
\begin{eqnarray}
W^n \vert P_0\rangle = \vert P (n)\rangle
\end{eqnarray}
We want $\vert P (n)\rangle \to \vert \pi \rangle$ as $n\to\infty$  and any subsequent 
transition should not change the ensemble. For this to happen 
the largest eigenvalue of $W$ must be real, non-degenerate\footnote{the equilibrium corresponds to a 
unique ensemble} and unity;
\footnote{once the system reaches equilibrium it should remain in equilibrium}
all other eigenvalues 
must be smaller than unity in modulus;
also the eigenvectors\index{eigenvector}
 of $W$ must form a complete set.
We can then express the initial ensemble\index{ensemble} in terms of the eigenvectors \index{eigenvector}of $W$. Upon repeated 
operations by $W$, only the eigenvector of the largest 
eigenvalue\index{eigenvalue} 
 survives. Hence equilibrium ensemble is the right eigenvector
of $W$ corresponding to the (largest) eigenvalue unity: 
\begin{eqnarray}
W\vert \pi \rangle & = & 
\vert \pi \rangle \ .
\end{eqnarray}
$\vert \pi\rangle$ is called the equilibrium or the invariant distribution of the 
matrix $W$. 
The Peron - Frobenius theorem\cite{peron,gantmacher}
ensures that 
such an asymptotic convergence to an equilibrium ensemble is possible
if the transition matrix\index{transition matrix}  $W$ has the following 
properties:
\begin{itemize}
\item
$W_{i,j} \ge 0 \ \forall \  i,j$ 
\item
$\sum_{j} W_{j,i} =1 \ \forall \  i$ 
\end{itemize}
A transition matrix $W$ is called \lq\ balanced\ \rq, if it obeys the above two conditions.
In additionn we demand the Markov chain to be ergodic: $(W^m)_{i,j} > 0\ \forall\ i,j\ {\rm and}\ 
m\  <\  \infty$. This ensures that every microstate ($\in \hat{\Omega}_{CS}$) is reachable
from every other microstate ($\in \hat{\Omega}_{CS}$) in a finite number of transition steps. 
If the transition matrix obeys a more restricted detailed balance condition, then the above 
two requirements are automatically fulfilled and hence is \lq\ balanced\ \rq; 
 detailed balance condition is sufficient
though not necessary for asymptotic convergence to equilibrium, see below. 

There are  several Monte Carlo\index{Monte Carlo} 
techniques in vogue, that do not satisfy  detailed 
balance\index{detailed balance} condition,
see {\it e.g.}~\cite{violatedb}, but nevertheless assures 
march to equilibrium.  
As we have seen above, the  (weaker) balance condition  suffices for asymptotic  equilibrium,  
see for  {\it e.g.}~\cite{vimmwd}.
\footnote{ 
In fact we can define a transition matrix $\widehat{W}$ called the 
reversal of $W$ \index{reversal of Markov matrix} \cite{JRNorris} or the $\pi$-dual of $W$ \cite{KSK}. 
It is given by,
\begin{eqnarray}
\widehat{W} & = & {\rm diag} (\pi)W^{\dagger}{\rm diag} (1/\pi)\nonumber
\end{eqnarray}
where diag$(\pi)$ denotes a diagonal matrix whose diagonal
elements are the components of the equilibrium vector 
$\vert \pi \rangle$, and $W^{\dagger}$ denotes the transpose of $W$. Both the 
transition matrices $W$ and $\widehat{W}$ have the same invariant distributions.
$\widehat{W}$ describes the time reversed markov evolution. 
If $W$ obeys the detailed 
balance condition then $\widehat{W} = W$. The matrix $\widehat{W}$ has been found 
useful in several recent studies  in nonequilibrium statistical mechanics
like fluctuation theorems, \index{fluctuation theorems} work-free energy relations, {\it etc.}, see
for {\it e.g.} \cite{Crooks}.}

\section{How do we calculate Monte Carlo\index{Monte Carlo!average} 
averages and error bars?}\label{MC_error_bars} 
 
We calculate  the required macroscopic 
property by averaging over the ensemble\index{ensemble} constructed as  per the 
Metropolis\index{Metropolis!algorithm} rejection technique described earlier.   
For example, the  energy  given by,
\begin{eqnarray}
E({\cal C})&=&-J\sum_{\left\langle i,j\right\rangle} 
S_i ({\cal C})S_j ({\cal C})\ ,
\end{eqnarray}
when the Ising\index{Ising} spin system is in the configuration 
${\cal C}$. Similarly,
the magnetization  in the microstate ${\cal C}$ is given by
\begin{eqnarray}
M({\cal C})&=& \sum_i S_i ({\cal C})\ .
\end{eqnarray} 
The above and any other macroscopic quantity of interest  
can be averaged over an 
ensemble generated by the Monte Carlo\index{Monte Carlo} algorithm. 

Let ${\rm C}_1,\ {\rm C}_2,\ ,\ \cdots {\rm C}_N$ be the $N$ successive 
microstates in the asymptotic part of a Markov chain\index{Markov!chain}
 generated by
the Metropolis algorithm.\index{Metropolis!algorithm} Let $x_i = x({\rm C}_i )$ be the corresponding values
of a macroscopic property of the system. $x$ can be $M,\ E,\ M^2,\ E^2 $
{\it etc.}
An estimate of the  average of $x$   is 
given by
\begin{equation}\label{average_eqn}
\overline{x}_N  = {{1}\over{N}}\sum_{i=1}^{N} x_i,
\end{equation}
which in the limit of $N\to\infty$ gives $\left\langle x\right\rangle$.
We call $\overline{x}_N$ as a finite sample estimate of 
$\left\langle x \right\rangle$. 
We can estimate from the ensemble, all the  macroscopic properties of the 
Ising\index{Ising} spin system like $\left\langle E \right\rangle$,
and $\left\langle M\right\rangle$; Employing fluctuation dissipation theorems
\index{fluctuation dissipation theorem} we can 
estimate specific heat\index{specific heat} given by
\begin{equation}
C_V= {{\partial E}\over{\partial T}}=
{{\left\langle E^2\right\rangle - \left\langle E\right\rangle ^2}\over{k_{{\rm B}} T^2}},
\end{equation}
and the magnetic susceptibility\index{susceptibility} given by
\begin{equation}
\chi
={{\left\langle M^2\right\rangle -\left\langle M\right\rangle ^ 2}\over{
k_{{\rm B}} T}}.
\end{equation} 
The Monte Carlo\index{Monte Carlo!estimate} estimate $\overline{x}_N$ is always associated with statistical
error\index{statistical error} since $N$ is finite. The standard deviation of $\overline{x}_N$ is 
the one-sigma confidence\index{statistical error!one-sigma confidence} 
interval  or the statistical error associated with
$\overline{x}_N$. We express  the statistical error\index{statistical error} 
employing the Central Limit Theorem\index{Central Limit Theorem}:  
we present the Monte Carlo\index{Monte Carlo} result as
\begin{equation}\label{error}
\overline{x}_N  \pm {{ \sigma (x)}\over{ \sqrt{N}}},
\end{equation}
where $\sigma ^2 (x)= \left\langle x^2 \right\rangle - \left\langle
x \right\rangle ^2$. 
The quantity $\sigma  (x)/\sqrt{N}$ is called the statistical 
error\index{statistical error!one-sigma confidence}, or one-sigma confidence interval. The key idea is that 
in the limit $N\to\infty$ the random variable $\overline{x}_N$
becomes Gaussian\index{Gaussian} distributed with mean  
$\left\langle x \right\rangle$ and variance\index{variance} 
$\sigma ^2 (x) /N$. Notice that asymptotically ($N\to\infty$)   
 the statistical error\index{statistical error} goes to zero as $N^{-1/2}$. 
In other words  the distribution of the  sample  
average   tends to a delta function in the 
limit $N\to\infty$.

In calculating the statistical error\index{statistical error!one-sigma confidence}, replace $\sigma ^2 (x)$ by the  
sample fluctuations\index{fluctuations}, denoted by $S^2 _N$ and  given by,
$S^2 _N =  \sum_{i=1}^N (x_i - \overline{x}_N )^2 /(N-1)$ .
Equation~(\ref{error}) for the statistical error\index{statistical error}
holds good only if the Ising\index{Ising} spin configurations sampled are all 
 uncorrelated.  
The configurations generated 
by the  Metropolis\index{Metropolis!algorithm} algorithm 
at successive MCSS\index{Monte Carlo!step per spin (MCSS)} are usually 
correlated.  
Hence the actual statistical error\index{statistical error} would be higher
than what is calculated employing Eq.~(\ref{error}).
I shall have more to say on this later. 

Before we go further and take up more advanced topics there are a few details 
and few issues we should attend to. 
 For example before taking average over the ensemble, we must ensure that 
the Markov\index{Markov!process} process has equilibrated, {\it i.e.,} 
$P({\cal C},t)\to P({\cal C})$, as $t\to\infty$ $\forall\ {\cal C}\ \in\ 
\Omega_{{\rm CS}}$.  
The Ising\index{Ising} system should have forgotten the arbitrary initial spin 
configuration. How do we find this?  First we must check if the average converges to a constant 
value: split the data into several 
bins and  calculate the average  for each bin. Discard the initial bins which 
give averages different from the latter. Another way is to
calculate the autocorrelation function and from it, the  correlation time
$\tau ^{\star}$ in units of MCSS\index{Monte Carlo!step per spin (MCSS)}. 
Discard data from the initial, say $10\ \tau ^{\star}$ 
MCSS\index{Monte Carlo!step per spin (MCSS)}. 

Another problem is related to the issue of ergodicity\index{ergodicity}. 
The system can get trapped
in a region of the configuration phase and not come out of it at all.
It is also possible that 
the dynamics is quasi-ergodic\index{quasi-ergodic}, in the 
sense that the system gets trapped in local minimum  and is unable to 
come out of it in any finite time  due to the presence of 
high energy barriers. 
A possible check to detect these is to carry out 
several simulations starting from different initial spin configurations 
and see if all of them give more or less the same results within 
statistical fluctuations\index{fluctuations}.  

The next question concerns  boundary conditions\index{boundary condition}. 
Periodic boundary conditions\index{boundary condition!periodic}
are often  employed: the lattice wraps around on itself to form a torus. 
Consider a two dimensional square lattice of sites. The first 
site in a row is considered as the right  nearest neighbour of the last site in the 
same row and the last site in a row is considered as the left nearest neighbour
of the first site in the same row.  The same holds for the 
top and bottom sites in each column. 
Periodic boundaries are known to give  least 
finite size effects\index{finite size effect}; we shall see about
finite size effects sometime later.
There are also several other
boundary conditions {\it e.g.}
rigid, skew periodic\index{boundary condition!skew periodic}, 
anti periodic,\index{boundary condition!anti periodic}
 anti symmetric\index{boundary condition!anti symmetric}, 
free edge\index{boundary condition!free edge}  {\it etc.,}
that are  implemented depending on the nature of the 
problem. 

\section{How does an Ising\index{Ising} spin system behave in the vicinity 
of phase transition\index{phase transition}?}\label{Ising_phase_transition}

To distinguish one phase from the other we need an 
order parameter\index{order parameter};  magnetization
$M$ is a good order parameter to study phase transition\index{phase transition} in 
ferromagnetic systems\index{ferromagnetism}. 
Usually, the 
order parameter is zero in  the high temperature disordered 
paramagnetic\index{paramagnetism} phase and non-zero in  the low temperature ordered 
ferromagnetic\index{ferromagnetism} phase. 
In the Ising\index{Ising} model 
simulation, we shall be 
interested in the  second order phase transition\index{phase transition!second order}. 

The magnetic phase transitions\index{phase transition} are characterized by  critical 
exponents\index{exponent!critical}. In the limit of $T\to T_c$, we have
\begin{eqnarray}
{{M(T)}\over{V}} & \sim & (T_c - T )^\beta\ \  {\rm for}\ \  T < T_c \ , \nonumber\\
     &            \nonumber\\
{{\chi (T)}\over{V}} & \sim & \vert T - T_c \vert ^{-\gamma}\ ,\nonumber\\
     &            \nonumber\\
{{C_V (T)}\over{V}} & \sim & \vert T-T_c \vert ^{-\alpha}\ . 
\end{eqnarray}
In the above $V$ is the total number of spins in the system.
 $\beta$, $\gamma$ and $\alpha$ are called the critical exponents\index{exponent!critical}.
For the two dimensional Ising\index{Ising} model system $\beta = 1/8$, 
$\gamma = 7/4$ and $\alpha = 0$.
In fact, the specific heat\index{specific heat} goes like
\begin{equation} 
{{C_V(T)}\over{V}}\sim 0.4995\times {\rm ln}\  (\ \vert\  T-T_c\  \vert\ ^{-\nu}  ). 
\end{equation}
The  correlation 
length at temperature $T$, denoted by $\xi (T)$,  which measures the typical linear size
of a magnetic domain,
goes like $\xi (T) \sim  \vert T-T_c \vert ^{-\nu}$, and the 
exponent $\nu =1$ for the 
two dimensional Ising\index{Ising} model. The critical exponents\index{exponent!critical} 
are universal in the sense  their 
values do not depend on the details of the model; 
they depend only on gross features
like dimensionality of the problem, the  symmetry\index{symmetry} 
of the Hamiltonian\index{Hamiltonian} {\it etc.} 
Indeed this is 
precisely why very  simple models like the  Ising\index{Ising} model are successful 
in describing phase transition\index{phase transition} 
in real systems. 

The correlation length   $\xi (T)$ is expected to be of the order of zero  
when $T >> T_c$. At very high temperature, the spins behave 
as if they are independent of each other. Spin-spin interactions 
are irrelevant 
compared to the thermal fluctuations\index{fluctuations}. Entropy wins over energy completely.
The macroscopic properties of the system are determined by 
entropic considerations only. As $T$ decreases, 
the spin-spin interaction 
becomes more and more relevant and the 
correlation length diverges as $T\to T_c$.
If the system studied is finite, the correlation length, at best, can 
 be  of the order of the 
linear dimension of the system. 
This brings us to the important topic of 
 finite size effects\index{finite size effect} and  finite size scaling\index{finite size scaling}.

\section{What is finite size scaling\index{finite size scaling} and how do we implement it?}
\label{finite_size_scaling}

At the outset,  we  recognize that 
{\it a  finite system can not exhibit phase transition\index{phase transition}}.
Consider a square lattice of size say $L\times L$. If the correlation length
$\xi (T) << L$, then for all practical purposes, a finite system can be 
taken as a good approximation to  an infinite system. 
In other words, our simulations on a finite lattice 
would give good results if the temperature
$T$ is not close to $T_c$, {\it i.e.} if $\xi (T) << L$.
But when $T$ is close to $T_c$, the correlation length becomes of the order of
$L$.  A finite system thus can not show any divergence of the specific heat or 
susceptibility; we would get a sort of
rounded peaks at best, instead of sharp divergence. Of course as $L$ becomes larger
and larger, the peak would become sharper and sharper. We present in 
Fig.~ (\ref{mag_fig}),  magnetization per spin, $\vert M\vert/V$ (where
$V=L^2$), 
 as a function of $J/k_{{\rm B}} T$ for two dimensional square 
lattices of size $L=2,4,8,16,32$ and $64$. 
It is clear from the figure that the transition becomes sharper as
the system size\index{system size} increases. Similar plots for  magnetic 
susceptibility\index{susceptibility}, 
are shown in Fig.~({\ref{sus_fig}). The peak is rounded and broad 
for smaller system sizes\index{system size}. As the system size\index{system size}
 increases we find the 
peak becomes sharper and its  height increases. The behaviour 
of  specific heat\index{specific heat} as a function of temperature for different 
system sizes\index{system size} are shown in Fig.~(\ref{cv_fig}).    
\footnote{These plots have been obtained from the computer program
ISING.FOR, written and tested by my colleague V. Sridhar, MSD, IGCAR, Kalpakkam.
The program can be obtained from him by an e-mail to vs@igcar.ernet.in\ .}
We recognize that the correlation length
for a finite $L\times L$ system can at best be 
$L$, when  $T$ is close to $T_c$.  
Hence we can write
\begin{eqnarray}
\xi (T)\  {}^{\ \ \sim}_{T\to T_c}\  \vert T-T_c\vert ^{-\nu} &=& L
\end{eqnarray}
From this  we get 
\begin{eqnarray}
\vert T-T_c\vert & \sim & L^{-1/\nu}
\end{eqnarray}
\begin{figure}[hp]
\centerline{\psfig{figure=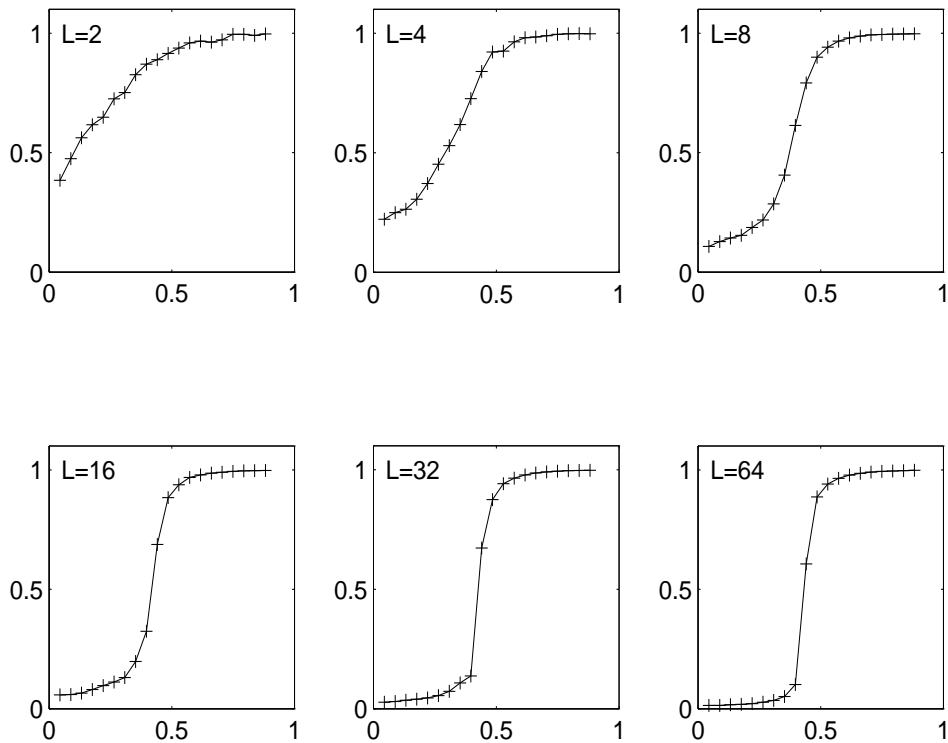,height=9.90cm,width=12.60cm}}
\caption{\small\protect{Magnetization per spin ( $\vert M\vert/L^2 ) $ 
 {\it vs.} $J/k_{{\rm B}} T$ 
for various 
\index{system size}system sizes $ L$.}}
\label{mag_fig}
\vskip 3mm
\hrule
\end{figure}
\begin{figure}[hp]
\centerline{\psfig{figure=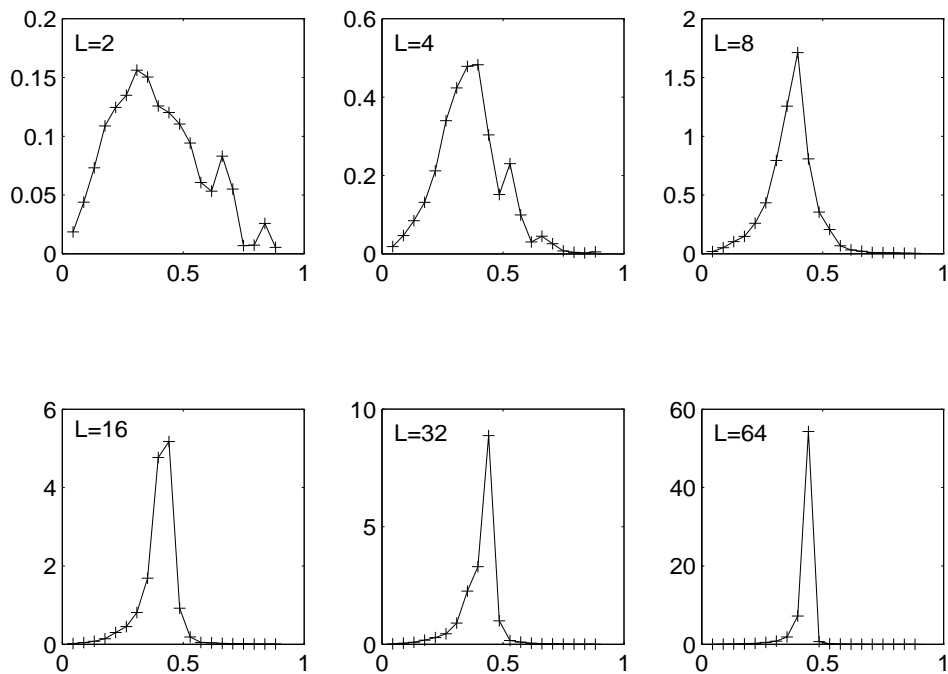,height=9.90cm,width=12.60cm}}
\caption{\small\protect{Magnetic susceptibility\index{susceptibility}
 per spin ($ \chi /L^2 $)  
{\textit vs.}  $ J/k_{{\rm B}} T $  for various system sizes $ L$ }}
\label{sus_fig}
\vskip 3mm
\hrule
\end{figure}

\begin{figure}[htp]
\centerline{\psfig{figure=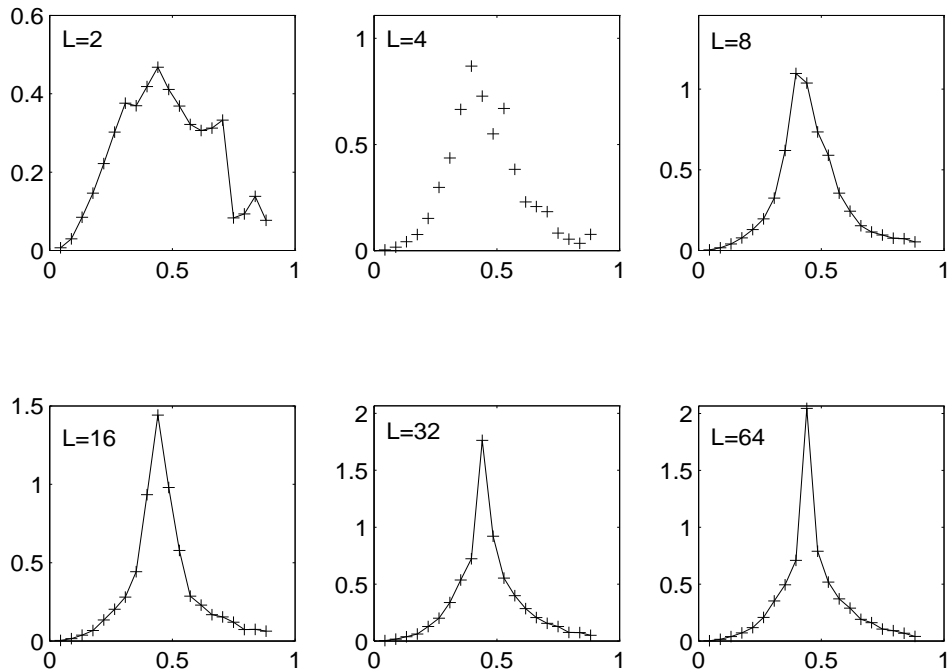,height=9.90cm,width=12.60cm}}
\caption{\small\protect{Specific heat per spin } ($C_V /L^2 $)  {{\textit vs.}} $ J/k_{{\rm  B}} T  $ 
 for various system
        sizes ${ L}$ }
\label{cv_fig}
\vskip 3mm
\hrule

\end{figure}
We can now express the magnetization, magnetic 
susceptibility\index{susceptibility}, and the specific
heat as power laws in $L$. 
\begin{eqnarray}
{{\vert M(T)\vert }\over{V}}\ \ \    {}^{\ \ \sim}_{T\to T_c}\ \ \   
(T_c -T )^{\beta}\ \ \ & {}^{\ \ \sim}_{L\to\infty}&
  \  L^{-\beta/\nu} {\rm for}\ T < T_c\nonumber\\
{{\chi (T)}\over{V}}\ \ \    {}^{\ \ \sim}_{T\to T_c}\ \ \   \vert T-T_c \vert ^{-\gamma}  
\ \ & {}^{\ \ \sim}_{L\to\infty}&
  \ L^{
\gamma /\nu }\nonumber\\
{{C_V (T)}\over{V}}\  \ \ {}^{\ \ \sim}_{T\to T_c}\   \ \ \vert T - T_c \vert ^{-\alpha}    
\ \ &{}^{\ \ \sim}_{L\to\infty}& \   L^{\alpha/\nu}\nonumber\\
\vert T_c (L) - T_c (L=\infty)\vert\  & {}^{\ \ \sim}_{L\to\infty}&\  L^{1/\nu}.     
\end{eqnarray}
The above finite size scaling\index{finite size scaling} is implemented as follows.
For a given $L$, find the  value of $T$ at which 
the specific heat\index{specific heat} peaks. Call that temperature
as $T_c (L)$. Calculate, for example 
$\vert M(T)\vert / V$ at $T=T_c (L)$
for different values of $L$. Plot $\ln \  [\vert M (L)\vert / V]$ 
against $\ln (L)$.  The slope 
of the best fit  straight line would 
give $-\beta/\nu$. 

So far so good; we know now how to assemble, employing random numbers,  a Markov chain of microstates
each produced from its predecessor through an attempted single flip. What happens 
when the flipping probability is very very small?\footnote{the flipping probability 
can become small for several reasons: (a) when the temperature is very low;
(b) when the system is in or very close to equilibrium (c) when the system is in a 
metastable state (d) when the system is very close to a critical state
{\it etc.} } 
The system 
has a natural tendency to remain stay put in a microstate 
step after step after step. Nothing happens to the dynamics for a 
very long time. But the Monte Carlo clock is ticking all the time. 
Can such a wastage of computer time be avoided? In other words, can we simulate
a very slow dynamics by a fast algorithm? The answer is yes; in the year 
1975, Bortz, Kalos and Lebowitz\cite{bortz}\index{Bortz}\index{Kalos}\index{Lebowitz}
 proposed an event-driven\index{Monte Carlo!event driven}
 algorithm
that precisely does this.  
They called their algorithm the n-fold way.\index{n-fold way}

\section{What is the n-fold way?}
\label{n-fold_way}

The n-fold way is an event - driven algorithm.\index{Monte Carlo!event driven} 
We ensure that an event happens at every
algorithmic time step. In the context of Ising\index{Ising}  model simulation, this 
means that at every MCS\index{Monte Carlo!step (MCS)} we have a new spin configuration.  
Let ${\rm C}_k$ denote the spin configuration at time $k$. As we have already 
explained, this configuration belongs to one of the microstates, say 
${\cal C}_i$ 
of the sample space\index{sample space} $\Omega_{{\rm CS}}$ underlying the closed system. 
\index{closed system} Define $\lambda_k$ as the probability 
for the system to persist in its current microstate ${\cal C}_i$ for one 
MCS. $\lambda_k$ is formally given by $\lambda_k=W_{i,i}=1-\sum_{j\ne i}W_{j,i}$, 
where $W$ is the 
transition matrix\index{transition matrix}  underlying either the Metropolis  or the 
Glauber dynamics.\index{Glauber dynamics} 
The probability that the system continues to be in the 
current microstate for the next $m$ MCS and makes a transition to a 
different microstate in the $(m+1)$-th step 
is given by the geometric distribution,
\begin{eqnarray}\label{geometric_distribution}
p(m) & = & \lambda_k ^m (1-\lambda_k) 
\end{eqnarray}
The first part of the n-fold way\index{n-fold way} consists of assigning a life time 
to the microstate ${\rm C}_k$. It is  
sampled  from its distribution, see  
Eq. (\ref{geometric_distribution}) 
by  a simple 
analytical inversion: 
\begin{eqnarray}\label{lifetime}
\tau ({\rm C}_k )  & = &   {{\ln \xi}\over{\ln \lambda_k}},
\end{eqnarray}
where $\xi$ is a random number (uniformly distributed in the range $0$
to $1$); we need the life time in terms of MCS; hence we calculate $m({\rm C}_k)=\lfloor \tau ({\rm C}_k)\rfloor$ 
where the RHS denotes the largest integer less than or equal to $\tau ({\rm C}_k)$.  
Thus in one shot we have advanced the Markov Chain by $m$ steps:
$$\{ \cdots\ ,\ {\rm C}_k={\cal C}_i ,\  {\rm C}_{k+1}\equiv {\cal C}_i,\  {\rm C}_{k+2}\equiv {\cal C}_i,\  
\cdots ,\  {\rm C}_{k+m}\equiv {\cal C}_i ,\ \cdots  \}.$$ The second part of the n-fold way consists of
determining ${\rm C}_{k+m+1}$.  
This can be obtained in principle from the 
the transition matrix $W$. 
The whole process is repeated and this constitutes 
the n-fold way\index{n-fold way}  of generating a 
Markov chain\index{Markov!chain} of microstates.  

I said  that $\lambda_k$ and ${\rm C}_k$ can  be obtained, in principle, from the 
transition matrix $W$. But in practice, we do not need the transition matrix; also 
it is not desirable nor feasible to work with the transition matrix since 
its size is rather large; as we have already seen, even for a small system of Ising 
spins on a $10\times 10$ square lattice, the associated transition matrix
would contain some $10^{60}$ elements!
The practical implementation of the n-fold way\index{n-fold way}  
proceeds as follows. 
All the spins in the system can be put into $n$ classes where $n$ is a small 
manageable number; Let $n_i$ 
be the number of spins in the class $i$ when the system is in a 
microstate ${\rm C}_k$ at time $k$. 
It is clear $\sum_{i=1}^{n}n_i = V$.
What is the value of $n$ and what are these n classes? 
Consider a two dimensional square lattice with periodic boundaries.
Let us consider a spin which is 
oriented {\it up} ($\uparrow$). Call this the reference spin. 
The local environment of the reference spin is  one of the following 
five types: (i) all the four nearest neighbours are up; (ii) three nearest 
neigbours are up and one  is down; (iii) two nearest neighbours 
are up and two are down (iv) one nearest neighbour is up and the other three are down and (v)
all the four nearest neighbour spins are down. There are thus five classes when the 
reference spin is up; When the reference spin is down, there are five 
more classes,  making the total number of classes $10$. Thus each of the 
$V$ spins in the system belongs to one or the other of these $n$ classes,  
hence the name n-fold way, where $n$ is $10$ in the present example.\index{n-fold way}  
Let us define a local 
energy for each spin; it is given by  the sum of its  interaction 
energies with its four neighbours and with an external field $B$. 
We can flip the spin and calculate the change in the local energy.
All the spins belonging to a class have the same local environment, same local energy and 
cause the same change in energy when flipped.  
Table (5) presents the ten classes and their common properties. 
Hence it is quite adequate, see below, if we consider
only these ten classes for purpose of obtaining 
$\lambda_k$ and ${\rm C}_{k+m+1}$. 
Let $\Delta E(i)$ be the increase in energy when a spin of the $i$-th
class is flipped, see the last column of 
Table (5).
This quantity is the same for all the spins belonging to the same class.
\begin{table}[th]\label{n-fold-way_table} 
\caption{\protect\small  The ten classes of the 
10-fold way\index{n-fold way}  for a two dimensional
square lattice with periodic boundaries}
\bigskip
\begin{center}
\begin{tabular}{|c|c|c|c|c|}
\hline
     &                      &                   &    &                   \\
Class & reference & number of  & Local energy & $~~\Delta E$~~ \\
      &  spin     & nearest neigbour &        &            \\
      &           &   up spins       &        &            \\
      &           &                  &        &            \\
\hline         
      &            &                  &        &             \\
1 & $\uparrow$     & 4 & $\mathbin{-}4J\mathbin{-}B$  & ~~$\mathbin{+}8J\mathbin{+}2B$~~  \\[2mm]
2 & $\uparrow$     & 3 & $\mathbin{-}2J\mathbin{-}B$  & $\mathbin{+}4J\mathbin{+}2B$      \\[2mm]
3   &$\uparrow$    & 2 & ~~~~~~~$\mathbin{-}B$       &~~~~~~~$\mathbin{+}2B$              \\[2mm]
4   &$\uparrow$    & 1 &  $\mathbin{+}2J\mathbin{-}B$    &$\mathbin{-}4J\mathbin{+}2B$    \\[2mm]
5   &$\uparrow$    & 0 & $\mathbin{+}4J\mathbin{-}B$    & $\mathbin{-}8J\mathbin{+}2B$    \\[2mm] 
6    &$\downarrow$ & 4 & $\mathbin{+}4J\mathbin{+}B$    & $\mathbin{-}8J\mathbin{-}2B$    \\[2mm]
7   & $\downarrow$ & 3 & $\mathbin{+}2J\mathbin{+}B$    & $\mathbin{-}4J\mathbin{-}2B$    \\[2mm]
8   &$\downarrow$  & 2 & ~~~~~~~$\mathbin{+}B$       &~~~~~~~$\mathbin{-2}B$              \\[2mm]
9   &$\downarrow$  & 1 & $\mathbin{-}2J\mathbin{+}B$  & $\mathbin{+}4J\mathbin{-}2B$      \\[2mm]
10  &$\downarrow$  & 0 & $\mathbin{-}4J\mathbin{+}B$   & $\mathbin{+}8J\mathbin{-}2B$     \\[2mm]
    &              &   &         &      \\
\hline
\end{tabular}
\end{center}
\end{table}
Let $p_i$ be the probability of accepting a spin flip in the $i$-th class.
It is given by,
\begin{eqnarray}
p_i  = \cases{ 
              {\rm  min} \Bigg( 1,\ \exp\bigg[ -\beta\Delta E(i)\bigg]\Bigg)  & for Metropolis \cr
                                                                            &                      \cr
                                                                            &                      \cr
              \Bigg( 1+\exp\bigg[  \beta\Delta  E(i)\bigg] \Bigg) ^{-1}     &  for Glauber    \cr      
                                                                                                       }
\end{eqnarray}
The value of $p_i$ depends on whether we we use the Metropolis dynamics or the Glauber 
dynamics.\index{Glauber dynamics} 
The probability of selecting a class is $n_i/V$. Thus
the probability of flipping a spin in the system is given by
$q=\sum_{i=1}^{10}n_i p_i/V$. 
This immediately gives us $\lambda_k = 1-q$.
Once we know $\lambda_k$ we can sample a life time $m$ from the 
geometric distribution as described earlier. We then select a class
from the distribution $\rho_i =  n_i p_i /(qV) : i=1,10$; then a spin is chosen 
randomly from the selected class and flipped to give ${\rm C}_{k+m+1}$.
Calculate the set $\{ n_j\ :\ j=1,\ 10\}$ for the new configuration. 
This kind of book-keeping is the time consuming part of the n-fold 
way; however the changes are confined to the class indices of the flipped spin and 
its four nearest neighbours. These changes can be carried out employing 
simple rules: If the flip is from up ($\uparrow$) to down ($\downarrow$), the class of the 
flipped spin increases by $5$ while  that of each of its four nearest neighbours  
increases  by $1$. If the flip is from down ($\downarrow$) to up ($\uparrow$), then the 
class of the flipped spin decreases by $5$ and that each of the four nearest neighbours 
decreases by $1$. Having updated the array $\{ n_j\}$ and indexed each spin by its 
class, we iterate the entire process  and 
get a Markov chain of microstates.
The macroscopic properties are calculated from the chain in the usual 
way described earlier. 

Alternately we can store the life times $\{ \tau ({\rm C}_i)\}$ of each of the microstates
belonging to the Markov chain $\{ {\rm C}_0 ,\ {\rm C}_1 ,\ {\rm C}_2 ,\ \cdots\}$ where 
a microstate ${\rm C}_i$ is obtained from ${\rm C}_{i-1}$ through a spin flip in the 
n-fold way. While averaging we weight each microstate with its corresponding 
life time:
\begin{eqnarray}
\left\langle O\right\rangle &=& {}^{\  {\rm Lim.}}_{N\to\infty}\ \overline{O}_N = 
 {{ \sum_{i=1}^{N} O({\rm C}_i) \tau ({\rm C}_i)}\over{\sum_{i=1}^N \tau ({\rm C}_i)}}
\end{eqnarray}
Instead of sampling the life time from its geometric distribution one can use the 
average life time of a microstate ${\rm C}_i$, given by,
\begin{eqnarray}
\overline{\tau}\ ({\rm C}_i) &=& {{1-q({\rm C}_i)}\over{q({\rm C}_i)}}
\end{eqnarray}
and employ it as weight while averaging. 
This completes the discussion on the 
n-fold way.\index{n-fold way}

The n-fold way\index{n-fold way}  is a particular case of a more general technique 
called the Monte Carlo with Absorbing Markov 
Chain (MCAMC) \cite{novotny}.\index{Monte Carlo!absorbing Markov chain} I shall not discuss 
these techniques here  and instead refer you to the 
beautiful recent review  article of Novotny\cite{novotny_review}.\index{Novotny}

The n-fold way is a fast algorithm that realizes a slow dynamics. As I said earlier,  
the dynamics of an equilibrium or near equilibrium system is slow; at very low temperatures 
the dynamics is slow; many a times the system gets stuck in a metastable state which drastically
slows down the dynamics. There exists another  phenomenon for which  the 
Metropolis/Glauber\index{Glauber dynamics} dynamics is very very slow: when the system is close to a critical 
state. This is called critical slowing down  - an important issue in the statistical mechanics 
of continuous phase transition. To this we turn our attention below. 

\section{What do we mean by critical slowing down\index{critical slowing down}?}
\label{critical_slowing_down}

A problem with Monte Carlo\index{Monte Carlo} simulation is that statistical error\index{statistical error} 
decreases with increase of sample size $N$ only as $N^{-1/2}$. This means that to get the numbers to 
just one extra decimal accuracy, we need to increase the sample size 
by a factor of hundred. Things are much worse.  The error reduction 
as $N^{-1/2}$ happens only if the members of the ensemble are 
independent. But successive Monte Carlo\index{Monte Carlo} configurations generated 
by Metropolis\index{Metropolis!algorithm} rejection technique are usually correlated~\cite{mkb}.
Such correlations\index{correlations} increase the statistical 
error\index{statistical error}. Let us see how this 
comes about. First notice that $\overline{x}_N$ given by
Eq.~(\ref{average_eqn}) is a random variable. The mean of $\overline{x}_N$ is  
$\left\langle x\right\rangle $. The variance\index{variance} of ${\overline x}_N$ is formally
given by,
\begin{eqnarray}
\sigma ^2 ( \overline{x}_N )& =&    {{ \sigma ^2 (x) }\over{N}}
    \left[ 1+ 2\sum_{k=1}^{N-1} \left( 1- {{k}\over{N}}\right) \gamma_k 
     \right]\ ,
\end{eqnarray}
where the correlation $\gamma_k$ is given by
\begin{equation}
\gamma _k ={{  \left\langle x_i x_{i+k} \right\rangle - 
 \left\langle x_i \right\rangle\left\langle x_{i+k}\right\rangle }
    \over{ \sigma ^2 (x) }}.
\end{equation}
In writing the above, we have assumed that the 
sequence $x_1,\ x_2,\ \cdots ,\ x_N$ is stationary. 
The correlations\index{correlations} are estimated from the Monte Carlo\index{Monte Carlo} Markov chain  as,
\begin{equation}
\gamma _k = {{N}\over{N-k}}  {{\sum_{i=1}^{N-k} \left( x_i - \overline{x}_N\right)
            \left( x_{i+k} - \overline{x}_N\right)}
     \over{
            \sum_{i=1}^{N} \left( x_i - {\overline x}_N \right) ^2 }}
\end{equation}
Let us now define the so called integrated correlation time as,
\begin{equation}
\tau^{\star} = \sum_{k=1}^{N-1} \left( 1 - {{k}\over{N}}\right) \gamma_k .
\end{equation}
We can now express the Monte Carlo\index{Monte Carlo} result as 
\begin{equation}\label{error_correlated}
\overline{x}_N \pm {{ \sigma (x)}\over{\sqrt{N} }}\times \sqrt{1+2\tau^{\star} }
\end{equation}
It is clear that if we have an uncorrelated data set, $\tau^{\star}=0$
and the expression for the 
statistical error\index{statistical error} given by Eq.~(\ref{error_correlated}) reduces to that
given by Eq.~(\ref{error}). The value of $\tau^{\star}$ is a measure of
the number of MCSS\index{Monte Carlo!step per spin (MCSS)} we need to skip  for getting a microstate uncorrelated 
to the present microstate, after the system has equilibrated.
The important point is that the correlation time $\tau^{\star}$ diverges as
$T\to T_C$; this means that finite sample Monte Carlo\index{Monte Carlo} estimates of the 
macroscopic properties of the system become unreliable when the  
temperature of the system is very close to the critical
value. The divergence of $\tau^{\star}$ as $T\to T_C$ is called 
{\it critical slowing down\index{critical slowing down}}. 

Thus we find that the statistical error\index{statistical error} depends on the correlation 
time. Is there a way to estimate the statistical error\index{statistical error} that does
not require explicit calculation of the correlation function
$\gamma_k$?  The answer is {\it yes} and we shall briefly discuss the 
 so called  blocking technique\index{blocking technique}, which is essentially a real space 
renormalization group technique applied to the one dimensional discrete 
space of MCSS\index{Monte Carlo!step per spin (MCSS)}.

\section{What is a blocking technique\index{blocking technique}\ ?}
\label{blocking_technique}

For a lucid account of the blocking technique\index{blocking technique}, see the  paper by 
Flyvbjerg and Petersen~\cite{fp}.  I shall confine myself to 
discussing the implementation of the technique. 
We are given a set of correlated data,
\begin{equation}
\Omega_0 = \{ x_{1}^{(0)} , x_{2}^{(0)} \cdots x_{N}^{(0)} \}. 
\end{equation} 
How do we calculate the statistical error\index{statistical error} associated with the sample mean, 
\begin{equation}
m={{1}\over{N}}\sum_{i=1} ^N  x_{i}^{(0)}  \ ?
\end{equation}  
Let $\hat{\Omega}_0$ denote the number of data points in the set 
$\Omega_0$. Note $\hat{\Omega}_0  = N$.  Calculate  the sample variance\index{variance} 
given by,
\begin{eqnarray}
\ & \  & \nonumber\\
S^2 (\Omega_0) & = & {{1}\over{\hat{\Omega}_0} }
\sum_{i=1}^{\hat{\Omega}_0} \left(  x_{i}^{(0)}\right) ^2 - \left( 
                 {{1}\over{\hat{\Omega}_0}}
\sum_{i=1}^{\hat{\Omega}_0} x_{i}^{(0)} \right) ^2  \ .\\
\ &\ &\nonumber
\end{eqnarray}
Let $\epsilon_0 ^2 = S^2 (\Omega_0)/(\hat{\Omega}_0 -1) $.  Then transform 
the data to another set of data $\Omega_1 = \{ x_{1}^{(1)} , x_{2}^{(1)}  , 
 \cdots x_{\hat{\Omega}_1 }^{(1)}\}$ which is half as large, {\it i.e. } 
$\hat{\Omega}_1 = \hat{\Omega}_0 /2 = N/2$. In general the equations that relate the data 
set $\Omega_{k}$  to $\Omega_{k-1}$ are given by,
\begin{eqnarray}
x_{i}^{(k)} & = & {{ x_{2i-1}^{(k-1)} + x_{2i}^{(k-1)} } \over{
              2}}\ \ \ \ \ i=1,\ 2,\ \cdots \hat{\Omega}_k\ ,\nonumber\\
\ &   \  & \nonumber\\
\hat{\Omega}_k &=& {{ \hat{\Omega}_{k-1} }\over{2}} \ \ \ \ k=1,\ 2,\ \cdots\ .
\end{eqnarray}
Calculate $\epsilon_{k}^2$ from the data set $\Omega_k$. We 
get a string of numbers $\epsilon^2 _0 , \epsilon^2 _1 , \epsilon^2 _2,
\cdots $. These numbers would increase with iterations and eventually
reach a constant value within statistical fluctuations\index{fluctuations} beyond 
certain number of iterations. In other words  
$\epsilon^2 _{l+k} = \epsilon^2 _l$ for all $k \ge  1$ and for some $l$. 
In any case the iteration has to stop when the number of data points in the 
set  becomes $2$. The desired statistical error\index{statistical error!one-sigma confidence} (one sigma confidence) 
of the calculated  mean $m$ is given by $\epsilon _l$. 

An immediate issue of concern is the implications of  critical
slowing down to Monte Carlo\index{Monte Carlo} simulation. The Metropolis\index{Metropolis!algorithm} 
dynamics slows down for $T$ close to $T_C$. The closer $T$ is to 
$T_C$, the slower does the dynamics become. As we have already
seen, this builds up temporal correlations\index{correlations} and Monte Carlo\index{Monte Carlo!estimate} estimates
of the macroscopic properties become unreliable. Let the 
divergence of the integrated correlation time $\tau ^{\star}$
with $T$ be expressed as 
\begin{eqnarray}
\tau^{\star} (T)\  {}^{\ \ \sim}_{T\to T_C} \ \vert T - T_c \vert ^{-\Delta}
\end{eqnarray}
We can use  finite size scaling\index{finite size scaling} arguments and write 
\begin{eqnarray}
\tau^{\star} (L) \ {}^{\ \ \sim}_{L\to \infty} \ L^{z}
\end{eqnarray}
where $z=\Delta /\nu$. The  dynamical 
critical exponent\index{exponent!dynamical critical} $z$ is greater than unity; in fact,  for the 
local update algorithms like the Metropolis\index{Metropolis!algorithm}, 
the value of $z$ is nearly $2$ and above. 
This tells us that for $T$ close to $T_C$, the Monte Carlo\index{Monte Carlo!error} 
error becomes large especially when the system size $L$ is large. 
This is indeed a major drawback of the Metropolis\index{Metropolis!algorithm} Monte Carlo\index{Monte Carlo} 
technique. Fortunately this problem of critical slowing down\index{critical slowing down} can be 
very efficiently overcome by employing the so called cluster algorithms\index{cluster algorithm}, 
wherein, a large cluster of spins is updated in a single step.
Swendsen and Wang~\cite{sw}\index{Swendsen}\index{Wang} derived a cluster algorithm  by mapping
the spin problem to a  bond  percolation\index{percolation!bond}   problem, based on the 
work of Kasteleyn and Fortuin~\cite{fk} and developed by 
Coniglio and Klein~\cite{ck}.\index{Coniglio}\index{Klein}\index{Kasteleyn}\index{Fortuin} 

\section{What is percolation\index{percolation} ?}
\label{percolation_problem}

Percolation\index{percolation}  is a familiar term we use in the context of brewing of coffee:
Hot water injected at one end percolates through packed coffee powder and 
is collected at the other end. In fact percolation\index{percolation}  was introduced in the 
year 1957 by Broadbent and Hammersley~\cite{BH} as a model for the 
transport of liquid through a porous medium, spread of a disease in a 
community and other related phenomenon.  For an excellent early 
discussion on percolation\index{percolation}  see the little book by Hammersley and 
Handscomb listed under~\cite{mcref2}.  Let us abstract the notion  
and consider $N\times N$ square lattice of sites. Let $0\le\  p\ \le\  1$ 
denote the probability of placing a bond between a pair of nearest 
neighbour sites.  Carry out the following experiment.  Select a pair 
of nearest neighbour sites; select a random number $\xi$; if $\xi \le p$
place a bond between the selected pair. Otherwise
do not put a bond. Carry out this operation independently on all 
the nearest neighbour pairs of sites in the lattice. 
When two nearest neighbour sites 
are connected by a bond they form a cluster of two sites. When another 
site gets bonded to one of these two sites then the cluster is of three 
sites, and so on.  At the end, you will have clusters of different sizes. 
If $p$ is small, typically, we expect the lattice to contain only  
small clusters.  If $p$ is close to unity, there will exist a cluster that 
extend from one end of the lattice to the other. Such a cluster is said 
to span the lattice. It is called a spanning cluster\index{spanning cluster}. Let $P_L (p)$ 
denote the probability that a spanning cluster\index{spanning cluster} exists on an $L\times L$ 
square lattice.  For a given $L$ it is clear that 
\begin{eqnarray}
P_L (p) = \cases{  0    \  \  {\rm for} \  & $p\to 0$\ ,\cr
                               &         \cr 
                  1   \   \ {\rm for} \    & $p\to 1$\ .\cr}
\end{eqnarray}
In fact, we find that the percolation\index{percolation!bond}  probability $P_L (p)$ changes 
sharply from zero to unity  at a critical value of $p=p_C$. The transition
becomes sharper as $N\to\infty$. In other words,
\begin{eqnarray}
P_L (p)\  {}^{\ \ \sim\ \ }_{L\to\infty}\   \cases{ 0  \ \   {\rm for} \  &  $p < p_C$\ ,\cr
                                                         &   \cr 
                  1   \  \ {\rm for} \   &  $p\ge p_C$\ . \cr}
\end{eqnarray}
Thus, there is a  geometric phase transition\index{phase transition!geometric}
  at $p=p_C$.  The value of
$p_C$ is $0.5$ for a two dimensional square 
lattice in the limit $N\to\infty$. 
Figure (\ref{percolation_fig}) depicts $P_L (p)$ versus $p$ for various values of
$L$.  
We can clearly see that the transition becomes sharper with increasing system size. 
Even for $L=50$ the transition is already quite sharp. The statistical error 
in any of the data points plotted in Fig. (\ref{percolation_fig}) does not exceed a 
maximum of $\pm 5\%$.  
\begin{figure}[htbp]
\centerline{\psfig{figure=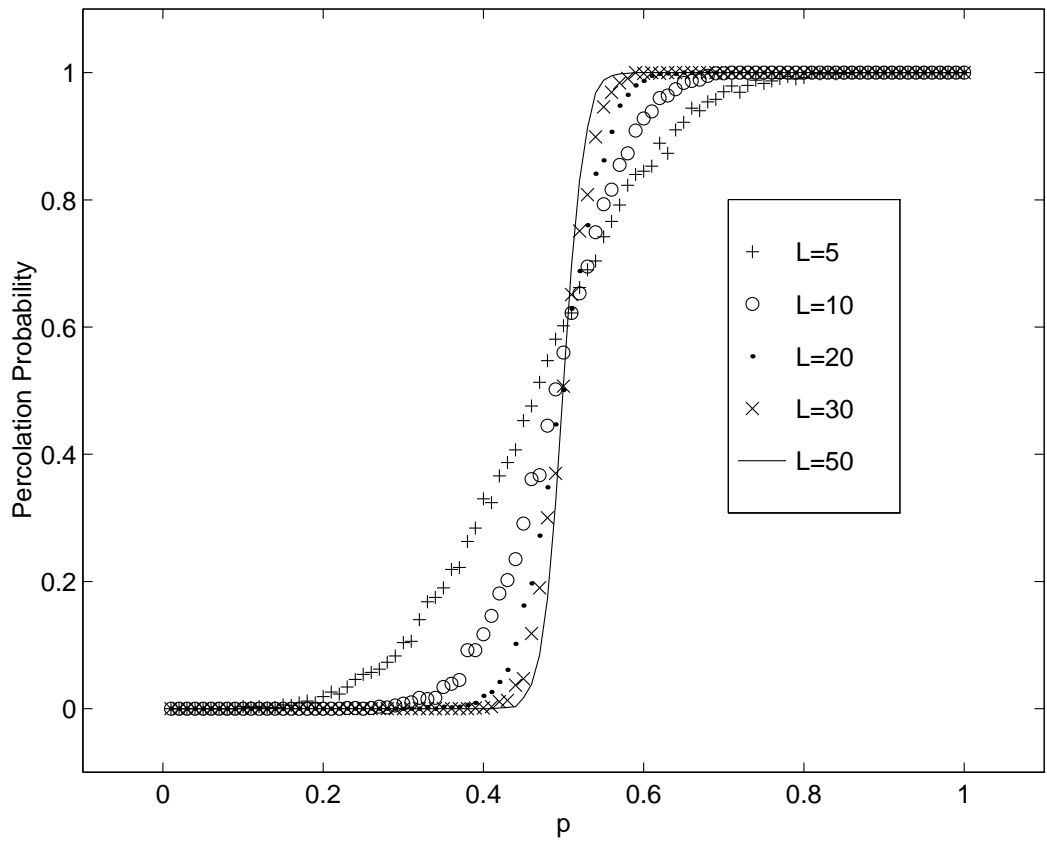,height=11.132cm,width=13.915cm}}
\caption{\small\protect{ $P_L (p)$ {\it versus} $p$ for bond percolation. The statistical 
error is less than $5\%$ for all the points. }}
\vskip 3mm
\hrule

\label{percolation_fig}
\end{figure}

What we have discussed above is called bond percolation\index{percolation!bond}, which is of relevance 
to  us in the context of discussions on Swendsen-Wang\index{Swendsen-Wang algorithm}
\index{Swendsen}\index{Wang} algorithm. 
There is also site percolation\index{percolation!site}. For more on percolation\index{percolation}  
read the delightful little book by Stauffer~\cite{stauffer}\index{Stauffer}. The most
interesting feature of the phenomenon of percolation\index{percolation}  
in the context of our present discussion is that unlike thermal
transition,  the geometric percolation\index{percolation}  transition is free from 
critical slowing down\index{critical slowing down}. Every sweep over the lattice generates 
an independent configuration and hence the correlation time is exactly
zero. This is precisely what Swendsen-Wang\index{Swendsen-Wang algorithm}\index{Swendsen}\index{Wang}
 algorithm makes use of.  

\section{What is Swendsen-Wang\index{Swendsen-Wang algorithm} algorithm? How do we implement it?}
\label{s_w_algorithm}

In the Swendsen-Wang\index{Swendsen-Wang algorithm}
  algorithm we start with a spin configuration
say ${\cal C}$. Select two neighbouring sites and call them $(i,j)$. 
If $S_i = S_j $, then the bond connecting the sites $i$ and $j$ becomes 
eligible for occupation. Such a bond is often referred to as {\it satisfied}
bond.  If $S_i \ne S_j$, then the bond is not a satisfied bond and 
hence is never occupied.  A satisfied bond is occupied with a probability 
$p$, to be specified later. This is carried out as follows. Select a 
random number (uniformly distributed between $0$ and $1$). If it is 
less than $p$ occupy the (satisfied) bond. Otherwise do not occupy the bond.   
Carry out this process for each pair of nearest neighbour sites  in the 
lattice. At the end you will have several clusters of sites connected by 
occupied  bonds.  Let us call these  
Kastelyn - Fortuin - Coniglio - Klein (KF-CK)\index{Kastelyn - Fortuin - Coniglio - Klein (KF-CK) cluster} clusters.\index{Coniglio}\index{Klein}\index{Kasteleyn}\index{Fortuin}
Figure~(\ref{FKCK_clusters}) depicts an example of KF-CK clusters on 
a $10\times 10$ square lattice.  Filled circles denote the up spins and 
open circles denote the down spins. Occupied bond 
is denoted by line connecting the neighbouring \ \lq like\rq\ 
spins. The clusters are shaded and the spanning cluster\index{spanning cluster} is shaded dark.
We have used a value of $J/k_{{\rm B}} T = 0.6$. The critical value,   
$J/k_{{\rm B}} T_C$, is $0.4487$.   
I must emphasize that these bonds are fictitious; 
\begin{figure}[th]
\centerline{\psfig{figure=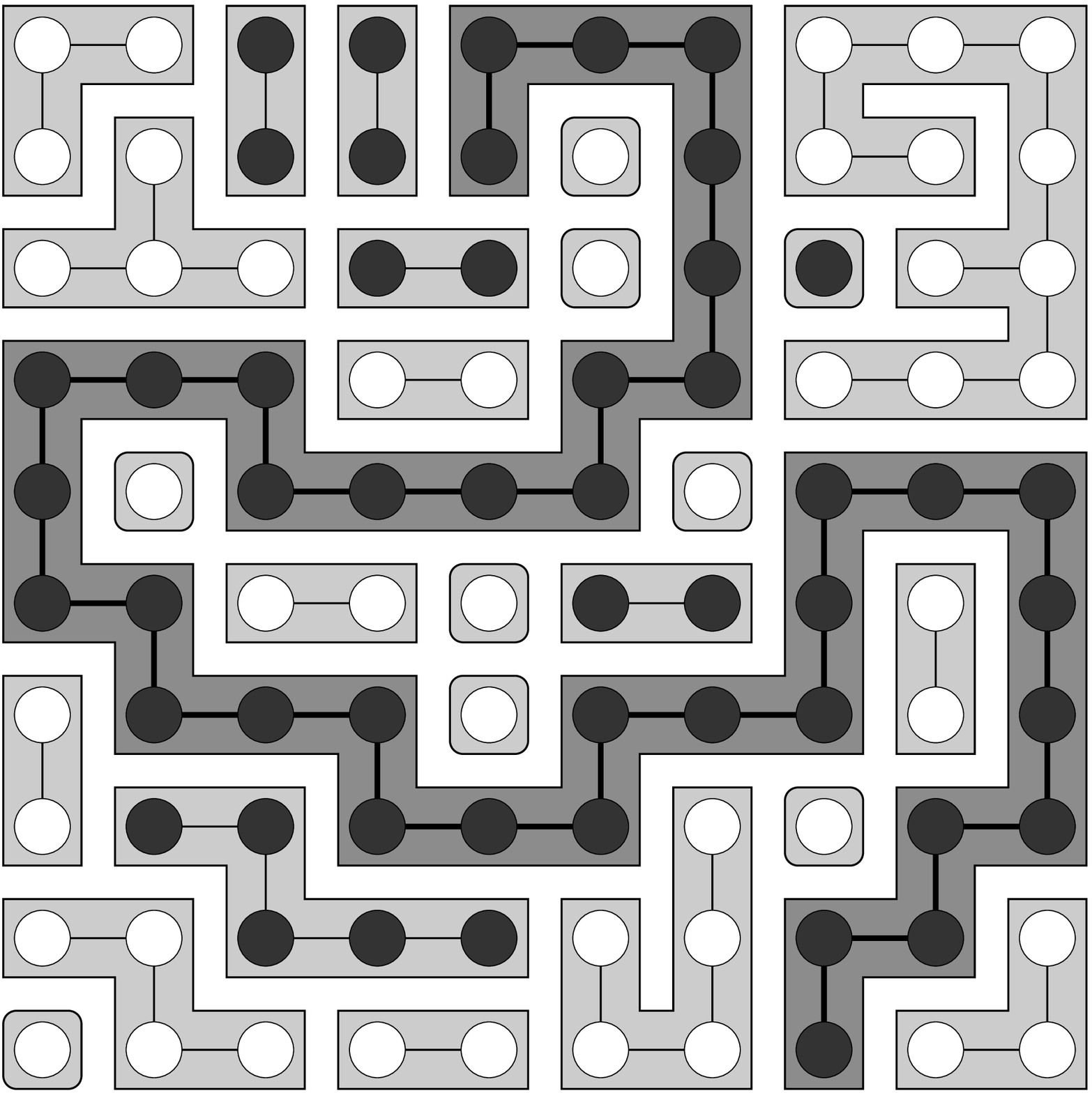,height=10.0cm,width=10cm}}
\caption{\small\protect{Kastelyn  - Fortuin - Coniglio - Klein
(KF-CK)\index{Kastelyn - Fortuin - Coniglio - Klein (KF-CK) cluster}
\index{Coniglio}\index{Klein}\index{Kasteleyn}\index{Fortuin} clusters on a  $10\times 10$ square lattice, for the 
Ising\index{Ising} spin system. 
Filled circles denote up spins;
open circles denote down spins. Satisfied and occupied bond is  
denoted by line connecting nearest neighbour \ \lq like\rq\ spins.
The spin-spin interaction strength $J/k_{{\rm B}} T$ is taken as $0.6$. 
The critical value $J/k_{{\rm B}} T_C$ is $0.4407$ for a square lattice.  
The clusters are shaded.
Spanning cluster is shaded dark.  
}}
\vskip 3mm
\hrule

\label{FKCK_clusters}
\end{figure}
there is no energy 
associated with them. They only serve to define a cluster.  Having thus
formed the KF-CK clusters\index{Kastelyn - Fortuin - Coniglio - Klein (KF-CK) cluster},
\index{Coniglio}\index{Klein}\index{Kasteleyn}\index{Fortuin} assign  to each cluster a spin variable 
$+1$ or $-1$ independently,  randomly and with equal probability. 
All the spins in a cluster acquire the spin value assigned to the 
cluster. Remove the fictitious  bonds.  What you get is a new  configuration 
of spins. Start all over again the process of constructing fresh 
KF-CK clusters\index{Kastelyn - Fortuin - Coniglio - Klein (KF-CK) cluster}.
\index{Coniglio}\index{Klein}\index{Kasteleyn}\index{Fortuin} In other words  iterate the  whole  process and generate 
a  chain of spin configurations; the process obeys detailed balance\index{detailed balance} 
and is ergodic; hence the configurations in the asymptotic part of 
the Markov\index{Markov!chain} Chain constitute a canonical ensemble\index{ensemble!canonical} of  microstates at 
the desired temperature. Note that  temperature enters into picture through
the bonding probability\index{bonding probability} $p=p(T)$.

\subsection{What is the appropriate value of the bonding 
probability $ p$ for a given $T$ ?}\label{bonding_probability}

The interaction energy associated with a pair of nearest neighbour spins
is $\epsilon_{i,j}=-JS_i S_j$. The energy is 
$-J$ when the two spins are aligned and  $+J$, when not. 
For convenience, let us 
set the minimum of the energy scale at zero by 
defining $\epsilon_{i,j}=-J(S_i S_j - 1)$. The Hamiltonian\index{Hamiltonian}
for the spin system can now be written as
\begin{equation}
H ({\cal C}) = -J \sum _{\left\langle i,j\right\rangle} 
\bigg[ S_i \left( {\cal C}\right) S_j \left( {\cal C}\right) -1\bigg]
\end{equation}
and the total energy ranges from $0$  to $+2JN_E$ where, $N_E$ is the total
number of pairs of nearest neighbour spins in the lattice.   
In the above expression for the Hamiltonian\index{Hamiltonian}, let us carry out
the sum over pairs of nearest neighbour  \lq like\rq\  spins and 
of nearest neighbour  \lq unlike\rq\  spins separately and get,
\begin{eqnarray}
H({\cal C}) & = & 2J\bigg[ N_E -B({\cal C})\bigg]
\end{eqnarray}
where $B$ is the number of satisfied bonds (number of pairs of nearest 
neighbour \ \lq like\rq\ spins in the spin  configuration (microstate) 
${\cal C}$).
The canonical partition function\index{partition function!canonical} is then given by,
\begin{eqnarray}
Z(\beta) & = & \sum_{{\cal C}} \exp\bigg[ - 2\beta 
            J\left( N_E - B\right)\bigg]\nonumber\\
         & = & \sum_{{\cal C}} q^{N_E - B}
\end{eqnarray} 
where $q=\exp (-2\beta J)$. Let us define $p=1-q$.
We can select randomly $b$ bonds from amongst 
$B$ satisfied bonds and occupy them. Number of ways of doing this is given
by $ \Omega (b,B) = B!/[b!(B-b)!]$. If we interpret $p$ as the probability of
occupying a satisfied bond, then the probability of having $b$ occupied 
bonds in a given a spin configuration ${\cal C}$ is 
$P(b,B({\cal C}))=\Omega(b,B({\cal C}))p^b q^{B({\cal C})-b}$. Notice that 
$\sum_{b=0}^B P(b,B) = (p+q)^B =1$.  Then the canonical partition 
function\index{partition function!canonical} can be written in a suggestive form, given by,
\begin{eqnarray}\label{canonical_partition}
Z(\beta) & = & \sum_{ {\cal C}} q^{N_E -B({\cal C})} (p+q)^{B({\cal C})} \nonumber\\
         & = & \sum_{ {\cal C}} q^{N_E - B}\sum_{b=0}^{B} \Omega (b,B) p^b q^{B-b}\nonumber\\
         & = & \sum_{{\cal C}} \sum_{b=0}^{B({\cal C})}
               \Omega (b,B)p^b q^{N_E - b}.
\end{eqnarray}
From the above it is clear that the correct value of the bonding 
probability that must be used in the Swendsen-Wang\index{Swendsen-Wang algorithm}\index{Swendsen}\index{Wang}
 cluster 
algorithm\index{cluster algorithm} is given by 
\begin{equation}\label{bondprob}
p= 1-q = 1-\exp(-2\beta J).
\end{equation} 

We can construct the same partition sum in a different way. Start with 
the lattice sans Ising\index{Ising} spins.
Put randomly $b$  bonds, each connecting a  pair
nearest neighbour sites.
Number of ways
of doing this is $\Omega(b,N_E)=N_E !/[b!(N_E -b)!]$. The 
probability of constructing a single bond is taken as $p=1-\exp(-2\beta J)$.  
We call the lattice
with  a specific arrangement of $b$ bonds as a  graph\index{graph} $G$. Let $N_c (G)$
denote the number of clusters in the graph\index{graph} $G$. 
Note that all the graphs\index{graph} having the same number of 
bonds need not have the same number of clusters.  
Consider now the graph\index{graph} $G$ 
having  vertices equal to the number of lattice sites or  Ising\index{Ising} 
spins,  $N_E$ edges (equal to the number of nearest neighbour pairs of 
spins) and $b$ bonds.  
We can 
now decorate the vertices of  $G$ with Ising\index{Ising} spins. While doing so  
we would like to be consistent so that 
the bonds we have placed in the lattice are occupied bonds. 
How many Ising\index{Ising} 
spin configurations are compatible with the  graph\index{graph} $G$?
It is clear that all the vertices in a given cluster must hold 
\ \lq like \rq \  spins ( all \ \lq up\rq \  $\uparrow$ 
or all \ \lq down\rq \  $\downarrow$). Hence there are  $2^{N_c (G)}$
spin configurations compatible with $G$. The statistical weight of
the  graph\index{graph} $G$ is $p^{b(G)} q^{N_E -b(G)} 2^{N_c (G)}$.  
We can calculate such weights
for all the graphs\index{graph} for a given $b$ and then for all $b$ ranging
from $0$ to $N_E$. 
The canonical 
partition function\index{partition function!canonical}, see Eq.~(\ref{canonical_partition}), can now be 
written equivalently as a sum  of the weights 
of all possible graphs\index{graph} and is  given by,
\begin{eqnarray}
Z(\beta) & = & \sum_G p^{b(G)}\ q^{N_E - b(G)}\  2^{N_c (G)}
\end{eqnarray}    

Thus the interacting Ising\index{Ising} spin problem  has been mapped on to a 
noninteracting correlated random bond percolation\index{percolation!bond}  problem,  
see~\cite{sw,fk,ck,stauffer,ckhu}. The spontaneous
magnetization in the Ising model\index{Ising} 
 corresponds to the percolation  probability in the 
percolation\index{percolation!bond}  problem; the magnetic
susceptibility\index{susceptibility} is analogous to the average number of spins per 
KF-CK 
cluster\index{Kastelyn - Fortuin - Coniglio - Klein (KF-CK) cluster}; 
\index{Coniglio}\index{Klein}\index{Kasteleyn}\index{Fortuin} energy and 
specific heat\index{specific heat} are like the number of occupied bonds and its 
fluctuations\index{fluctuations}; the 
correlation length in Ising\index{Ising} system corresponds to the linear size of
cluster in the correlated random bond percolation\index{percolation!correlated random bond}  model. 
Purely from a practical point of view the Swendsen-Wang\index{Swendsen-Wang algorithm}
\index{Swendsen}\index{Wang} technique 
based on KF-CK clusters\index{Kastelyn - Fortuin - Coniglio - Klein (KF-CK) cluster}
\index{Coniglio}\index{Klein}\index{Kasteleyn}\index{Fortuin}  has emerged as a powerful tool in Monte 
Carlo simulation of large systems.  
For a 
$512\times 512$ lattice system the value of 
$\tau^{\star}$ can be made as small as ten  which is some ten  thousand times 
smaller than that for the single flip Metropolis\index{Metropolis!algorithm} algorithm. This completes
the discussion on Swendsen-Wang\index{Swendsen-Wang algorithm} algorithm. 

For implementing the cluster update Monte Carlo\index{Monte Carlo!cluster update} 
we need a fast algorithm to 
identify the clusters. An ingenious technique of
 cluster counting\index{cluster counting} was 
proposed by Hoshen and Kopelman~\cite{hoshen_kopelman}.\index{Hoshen-Kopelman}
The technique is straight forward, easy to implement and  quite
fast. 

\section{What is the basis of  the Hoshen-Kopelman\index{Hoshen-Kopelman} 
cluster counting\index{cluster counting}
algorithm?}\label{Hoshen_Kopelman}
 
The general principle of the Hoshen-Kopelman\index{Hoshen-Kopelman} algorithm is as follows. 
Consider an  arbitrary dimensional lattice with arbitrary coordination
number\index{coordination number}. The lattice sites are occupied by the relevant species -
like atoms or spins. Lattice sites  linked by suitably defined
neighbourhood conditions constitute a cluster. The linkage 
can be due to presence of same type of atoms on neighbouring sites (site 
percolation\index{percolation!site} ) or the presence of a bond between two neighbouring sites (bond
percolation\index{percolation!bond}  ) or any other definition dictated by the nature of the 
problem under investigation. In a lattice there can be many clusters.  
Ideally one can assign distinct labels
to the clusters and associate with these labels the corresponding cluster 
sizes.  
 A straight forward procedure
would be to scan the lattice row by row and then layer by layer. The first 
time we hit a \ \lq relevant\rq\ site we assign a label, say $1$ 
to the site and associate with the label a number $LL(1)=1$. If the next 
site in the row is linked to this site through the neighbourhood relation,
then it is assigned with the same label and the array $LL$ is updated,
$LL(1)=LL(1) +1$ signifying that the cluster in its process of
\ \lq growth\rq\  has acquired one more site. 
On the other hand if the next site is 
not linked to the site previously examined, then it is assigned with a 
new label, say $2$ and $LL(2)=1$. The process of 
labelling the sites  and updating the array $LL$ is continued.
 A conflict starts
when we go down the rows or the layer;  we come across a site, say X, whose 
neigbouring sites to which this site is linked carry different labels,
say $l_1$, $l_2$, $l_3$, $\cdots$, $l_n$. Physically this means that the 
present site under examination is the link through which the clusters 
seemingly carrying different labels, coalesce. In other words, the different
clusters coalesce at this site to form a big cluster.  Therefore we would like 
to assign a single label to all the sites belonging 
the coalesced big cluster. This would mean that we should go over all the 
previous sites examined and relabel them; the associated $LL$ also should be
changed accordingly. In other words, 
all the sites carrying the labels ($l_1$, $l_2$,
$\cdots$ ,\ $l_n $) should be assigned a common label, say $l=l_k =
{\rm min}(l_1 ,\ l_2 ,\  
\cdots ,\ l_n)$  where $1\le k\le n$. 
Also we must set $LL(l)=LL(l_1)+LL(l_2)+\cdots +LL(l_n) + 1$.
This kind of backward relabelling is the 
most time consuming part of the algorithm.
Hoshen and Kopelman\index{Hoshen-Kopelman} avoided  the backward relabelling completely. The 
current site X under examination is assigned the label $ l = l_k$; 
However we set 
$LL(l_i) = -l,\ \forall\ \  i=1,n\ \  {\rm and}\ \  i \ne k $. 
Therefore 
$LL(j)$ denotes the size of the cluster only if it is positive.
If it is  negative then  $-LL(j)$ 
 denotes the label of the cluster  
it would start belonging to from now. A label $k$ is proper if $LL(k)\  >\ 0$;
if $LL(k) < 0$ then the label $k$ is called improper. When we come across
 a site  whose neighbours carry improper labels then by examining the 
corresponding entries in the array $LL$, we can find the proper label.
All these issues  would become clear when we consider a concrete example, 
in bond percolation\index{percolation!bond} .  

Consider a two dimensional square lattice. The aim is to assign a cluster 
label to each site so that at the end of scanning we have sites 
having the same 
label belonging to a cluster; we shall have different clusters
having distinct cluster labels\index{cluster label}.
To this end we  decide to scan the 
lattice sites from left to right in a row and then go to the next row below.
 We start with the first row, see Fig.~\ref{hkps1}. We assign cluster
label \ \lq $1$\rq\  to the first site and set $LL(1) =1$. 
The next site 
is bonded to the previous site; hence  we assign the same 
cluster label\index{cluster label} \ \lq $1$\rq\  to this site and update $LL(1) = LL(1) + 1$. 
The next site (the third site in the first row)  
is not bonded to the previous site; hence we assign  
the next new label \ \lq $2$\rq\  to this site 
and set $LL(2) =1$. We proceed in the 
same fashion assigning labels and updating the $ LL$ array.
We complete 
the scanning of the first row. We find, see Fig.~\ref{hkps1},  
the first row assigned with  labels 
$1 - 4$;  
Also  $LL(1)=2$, $LL(2)=1$, $LL(3)=3$, and $LL(4)=1$. 
Then we start scanning the 
second row.
\begin{figure}[tp]
\centerline{\psfig{figure=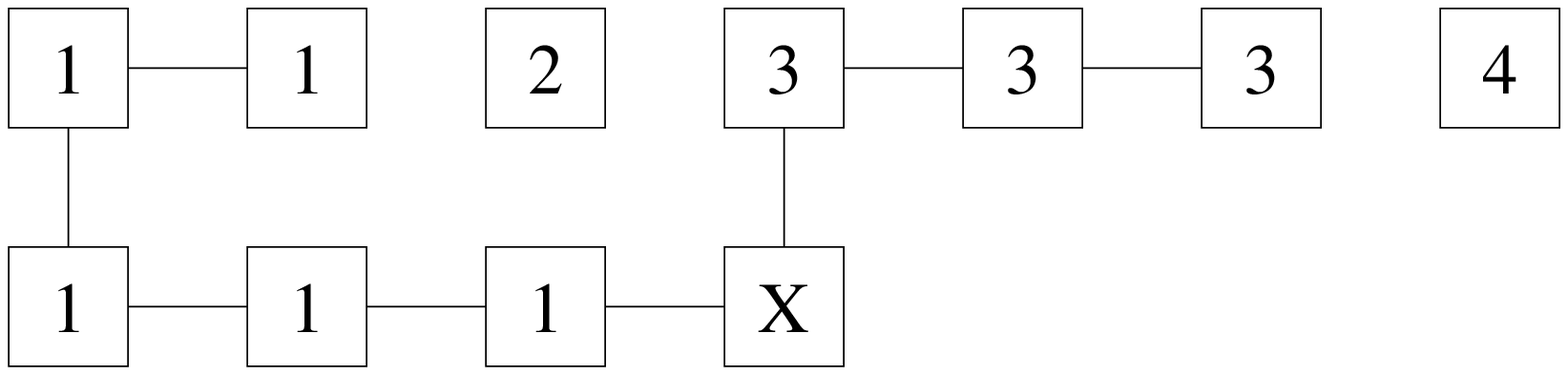,height=1.4in,width=5.54in }}
\caption{\small\protect{Hoshen-Kopelman\index{Hoshen-Kopelman} cluster counting\index{cluster counting} algorithm:
We have to assign a cluster label\index{cluster label} to the site marked X. 
Before labelling the site X, we have 
$LL(1)=5$, $LL(2)=1$, $LL(3)=3$, $LL(4)=1$.
The conflicting labels are $l_1 =1$ and $l_2=3$. We select
the lower of the two labels,  namely $l_1=1$
 as label for the site X. We update 
$LL(1) = LL(1) + LL(3) + 1 = 5+3+1 = 9$.
We set $LL(3) = -l_1 = -1$. 
Thus the label \ \lq  $3$\rq\   has become improper. The 
other  entries in the array $LL$ remain the same. The two clusters
1 and 3 coalesce.   
       }} 
\label{hkps1}
\vskip 3mm
\hrule
\end{figure}
We consider the first site in the second row and 
check if it is bonded to the site at its
top. If yes, then it assigned with the 
cluster label\index{cluster label} of the top site. If not, it is assigned
with the  next new cluster label\index{cluster label}. In Fig.~(\ref{hkps1}) 
the first site in the second
row is assigned with the cluster label\index{cluster label} \ \lq 1\rq\ . 
While scanning the second row, we need to check if the site under 
consideration is bonded to the site at its left (the previous 
site examined) and/or to the site at its top. If only one of these
two sites is bonded to the current site then cluster labelling\index{cluster label} 
and updating of the array LL is carried out exactly the  same way as described 
earlier. You may come across a situation when the 
current site is bonded to both the sites (site at left and at top)
with labels, say $l_1$ and $l_2$. If $l_1 \ne l_2$, there is 
a label conflict. What is the cluster label\index{cluster label} we should assign to the 
current site, $l_1$ or $l_2$~? Fig.~(\ref{hkps1}) illustrates an example 
where the current site is marked X.   
Physically this conflict implies that what we have considered 
until now as two different clusters are in actuality one and the same.
The current site is the link. Such a cluster 
label conflict can be resolved by a second pass
or by a backward relabelling scheme. 
But then this would be time consuming. The ingenuity
of the Hoshen-Kopelman\index{Hoshen-Kopelman} algorithm 
lies in the fact it avoids completely 
the second pass or backward relabelling. 
To understand the algorithm see Fig.~(\ref{hkps1}). 
The current site marked X is linked to the 
site at its left (with label $l_1 =1$) and also to the 
site at its top (with label $l_2 =3$). 
Thus $l_1 (=1)$ and $l_2 (=3)$ are the conflicting labels. 
Let $L=min(l_1 , l_2)$. 
In the example of  Fig.~(\ref{hkps1}), 
$L=l_1 =1$. We assign the label $L=l_1 =1$ to the current site,  
update $LL(L) = LL(L) + LL(l_2) +1$ and set $LL(l_2)=-L$. In the 
example considered, the current site is assigned label \ \lq  $1$\rq\ ~; 
$LL(1) = 5+3+1=9$ and $LL(3)=-1$. The label 
$3$ has now  become improper since $LL(3)$ is negative. 
Before labelling the site X, the clusters 
having labels \ \lq 1\rq\ and \ \lq 3\rq\  were 
distinct. While examining the site X 
we discover that the two clusters are linked
through the site X and hence are not distinct. 
The two clusters, so to say, coalesce. 
We give the lower cluster label\index{cluster label} ($L=1$) to the coalesced cluster, 
store in the array $LL(L)$ the size of
the coalesced cluster and declare the cluster label\index{cluster label} $3$ 
as improper by setting
$LL(3) = -1$.   $LL(3)$ takes the role of a pointer.   
After labelling the 
current site and updating the LL array  we continue with the 
scanning.  
\begin{figure}[tp]
\centerline{\psfig{figure=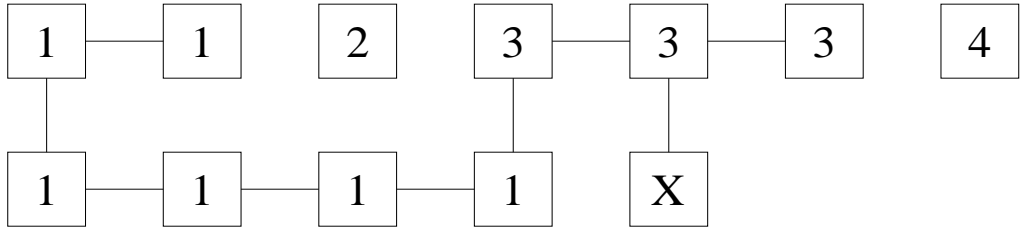,height=1.4in,width=5.54in }}
\caption{\small\protect{Hoshen-Kopelman\index{Hoshen-Kopelman} cluster counting\index{cluster counting} algorithm:
We have to assign a cluster label\index{cluster label} to the site marked X. 
Before labelling the site X, we have 
$LL(1)=9$, $LL(2)=1$, $LL(3)=-1$, $LL(4)=1$. The site X is bonded to the
site above with label \ \lq $ 3$\rq\ . 
Hence we should assign label \ \lq $3$\rq\   
to the site X. But label \ \lq $3$\rq\  is improper
since $LL(3)=-1$. We then check the label \ \lq $1$\rq\ . It is proper since
$LL(1)=9 > 0$. Hence
the root label of $3$ is $1$. {\it i.e.} $l^R _{3} = 1$. 
We assign label\ \lq $1$\rq\  
to the site X and update $LL(1)=LL(1) + 1 = 9+1 =10$.
       }}
\vskip 3mm
\hrule
\vskip 3mm
\label{hkps2}
\end{figure}
There may arise a situation when the current site is bonded to a site
with an improper label, say $l_1$. In other words we find 
$LL(l_1) = - l_2 < 0$. In such a situation we examine 
if $l_2$ is a proper label. If yes, 
then the root label corresponding to $l_1$ is $l_2$. Let 
us denote the root label of $l_1$  by the symbol $l^R _{l_1}$. 
Thus $l^R _{l_1} = l_2$. If on the other hand we find that 
$l_2$ is again an improper label
with $LL(l_2) = - l_3 < 0$,  the search for the root label continues until
we get a proper root label $l^R _{l_1}$. This situation is depicted 
in Fig.~(\ref{hkps2}). The site marked $X$ is bonded to the site 
above with label $l_1 = 3$. But the label $l_1 =3$ is an improper label
since $LL(l_1 =3) = -l_2 =-1 < 0$. The label $l_2=1$ is a proper label
and hence $l^R _{3} =1$. Hence we assign the label $1$
to the current site, update $LL(1) = LL(1) + 1= 10$. 
If there arises a cluster\index{cluster label} label conflict, it is resolved exactly
the same way as described earlier except that the resolution 
is done after searching for the corresponding proper  root labels.
A situation of this type is depicted in Fig.~(\ref{hkps3}). 
We have to assign a cluster label\index{cluster label} to the site marked X.  Before
labelling the 
site X, we have
$LL(1)=10$, $LL(2)=1$, $LL(3)=-1$, $LL(4)=3$, $LL(5)=-4$, $LL(6)=5$.
The conflicting labels are $l_1=6$ and $l_2=5$. The label
$l_1=6$ is a proper label. The label $l_2=5$ is however  an improper label,
since $LL(5)=-4$. Hence we examine the label $4$. It is proper since $LL(4) = 3 > 0$. 
Therefore $l^R _{l_1}=l^R _6 = 6$ and $l^R _{l_2}=l^R _5 = 4$. 
We assign the smaller label 
\ \lq $4$\rq\  to the site, X  update $LL(4) = LL(4) + LL(6) + 1 = 3+5+1 = 9 $, 
and set $LL(6)=-5$.  The label \ \lq 6\rq\ has now become improper. 
In this fashion
we complete scanning the entire cluster. In the end we shall have
an array   $\{ LL(i) : i=1,\ 2,\ \cdots\}$. The positive entries
in this array denote the sizes of the different clusters present in the 
system.
The Hoshen-Kopelman\index{Hoshen-Kopelman} algorithm is quite general and can be easily 
employed to carry out cluster counting\index{cluster counting} in a general $d$- dimensional lattice
with arbitrary coordination number\index{coordination number}. The boundary conditions 
can also be easily taken care of in cluster counting\index{cluster counting}. In the examples 
discussed above we have not considered any boundary conditions. 
Babalievski~\cite{FB}\index{Babalievski} has extended the Hoshen-Kopelman\index{Hoshen-Kopelman} algorithm to random and 
aperiodic lattices. Hoshen, Berry\index{Berry} and Minser\index{Minser}~\cite{HBM} have proposed an
enhanced Hoshen-Kopelman\index{Hoshen-Kopelman} algorithm that enables efficient
calculation of cluster spatial moments, radius of gyration of each
cluster, cluster perimeter {\it etc.}  Indeed more recently 
Ahmed Al-Futaisi\index{Ahmed Al-Futaisi} and Patzek\index{Patzek}
 have extended the Hoshen-Kopelman\index{Hoshen-Kopelman} 
algorithm  to non-lattice environments~\cite{aafp}. We can say that  
the Swendsen-Wang\index{Swendsen-Wang algorithm}\index{Swendsen}\index{Wang} 
cluster algorithm\index{cluster algorithm}  
coupled with the Hoshen-Kopelman\index{Hoshen-Kopelman}  counting constitutes
a major step forward in Monte Carlo\index{Monte Carlo} simulation.   
\begin{figure}[bh]
\centerline{\psfig{figure=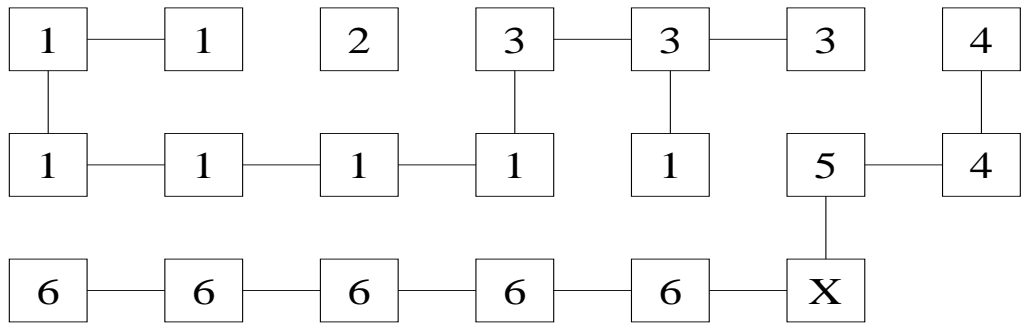,height=1.8467in,width=5.54in }}
\caption{\small\protect{Hoshen-Kopelman\index{Hoshen-Kopelman} cluster counting\index{cluster counting} algorithm:
We have to assign a cluster label\index{cluster label} to the site marked X. 
Before labelling the site X, we have  
$LL(1)=10$, $LL(2)=1$, $LL(3)=-1$, $LL(4)=3$, $LL(5)=-4$, $LL(6)=5$.
The conflicting labels are $l_1=6$ and $l_2=5$. The label
$l_1=6$ is a proper label since $LL(6)=5$. 
The label $l_2=5$ is an improper label,
since $LL(5)=-4$. Hence we examine the label $4$. The label \ \lq  $4$\rq\ 
is a proper label since $LL(4) = 3 > 0$.  
Therefore the root labels are  $l^R _{6} =6$ and $l^R _{5} =4$. We assign label 
\ \lq $4$\rq\  (the smaller of the two root labels) 
to the site X, update $LL(4) = LL(4) + LL(6)+  1 = 3+5+ 1 = 9$, 
and set $LL(6)=-5$. The label \ \lq 6\rq \ has become improper.
       }} 
\label{hkps3}
\vskip 3mm
\hrule
\end{figure}
 
\section
{What are the improvements to the Swendsen-Wang\index{Swendsen-Wang algorithm}\index{Swendsen}\index{Wang} 
cluster algorithm\index{cluster algorithm}?}\label{cluster_more}

\subsection{Wolff algorithm}\label{wolff_algorithm}
 
In the Swendsen-Wang\index{Swendsen-Wang algorithm} algorithm, all the clusters, big or small are grown and 
updated. Small clusters do not influence the 
critical slowing down\index{critical slowing down} and
hence computing efforts toward growing them are a waste. This is somewhat 
eliminated in the algorithm proposed by U. Wolff\index{Wolff algorithm}~\cite{wolff}. 
A site, say \ \lq$i$\rq\ is chosen randomly and a single cluster is 
grown from it as follows. 
If a  neighbouring spin has the same orientation 
then the bond to it is \ \lq satisfied\rq\  and hence is occupied
with a 
probability $p$ given by Eq.~(\ref{bondprob}): 
Select a random number; if it is less than $p$ 
occupy the bond; if not do not occupy. 
This process
is repeated on all the neighbours of the site $i$ and
 further on all the neighbours of the 
new sites that get bonded. 
The set of bonded sites constitutes a cluster and the cluster grows.   
The growth of the cluster would eventually terminate. 
Then all the spins in the cluster are flipped; the bonds are removed; we have
a new spin configuration. 
The whole process is 
iterated. The measurement of Monte Carlo\index{Monte Carlo!step per spin (MCSS)} time becomes problematic
in the Wolff\index{Wolff algorithm} algorithm since the number of spin-flips is different
in different steps. Usually the time advanced in a cluster flip is taken
as the ratio of the number of spins flipped to the 
total number of spins in the system.  

The Wolff\index{Wolff algorithm} algorithm reduces significantly the correlation times and the 
dynamical critical exponent\index{exponent!dynamical critical} $z$. Hence for large lattices the 
Wolff\index{Wolff algorithm} algorithm would prove much better than the local update
Metropolis\index{Metropolis!algorithm} algorithm. 

An important point about the  Swendsen-Wang\index{Swendsen-Wang algorithm}/Wolff\index{Wolff algorithm}  
\index{Swendsen}\index{Wang} algorithm is that a KF-CK cluster\index{Kastelyn - Fortuin - Coniglio - Klein (KF-CK) cluster}  
\index{Coniglio}\index{Klein}\index{Kasteleyn}\index{Fortuin}
percolates at the Ising\index{Ising} critical point.
Indeed one can show that the spin-spin pair correlation function 
$\left\langle S_i S_j\right\rangle $ is equal 
to $\left\langle \gamma_{ij}\right\rangle $ - the probability that 
the sites $i$ and $j$ belong to the same cluster. 
Here $\gamma _{ij}$ is an indicator 
function\index{indicator function}: it is unity if the sites belong to 
the same cluster and zero otherwise. The average in the above 
is taken over all spin and bond configurations. 
This property suggests that there is a natural tendency 
for the system to self organize itself to criticality\index{self organized criticality}; this tendency 
for selforganized criticality can be captured in an algorithm.  

\subsection{Invaded cluster algorithm}\label{invaded_cluster}

Machta {\it et al}~\cite{mclsc} proposed
\index{Machta}
the  invaded cluster algorithm\index{invaded cluster algorithm}
to calculate the equilibrium critical points. This technique is 
based on the idea of invasion 
percolation~\cite{invasion-percolation}\index{percolation!invasion} .  
In invasion percolation\index{percolation!invasion}  we assign 
random numbers independently to the bonds of the lattice.
Growth starts from one or several sites. 
At each step a cluster grows by the addition of 
the perimeter bond which has the smallest random number. 
The growth process stops when a cluster 
percolates. It is clear that the invasion percolation\index{percolation!invasion}  
is a self organized critical\index{self organized criticality}  phenomenon. 
In the  algorithm of Machta {\it et al}~\cite{mclsc}\index{Machta}
 we grow clusters 
from every site of the lattice and  
along the \ \lq satisfied\rq\  bonds only. Bonds are occupied in a random 
order until an appropriate stopping condition is fulfilled. The growth
process is stopped and each cluster is flipped randomly and independently
with equal probability. This gives rise to a new spin configuration;
in the new configuration re order the satisfied bonds randomly and 
start growing invaded clusters. Iterate.  For 
example the stopping condition may be a requirement on the size of the 
largest cluster. The stopping rule can be devised  suitably so that 
the Invaded Cluster algorithm simulates the system at critical point of the 
model. 
Thus the occupied 
bonds are determined 
by the stopping rule. \footnote{Note that in the 
Swendsen-Wang\index{Swendsen-Wang algorithm}\index{Swendsen}\index{Wang} algorithm it is the temperature
that determines the occupation of satisfied bonds. See sections (\ref{s_w_algorithm}) and 
(\ref{bonding_probability}).} 
 Hence the invaded cluster algorithm\index{invaded cluster algorithm} has a 
natural feed back mechanism 
which ensures that  the desired equilibrium critical 
regions are always  reached. The phase transition 
temperature is an output. For more details see~\cite{mclsc}.

\subsection{Probability changing cluster algorithm}\label{pcc_algorithm}

Tomita\index{Tomita} and Okabe\index{Okabe}~\cite{to} have proposed an  
algorithm 
wherein the system evolves to critical regions 
but asymptotically remains still a canonical
ensemble\index{ensemble!canonical}. They call this a  probability changing 
cluster (PCC)  algorithm. 
The bonding probability\index{bonding probability} is adjusted at the end of 
every step or a group of steps so that the 
spin system is driven to criticality. We start 
with an Ising\index{Ising} spin configuration, and a 
certain value of the bonding probability\index{bonding probability} $p$.  
As in the Swendsen-Wang\index{Swendsen-Wang algorithm}\index{Swendsen}\index{Wang} algorithm,  we grow 
the KF-CK
 clusters\index{Kastelyn - Fortuin - Coniglio - Klein (KF-CK) cluster};
\index{Coniglio}\index{Klein}\index{Kasteleyn}\index{Fortuin} then  
check if there is present
a spanning cluster\index{spanning cluster} ($k=1)$ or not $(k=-1)$. 
Carry out the cluster-spins-flipping 
exactly like you did in the Swendsen-Wang\index{Swendsen-Wang algorithm}\index{Swendsen}\index{Wang} 
algorithm. For the next step change 
$p$ to $p-k\delta p$. Iterate.  
It is clear that 
in the limit $\delta p \to 0$, the PCC ensemble 
is identical to the canonical ensemble\index{ensemble!canonical}. 
Hence the algorithm not only drives 
the system to criticality and hence the 
critical points can be estimated  but also
the physical quantities like magnetization, 
susceptibility\index{susceptibility}, specific heat\index{specific heat}, {\it etc} can  be 
evaluated. For more details see~\cite{to} 

\subsection{More on clusters and some discussions}\label{more_clusters_discussions}

I think
it is rather clear now why the Swendsen-Wang\index{Swendsen-Wang algorithm}\index{Swendsen}\index{Wang} algorithm in particular and 
several other cluster
algorithms (like the Wolff\index{Wolff algorithm}, invaded  cluster\index{invaded cluster algorithm} 
and the probability changing cluster 
algorithms, discussed above) in general 
are very effective in speeding up the computations;
this  stems from a very simple observation   that a  
KF-CK cluster\index{Kastelyn - Fortuin - Coniglio - Klein (KF-CK) cluster} 
\index{Coniglio}\index{Klein}\index{Kasteleyn}\index{Fortuin}
contains correlated spins; if a spin 
is flipped then all the other spins 
in the same KF-CK cluster\index{Kastelyn - Fortuin - Coniglio - Klein (KF-CK) cluster} 
\index{Coniglio}\index{Klein}\index{Kasteleyn}\index{Fortuin}
have a tendency to flip. It is precisely 
this tendency for coherent flipping that is exploited 
when we flip all the spins in a cluster in one shot thus speeding up the 
dynamics. By the same token,  in a physical phenomenon if a  
KF-CK cluster\index{Kastelyn - Fortuin - Coniglio - Klein (KF-CK) cluster}
\index{Coniglio}\index{Klein}\index{Kasteleyn}\index{Fortuin}
ceases to represent correlated spins, then the  
Swendsen-Wang\index{Swendsen-Wang algorithm}\index{Swendsen}\index{Wang} and the related algorithms   would fail.  
An example is the temperature
driven first order phase transition\index{phase transition!first order} studied by 
Gore\index{Gore} and Jerrum\index{Jerrum}~\cite{gj};
they show that the Swendsen-Wang\index{Swendsen-Wang algorithm}\index{Swendsen}\index{Wang}
 cluster algorithm is not efficient.
The so called super critical slowing down\index{super critical slowing down} of dynamics in a first order
phase transition\index{phase transition!first order} has its origin in the nature of the canonical
distribution: it peaks at two phases and the system switches 
from one to the other through a rare\index{rare events} \ \lq  interface\index{interface} event\rq\  . 
The correlation time $\tau^{\star}$ is proportional to the residence time
in the phases. The \ \lq interface\index{interface}\rq\  regions are poorly sampled
in the local update Monte Carlo\index{Monte Carlo} algorithm. 
To handle such super critical slowing down\index{super critical slowing down},
we need to go beyond Boltzmann\index{Boltzmann} sampling,  
like multicanonical Monte Carlo\index{Monte Carlo!multicanonical}. I shall 
say something  on these issues later.    

For another example consider the following disordered Ising\index{Ising} Hamiltonian\index{Hamiltonian}
\begin{eqnarray}
H & = & - J \sum_{\left\langle ij\right\rangle} \left(
         \epsilon_{ij}S_i S_j  - 1\right)\ ,
\end{eqnarray}
where $\epsilon_{ij} (= \pm 1)$ is random quenched disorder\index{quenched disorder}. If there 
exists a closed path such that that the product of 
the disorder parameter $\epsilon$
over the closed path is -1, then the Hamiltonian\index{Hamiltonian} is said to contain 
frustration\index{frustration}. A naive application of 
Swendsen-Wang\index{Swendsen-Wang algorithm}\index{Swendsen}\index{Wang} 
 cluster algorithm  to a fully frustrated 
Ising\index{Ising} system will not lead to any improvement whatsoever~\cite{acp}; it is because,   
in a frustrated Ising\index{Ising} system,  
a KF-CK cluster\index{Kastelyn - Fortuin - Coniglio - Klein (KF-CK) cluster}
\index{Coniglio}\index{Klein}\index{Kasteleyn}\index{Fortuin} does not contain
 correlated spins. 
One needs to modify the cluster 
definition to handle these situations. 
Several ideas have been proposed to this end and 
we refer to some of them in~\cite{franzes}.

We need to carry out a series of 
Monte Carlo\index{Monte Carlo} simulations at different temperatures to obtain the 
temperature dependence of the desired thermodynamic variable.
This can be time consuming.  It would be desirable to
have a technique that would give the temperature dependence 
of a  macroscopic variable from a single Monte Carlo\index{Monte Carlo} run. Also,
in the standard Metropolis\index{Metropolis!algorithm} Monte Carlo\index{Monte Carlo} technique, 
it is rather difficult to calculate entropy\index{entropy} 
(or free energy\index{free energy}), since 
the entropy\index{entropy} is a function of the probability with which a 
microstate occurs in a canonical ensemble\index{ensemble!canonical}. The 
energy distribution in the canonical ensemble\index{ensemble!canonical}, 
$P(E)\propto D(E)\exp (-\beta E)$,
is the product of density of states\index{density of states} $D(E)$ which 
usually increases with increase of
energy rather rapidly and an exponential\index{exponential} which 
decreases rather sharply with increase of energy
at low temperatures. As a result, microstates 
with very low or high energies get
sampled rather poorly  in Metropolis\index{Metropolis!algorithm} algorithm. 
If the energy minima are separated
by high barriers, the simulation becomes difficult since 
the system becomes 
quasi  ergodic\index{quasi-ergodic}. 
Several techniques have been proposed addressing 
these and related issues. I shall discuss
some of them in the remaining part of the talk.  
First let us consider the so called single  histogram\index{histogram}  technique
often used in the context of simulation of critical systems.  

\section{What is an histogram technique\index{histogram} ? How do we  implement it?}
\label{histogram_technique}

Consider a canonical ensemble\index{ensemble!canonical} of spin configurations 
generated at  inverse temperature  $\beta^{\star}$.
The idea behind histogram technique\index{histogram}  is to calculate from this ensemble
the macroscopic properties of the system
at  $\beta \ne \beta ^{\star}$. 
 Let me illustrate 
the technique by considering energy histogram\index{histogram}. 

Accordingly, let  $h_i ( \beta ^{\star})$ denote 
the histogram\index{histogram}  of the energy obtained from a 
canonical ensemble\index{ensemble!canonical} of spin configurations generated at $\beta ^{\star}$. 
In other words, $h_i ( \beta ^{\star})$ is the number of spin 
configurations with energy
in the interval $\Delta E$ around $E_i$. Let $P(E_i , \beta ^{\star})$ 
denote the 
probability  
density that the system takes an energy $E_i$. 
It is clear that $P(E_i , \beta ^{\star})$
 is proportional to $h_i (\beta ^{\star})$. Formally we  have
\begin{equation}
P(E_i , \beta ^{\star}) ={{1}\over{Z(\beta ^{\star}) }}   
D(E_i ) \exp\left( -  \beta ^{\star} E_i \right),
\end{equation}
where $D(E_i)$ is density of states\index{density of states} at energy $E_i$ 
and is independent of temperature.  Therefore 
we have
\begin{equation}\label{dos}
D(E_i) \propto Z(\beta ^{\star}) h_i ( \beta ^{\star}) 
\exp \left( \beta ^{\star} E_i \right).
\end{equation}
 Let $P(E_i , \beta )$ denote the probability 
density of energy at $\beta  \ne \beta ^{\star}$. Formally
we have
\begin{equation}
P(E_i , \beta   ) ={{1}\over{Z(\beta )}}  
D(E_i) \exp\left( - \beta E_i \right) .
\end{equation}
In the above, we substitute for $D(E_i)$ from Eq.~(\ref{dos}) and  get
\begin{equation}
P(E_i , \beta )\propto  {{Z(\beta ^{\star})}\over{Z(\beta )}} 
h_i ( \beta ^{\star}) 
\exp\left[ - \left( \beta -\beta ^{\star}\right) E_i 
   \right] .
\end{equation}
Imposing the normalization condition, we get
\begin{equation}
P(E_i , \beta )= {{ h_i ( \beta ^{\star})
             \exp\left[ - \left( \beta - \beta ^{\star}\right)  E_i  
                   \right] }
               \over{ 
                  \sum_j h_j ( \beta ^{\star})
\exp\left[ -\left( \beta - \beta ^{\star} \right) E_j  
                      \right]}}\ .
\end{equation}
Thus from a single Monte Carlo\index{Monte Carlo} run 
at $\beta ^{\star}$ we  obtain $P(E, \beta )$ for various 
values of $\beta  \ne \beta ^{\star}$. 
Any energy dependent  macroscopic quantity as a function of
$\beta $ can now be obtained from the knowledge of $P(E,\beta )$. 

Another way of saying the same thing is as follows. We sample a microstate
${\cal C}$ from a probability distribution  
$\exp [-\beta^{\star}E({\cal C})]/Z(\beta^{\star})$. We say that 
the microstate  ${\cal C}$ 
is generated with a weight $\exp [-\beta^{\star}E({\cal C})]$.
We are 
interested in evaluating the average of a macroscopic quantity $O({\cal C})$
over a canonical ensemble\index{ensemble!canonical} at $\beta \ne \beta^{\star}$ {\it i.e.} from 
a distribution $\exp [-\beta E({\cal C})]/Z(\beta)$. Hence we unweight 
the sampled microstate and then reweight it at the 
desired temperature; the unweighting - reweighting\index{reweighting}  
factor for the configuration ${\cal C}$ is
\begin{equation}
{{\exp[-\beta E({\cal C})]}\over{\exp [-\beta^{\star}E({\cal C})]}}
\end{equation}
Therefore,
\begin{eqnarray}
\left\langle O\right\rangle = {}^{Lim}_{N\to\infty} {\overline O}_N    
 &  = &  {{
             \sum_{i=1}^{N} O( {\cal C}_i ) \exp\left[ - 
               \left( \beta -\beta^{\star}\right) E({\cal C}_i )\right]  }
              \over{
             \sum_{i=1}^{N}  \exp\left[ - 
               \left( \beta -\beta^{\star}\right) E({\cal C}_i)\right]  }}
\end{eqnarray}
where the sum runs over the  microstates sampled from a canonical 
ensemble at $\beta^{\star}$. 
If $O(C_i) = O(E(C_i))$, then,
\begin{eqnarray}
\left\langle O\right\rangle & = &
    {{
             \sum_i O( E_i ) h_i (\beta ^{\star})\exp\left[ -
               \left( \beta -\beta^{\star}\right) E_i \right]  }
              \over{
             \sum_i h_i (\beta ^{\star})  \exp\left[ -
               \left( \beta -\beta^{\star}\right) E_i \right]  }}
\end{eqnarray}
where the sum runs over the energy bins. The above is identical
to energy histogram\index{histogram}  reweighted average. 

The histogram\index{histogram}  technique is one of the oldest techniques
 proposed, see for example~\cite{mcds,tv1}.  
But the technique became popular after the publication of   
a  paper by Ferrenberg\index{Ferrenberg} and Swendsen~\cite{fs}\index{Swendsen} in 1988. 
This technique is often referred to
as  Ferrenberg-Swendsen reweighting\index{reweighting}\index{Ferrenberg}\index{Swendsen} 
technique and is used in almost all Monte Carlo\index{Monte Carlo}
calculations of statistical physics problems, especially the ones 
dealing with the phenomenon of phase transition, see for example \cite{vsm}.\index{phase transition} 

In the single histogram\index{histogram}  technique, the estimated $P(E,T)$ is accurate 
only for $T$ close to the reference temperature $T^{\star}$. By 
generating many histograms\index{histogram}  that overlap we can widen the range of $T$.
This is called the  multi histogram\index{histogram}  technique~\cite{multihistogram}. 
It is also clear
that we can increase the range  of $T$ by directly estimating  the density
of states\index{density of states} $D(E)$.  
Multicanonical\index{Monte Carlo!multicanonical} sampling~\cite{multicanonicalsampling} 
is an early technique proposed that precisely does this.
Multicanonical sampling is a very general technique useful and often 
the method of first choice for a variety of problems that include
critical slowing down\index{critical slowing down} 
near second order phase transition\index{phase transition!second order} points, 
nucleation\index{nucleation}  in first order phase transition\index{phase transition!first order}, 
and trapping in the metastable
minima in systems with rugged energy landscapes. Originally this technique 
was proposed to tackle the so called super critical slowing down\index{super critical slowing down} in the 
first order phase transition\index{phase transition!first order} and calculation of the 
interfacial energy\index{interfacial energy} 
for large systems.
This problem of super critical slowing down\index{super critical slowing down} 
 arises due to the fact that the Boltzmann\index{Boltzmann} factor
suppresses configurations dominated by interfaces\index{interface} between ordered and 
disordered phases. As a consequence, the quality of the local
update Metropolis\index{Metropolis!algorithm} Monte Carlo\index{Monte Carlo} results deteriorates exponentially with 
increase of the system size\index{system size}.   
 Entropic sampling~\cite{entropicsampling}\index{Monte Carlo!entropic sampling}
  is a technique conceptually
equivalent to multicanonical\index{Monte Carlo!multicanonical} Monte Carlo. 

\section{What is the basic idea behind entropic sampling\index{Monte Carlo!entropic sampling} ?}
\label{entropic_sampling}

The canonical partition function\index{partition function!canonical} is given by,
\begin{eqnarray}
Z(\beta)&=&\sum_{{\cal C}} \exp [-\beta E({\cal C})]\ ,\nonumber\\
        & &                                            \nonumber\\
        &=& \sum_E \exp\left[ {{S(E)}\over{k_{{\rm B}}}} - \beta E\right]\ ,
\end{eqnarray}
where $S(E)$ is the microcanonical entropy\index{entropy}\index{microcanonical entropy}. 
The probability for the macroscopic system to have an energy $E$ in the 
canonical ensemble\index{ensemble!canonical} is given by,
\begin{eqnarray}\label{Boltzmann-distribution}
P(E) & \propto &  \exp\left[ {{S(E)}\over{k_{{\rm B}} }} - \beta E\right] \ .
\end{eqnarray}
The Metropolis\index{Metropolis!algorithm} algorithm helps you sample from the above Boltzmann\index{Boltzmann}
distribution and construct a canonical 
ensemble\index{ensemble!canonical} of microstates. 
Instead, if we want 
to sample from an arbitrary distribution 
\begin{eqnarray}\label{energypdf}
P_{\alpha (E)} (E) &\propto  &\exp \left[ {{S(E)}\over{k_{{\rm B}} }} -   \alpha (E)\right]\ ,
\end{eqnarray}
it is sufficient to impose, 
besides ergodicity\index{ergodicity}, a detailed  balance condition  
given  by
\begin{eqnarray}\label{equiv_balance}
{{ W(B\leftarrow A)}\over{W(A\leftarrow B)}} 
&=& {{ P_{\alpha (E)}I \left( E \left( B\right)\right)}\over{
                 P_{\alpha (E)}  \left( E \left( A\right)\right)}}\nonumber\\ 
&  & \nonumber\\
& = & \exp \bigg[  - \bigg\{   \alpha  \left( E(B)\right)  - 
\alpha \left( E(A) \right) \bigg\} \bigg] \ .
\end{eqnarray}
This implies that a trial configuration 
${\cal C}_t$ obtained by a local updating of the 
current configuration ${\cal C}$, is accepted with a probability,
\begin{eqnarray}\label{metro_entropy}
p & = & {\rm min} \bigg( 1, \exp\bigg[ - \bigg
\{  \alpha \left( E\left( {\cal C}_t \right)\right) -
 \alpha \left( E\left( {\cal C} \right)\right)
\bigg\}\bigg]\bigg)\ .
\end{eqnarray}
It is clear that if $\alpha (E) = \beta E$, 
Eq. (\ref{energypdf}) is the same as Eq. (\ref{Boltzmann-distribution}) and 
we recover the conventional Boltzmann\index{Boltzmann} 
sampling. For other choices of the function $\alpha (E)$, 
we get what
one may call 
 non-Boltzmann\index{Boltzmann}\index{non-Boltzmann sampling}
 sampling. The fact that non-Boltzmann\index{Boltzmann} sampling can  
be a legitimate alternative to Boltzmann\index{Boltzmann} sampling 
 has been recognized since the early days 
of Monte Carlo\index{Monte Carlo} practice.
However the practical significance of non-Boltzmann\index{Boltzmann} sampling was realized, 
only in the middle of nineteen seventies  for the 
first time, by Torrie\index{Torrie} and Valleau~\cite{tv1,TV}\index{Valleau}
who proposed the  umbrella sampling\index{umbrella sampling} technique, 
a forerunner to all the non-Boltzmann\index{Boltzmann}\index{non-Boltzmann sampling} sampling technique
like the entropic sampling\index{Monte Carlo!entropic sampling}, 
multicanonical Monte Carlo\index{Monte Carlo!multicanonical} 
and its several and recent variants, see below. 
 
Entropic
 sampling\index{Monte Carlo!entropic sampling} obtains when we set
$\alpha (E) =  S(E)/k_{{\rm B}}$  
that renders $P_{\alpha (E)} (E)$, 
see Eq.~(\ref{energypdf}),  the same for all $E$. But then, the 
entropic function is not known {\it \`a priori}. An algorithm to
build  $\alpha (E)$  iteratively on the basis of microstates visited
during short \ \lq learning\rq\  runs is usually recommended, see below.  

The key point is that  entropic sampling\index{Monte Carlo!entropic sampling} gives  an ensemble of 
microstates with energies uniformly distributed 
over its range. We can say that 
the algorithm generates a simple random walk\index{random walk} in energy space, and hence the 
presence of energy barriers does not affect the evolving 
Markov\index{Markov!chain} chain of microstates.  
Since unweighting - reweighting\index{reweighting} is done while averaging, even a 
reasonably approximate estimate of the  entropic  function 
$\alpha (E)$ would suffice.
The idea is that the  closer $\alpha (E)$ is to  
 $S(E)/k_{{\rm B}}$,
the flatter shall be the energy histogram\index{histogram}  it generates. 

Let $(E_{{\rm min}}, E_{ {\rm max}})$ be the energy 
range over which you want to 
generate microstates with uniform energy distribution\index{uniform distribution}. 
For example this range could include
the region of the sample  space where the probability is small like the 
interface\index{interface} region in a first order phase
transition\index{phase transition!first order}. Divide the range into a number of 
equal width energy bins. Let $E_i$ be the energy
of the $i-$th bin. In the first iteration 
set $\alpha_i = \alpha (E_i) = 0\  \forall\   i$. Carry out 
a small Monte Carlo\index{Monte Carlo} run, of say $N$  MCSS\index{Monte Carlo!step per spin (MCSS)}. 
The probability of acceptance of a trial
state constructed from locally changing the current state 
is given by Eq.~(\ref{metro_entropy}).
Since in the first iteration, $\alpha_i = 0\  \forall\  i$, 
every trial move gets accepted. But this will
not be so in subsequent iterations. From the microstates 
visited by the system in the first iteration stage,
calculate the histogram\index{histogram}   $\{ h_i \}$, where  
the number of microstates falling in the $j-$th energy bin
is denoted by $h_j$. At the end of the first iteration, 
calculate $\{ \alpha_i \}$ for the next iteration, as 
per the recursion\index{recursion} given below,
\begin{eqnarray}\label{recursion_alpha}
\alpha ^{(k+1)}_i  =\cases {     
\alpha ^{(k)}_i  \   &  if \    $h_i = 0 $\ , \cr & 
                    \cr
                    \alpha ^{(k)}_i +   
                       {{1}\over{k_{{\rm B}}}} \ln \left[ h_i\right]  & otherwise. \cr}
\end{eqnarray}  
where the superscript $(k)$ denotes the iteration 
index.

 Thus, starting from an initial
array of $\{\alpha ^{(0)}_i = 0\  \forall\  i\ \}$ 
in the zeroth iteration,  we successively calculate 
$\left\{ \{ \alpha ^{(k)}_i\}\ : k=1,\ 2,\ \cdots\right\}$. 
We  continue the iterations 
until we find that the histogram\index{histogram}  of energy accumulated from the
microstates visited during the iteration is approximately flat in the 
desired energy  
$(E_{min},\ E_{max})$.  
After the \ \lq learning\rq\   runs are over,
we carry out a long Monte Carlo\index{Monte Carlo} run, employing the 
set $\{ \alpha_i \}$, for calculating 
the macroscopic properties. We get an
{\it entropic}  
ensemble\index{ensemble!entropic}  of microstates $\{ {\cal C}_i \ :\ i=1,N\}$. 
Calculate the canonical ensemble\index{ensemble!canonical} average  of a macroscopic property
say, $O({\cal C})$
at the desired temperature $T=1/[k_{{\rm B}} \beta]$, 
from the entropic  ensemble\index{ensemble!entropic}, by unweighting (divide by 
$\exp [-\alpha(E( {\rm C}_i))]$) followed by  reweighting (multiply by $\exp [\-\beta E( {\rm C}_i)]$\index{reweighting},
\begin{equation}
\left\langle O  \right\rangle_{\beta}
 = {}^{ \ Lim.}_{N\to\infty} \ \ {\overline O}_N (\beta)=  
{{
                                \sum _{i=1}^{N} O( {\cal C}_i ) 
                          \exp \left[  -\beta E({\cal C}_i)  
                             +\alpha ( E ( {\cal C}_i ))  \right]}
                   \over{
                                \sum _{i=1}^{N}  
                          \exp \left[  -\beta E({\cal C}_i)  
                             +\alpha ( E ( {\cal C}_i ) )  \right]}}.
\end{equation}
Let $\xi (\beta)$ denote the maximum of 
$\{ -\beta E({\cal C}_i )  +\alpha (E({\cal C}_i))\ :\ i=1,N\}$.
While reweighting\index{reweighting}, we have found it necessary to employ the modified 
but equivalent formula below, when the system size\index{system size} is very large.
\begin{equation}
\left\langle O  \right\rangle_{\beta}
 = {}^{\  Lim.}_{N\to\infty} \ \ {\overline O}_N (\beta)=  
{{
                                \sum _{i=1}^{N} O( {\cal C}_i ) 
                          \exp \left[  -\beta E({\cal C}_i)
                             +\alpha ( E ( {\cal C}_i ))
                             -\xi (\beta)
                          \right]}
                   \over{
                                \sum _{i=1}^{N}  
                          \exp \left[  -\beta E({\cal C}_i)
                             +\alpha ( E ( {\cal C}_i ) )
                             -\xi (\beta)   
                              \right]}}.
\end{equation}

Let me repeat: since unweighting -  reweighting\index{reweighting} is any way 
carried out, there is no need to estimate 
the entropic function $\{ \alpha_i\}$
accurately. Suffice if the algorithm ensures a relatively 
flat histogram\index{flat histogram}  of energy in the desired range. 
It is a good idea
to increase the number of MCSS\index{Monte Carlo!step per spin (MCSS)} in an \ \lq $\alpha$ update  
iteration\rq\  as the energy range covered becomes larger. 
The important point is that in one shot 
we get $\langle O\rangle$ at \ \lq all\rq\   
temperatures, in addition to sampling of rare events\index{rare events}. 

I said  earlier that in principle  entropic sampling\index{Monte Carlo!entropic sampling}~\cite{entropicsampling}
(described above)  is the same as  
multicanonical\index{Monte Carlo!multicanonical}
sampling technique~\cite{multicanonicalsampling} proposed earlier. 

\section{How is entropic sampling\index{Monte Carlo!entropic sampling} related 
to multicanonical\index{Monte Carlo!multicanonical} sampling ? }
\label{entropic_multi_canonical}

That entropic sampling\index{Monte Carlo!entropic sampling} is in principle the same as 
multicanonical\index{Monte Carlo!multicanonical} sampling  has been shown lucidly by 
Berg, Hansmann and Okamoto~\cite{BHO}.\index{Berg}\index{Hansmann}\index{Okamoto}  
The probability of energy in multicanonical\index{Monte Carlo!multicanonical} sampling technique is 
usually parametrized as,
\begin{eqnarray}
 P(E) & \sim & \exp\left[ {{S(E)}\over{k_{{\rm B}} }} - 
\tilde{\beta}(E)\ E+\tilde{\alpha}(E)\right]\ .
\end{eqnarray}
The parameter $\tilde{\beta}$ in the above is given by the derivative of the 
microcanonical entropy\index{entropy}\index{microcanonical entropy},
\begin{eqnarray}
\tilde{\beta}(E) & = & {{1}\over{k_{{\rm B}}  T(E)}} = 
{{1}\over{k_{{\rm B}} }}{{\partial S}\over{\partial E}}.
\end{eqnarray}
It is precisely because of this the technique is named  
multicanonical sampling\index{Monte Carlo!multicanonical}: a microcanonical temperature
\index{microcanonical temperature} is defined for
each energy. 
The second parameter $\tilde{\alpha} (E)$ can be 
expressed as 
\begin{eqnarray}
\tilde{\alpha}(E) &=& \tilde{\beta}(E)\ F(E)\ ,
\end{eqnarray}  
where $F(E)$ is  microcanonical free energy\index{free energy}.
  
The two parameters $\tilde{\alpha}(E)$  and $\tilde{\beta} (E)$ are related: 
\begin{equation}
\tilde{\alpha} (E) =- {{S(E)}\over{k_{{\rm B}} }} + \tilde{\beta}(E) E\ .
\end{equation}
Differentiating the above with respect to $E$, we get,
\begin{equation}\label{entropy_multicanonical}
{{d\tilde{\alpha }}\over{dE}} = E {{d\tilde{\beta}}\over{dE}} \ .
\end{equation}
We have to build up the  function $\tilde{\beta}(E)$
recursively like we built  $\alpha (E)$ 
in  entropic sampling\index{Monte Carlo!entropic sampling}. 
It is clear that the functions $\alpha (E)$ of entropic sampling,
$\tilde{\beta}(E)$ and $ \tilde{\alpha}(E)$   of 
multicanonical\index{Monte Carlo!multicanonical} sampling are related. Note that $\alpha (E) = S(E)/k_{{\rm B}}$. Hence,
\begin{eqnarray}
\alpha (E) &=& \tilde{\beta}(E) E - \tilde{\alpha}(E)\nonumber\\
           &  &                                      \nonumber\\
{{d\alpha}\over{dE}}& =& \tilde{\beta}(E).
\end{eqnarray}
The recursion\index{recursion} for updating $\tilde{\beta}(E)$ in  
multicanonical sampling \index{Monte Carlo!multicanonical}
Monte Carlo\index{Monte Carlo} can be derived from the recursion\index{recursion}, see Eq.~(\ref{recursion_alpha}), 
 for updating $\alpha (E)$ in the 
entropic sampling Monte Carlo\index{Monte Carlo!entropic sampling} and is given by, 
\begin{eqnarray}
\tilde{\beta}^{(k+1)}_i &=& \tilde{\beta}^{(k)}_i + {{1}\over{k_{{\rm B}}\ \Delta E}}
\times \cases {\ln(h_i) & if $h_i \ne 0$ and $h_{i-1} = 0$\cr
                       &                                 \cr
              \ln(1/h_{i-1}) &  if $h_i =0$ and $h_{i-1}\ne 0$\cr
                            &                                \cr
               0 & if $h_i = h_{i-1} = 0$\cr
                 &                        \cr
                  \ln \left[ {{h_i}\over{h_{i-1}}}\right] & if $h_{i}\ne 0$ and
                  $ h_{i-1} \ne 0$ }
\end{eqnarray}
where $\Delta E = E_i - E_{i-1}$,  the suffix $i$ denotes the energy bin and 
$\{ h_i \}$  the energy histogram\index{histogram} . The corresponding 
$\{ \tilde{\alpha}_{i}^{(k+1)
}\}$ follows immediately,
\begin{eqnarray}
\tilde{\alpha}_{i-1} ^{(k+1)} &=& \tilde{\alpha}_{i} ^{(k+1)}
                            -E_i \left( \tilde{\beta}_{i}^{(k+1)} -
                            \tilde{\beta}_{i-1}^{(k+1)}\right)
\end{eqnarray}
In the above we set  $\alpha_{imax} =0$ where $imax$ is the index of the 
bin of highest energy.  

Indeed, one can say that the basic idea behind 
 multicanonical\index{Monte Carlo!multicanonical} Monte Carlo (and hence 
entropic sampling\index{Monte Carlo!entropic sampling}) of the nineteen nineties is the same as that of the 
umbrella and adaptive 
umbrella  sampling\index{umbrella sampling} of the 
nineteen seventies. But then it is the work on 
multicanonical\index{Monte Carlo!multicanonical} sampling and its 
remarkable success
in the accurate evaluation of the interface\index{interface} tension in the 
first order phase transition\index{phase transition!first order}
that made popular such non-Boltzmann\index{Boltzmann}\index{non-Boltzmann sampling}
 sampling  cum iterative techniques 
and  initiated  large scale applications 
and several new  developments.  
These include 
flat/broad histogram\index{flat histogram}\index{broad histogram}  sampling~\cite{flathistogram}, 
transition matrix\index{transition matrix} Monte Carlo\index{Monte Carlo}~\cite{transitionmatrix},
exchange Monte Carlo\index{Monte Carlo!exchange}~\cite{exchange_MC},
simulated tempering~\cite{simulated_tempering_MC}\index{simulated tempering}
and
cluster hybrid algorithms~\cite{hybrid}.
For a comprehensive review 
see~\cite{Wang_review}. Multicanonical\index{Monte Carlo!multicanonical} sampling and its 
variants have been employed in a variety of studies that include
phase coexistence in Ising\index{Ising} models~\cite{muca_ising}, 
Potts\index{Potts spin} spin models~\cite{muca_potts}
liquid-vapour~\cite{muca_liquid_vapour} and 
solid - solid~\cite{muca_solid_solid} models, systems with complex free 
energy landscape~\cite{muca_spin_glass}, protein 
folding~\cite{protein_folding} 
and Nematic - Isotropic transition 
in bulk and confined liquid crystal systems~\cite{muca_N-I_transition} 
{\it etc., } to name a few. 

Spanning of  energy space in  
entropic\index{Monte Carlo!entropic sampling} / 
multicanonical\index{Monte Carlo!multicanonical} 
Monte Carlo algorithms and for that matter in  broad  / flat 
histogram\index{flat histogram}\index{broad histogram}  methods, is achieved through  a pure diffusive 
process on the one dimensional energy axis. 
Larger the system, wider is its energy range and hence 
longer it shall take for the system to diffuse from one end of the 
energy range to the other. The diffusive evolution of the Markov\index{Markov!chain} chain
is a consequence of the penalty imposed when the system attempts 
to visit  oft-visited energy bins. Also  
the penalty  in a learning run  
depends on the microstates 
visited during the immediate preceding learning run. Perhaps increasing
the penalty at least during the initial stages  
and applying the penalty  continuously during the stochastic 
dynamical evolution of the Markov\index{Markov!chain} chain would speed up the convergence. 
To this end 
Wang and Landau\index{Landau}\index{Wang}\index{Wang-Landau method}~\cite{FWDPL} 
proposed a simple, easy-to-understand and easy-to-implement 
algorithm that substantially improves the performance of the 
 multicanonical\index{Monte Carlo!multicanonical} Monte Carlo 
simulation.

\section{What is the Wang - Landau\index{Wang-Landau method}\index{Wang}\index{Landau} algorithm?}
\label{wang_landau}

In the Wang\index{Wang} - Landau algorithm\index{Wang-Landau method}~\cite{FWDPL}, the entropic function 
$\alpha (E)$ is directly and continuously updated after every MCS during the  
stochastic dynamical process\index{stochastic process}.  
The energy 
histogram\index{histogram}  is accumulated but not used in the updating of 
the entropic function. 
Instead the updating of the entropic function is 
carried out as follows. 
 We define a convergence factor $f$ and set it at
$f=f_0>0$ in the beginning. 
Also we set $\alpha_i = \alpha (E_i ) = 0 \ \forall\  i$ in the beginning. 
Start with an
arbitrary microstate and construct a trial microstate by selecting a 
spin randomly and flipping it. The acceptance/rejection of the trial state
is made on the basis of Eq.~(\ref{metro_entropy}). If  the 
energy of the next microstate lies in the $i$-th bin, we set  
$\alpha_i=\alpha_i + f$. The histogram\index{histogram}  of visited sites is also 
updated;
continue the evolution until the histogram\index{histogram}  becomes reasonably flat: Let
\begin{eqnarray}
\left\langle h\right\rangle & = & {{1}\over{M}}\sum_{i=1}^{M} h_i
\end{eqnarray}
denote the average number of entries in the histogram where $M$ is the number of energy
bins. We say the histogram is flat if 
\begin{eqnarray}
h_i & \ge\  \epsilon\ \left\langle h\right\rangle \ \ \forall\ \ i
\end{eqnarray}
where $\epsilon$ is taken between $0.75$ and $0.95$ depending on the problem. 
Thus the entropic function and the histogram are updated after every
MCS. The energy  histogram 
\index{histogram}  simply plays a passive role 
and serves as a tool to monitor the convergence. 
Then we reset $f=f_1=f_0/2$ and proceed. A fresh histogram\index{histogram}  is accumulated
and tested for flatness. In the next stage $f=f_2=f_1 /2$. 
 It is clear the convergence factor $f$ decreases 
exponentially to zero as the number of iterations increases. In fact after 
twenty five iterations the convergence factor becomes as small as 
$3\times 10^{-8}\ f_0$ whence we stop the simulation process. $f_0$ is usually
taken as unity.  The resultant 
entropic function will be very close to the true one when 
normalized on the basis
of known information about the system under simulation.
For example, in the Ising\index{Ising} model, the total number of microstates
is $2^V$ where $V$ is the  number of Ising\index{Ising} spins in the system
under investigation. Also the ground state of the Ising\index{Ising} 
model is doubly degenerate and this can also be used for normalization
purposes. From the normalized entropic function we can calculate the free energy\index{free energy} 
and all other desired macroscopic properties of the system under simulation.

We can employ the unnormalized entropic function in a final production run 
(when we do not update the entropic function nor do we  monitor the energy histogram)
and construct an adequately large entropic ensemble, like we did in the production run
under entropic/multicanonical Monte Carlo, see section (23). From the entropic ensemble
we can calculate the desired macroscopic properties at all temperatures through  
unweighting and reweighting. 
The Wang - Landau\index{Wang-Landau method}\index{Wang}\index{Landau}
method has been applied to Ising\index{Ising} and Potts spin\index{Potts spin} models~\cite{FWDPL,cyyo},
quantum mechanical models~\cite{mtswfa}, glassy systems~\cite{yoytcy} 
polymer films\cite{jain}, protein folding\cite{ratore}, 
models with continuous degrees of freedom~\cite{sdp} and generalized to 
reaction coordinates\cite{calvo}.
 
There are a few difficulties with  
the Wang-Landau algorithm \index{Wang-Landau method}\index{Wang}\index{Landau}
and there are a few issues that remain to be understood.
The statistical accuracy of the results can not be improved beyond 
a certain value, no matter how much additional calculations you perform.
The relation between the convergence factor and the errors in the 
estimates of the macroscopic quantities is not clear; also it is not
clear how flat the energy histogram\index{histogram}  
should be for accurate evaluation
of the macroscopic quantities.  
The algorithm does not satisfy 
detailed balance\index{detailed balance}  condition until the 
convergence factor $f$ decays close to zero. 
Also during the late stages 
of iteration, the visited microstates do not contribute to the estimate
of the entropic function since the convergence  factor $f$ 
is very small; this  means that 
a lot of  information is generated but not used. 
Consider the situation when we employ the Wang-Landau algorithm 
to estimate the 
density of states in a restricted range of energy. 
What should we do when the transition
takes the system to a microstate whose energy 
falls outside the range? Ofcourse we reject 
the transition; we do not update either the entropy function or 
the energy histogram. We proceed
with the dynamics and randomly select the next  trial microstate.
 When we do this we find  the calculated 
density of states in the boundary energy bins
to be erroneous\cite{FWDPL,SBM}. However if we 
update these  two quantities in the current energy bin,
the errors disappear\cite{SBML}. The reason for 
this is not clear. See also \cite{wlimprovement} for some 
recent 
effort toward understanding and updating of the Wang-landau algorithm.

The Wang-Landau algorithm seems to hold great promise. 
It would indeed worth the effort to put 
the algorithm on a more rigorous pedestal and devise means of improving its 
performance through new schemes or by linking it to other existing schemes.
For example, already Wang - Landau algorithm has been combined with the 
$n$-fold way \cite{SBM} for estimating tunneling times in 
Heisenberg spin models and with multibondic method \cite{WL_multibondic}
for general discrete models.
\section{Jarzynski's equality}\label{Jarzynski}\index{Jarzynski C.}
Estimation of entropy \index{entropy} or free energy \index{free energy} 
from simulation  or  
experimental data has always remained a tricky problem. A quantity like 
energy or magnetization of a macroscopic system can be calculated 
as an average over a reasonably large sample of microstates drawn from an 
equilibrium ensemble. However, for estimating entropy or free energies,   we have to 
necessarily consider all the microstates accessible to the 
equilibrium system and this number  is indeed very
large.  
An early ingenious
method of calculating entropy from  data on dynamics  
was proposed by S. K. Ma \cite{S_K_Ma}, \index{Ma S. K. } and it 
is based on counting of coincidence (or repetition) of states along a 
long phase space 
trajectory. We have already seen that umbrella 
sampling \index{Umbrella sampling} 
and its variants and  later improvements like entropic sampling, multicanonical sampling, and 
Wang-Landau algorithm are some of the useful  Monte Carlo techniques  
that can be adapted  for computation of entropy and 
free energy. In fact, the  very formulation of thermodynamics and the definition 
of entropy and free energies  provide 
us with a method of calculating these quantities, see below. 

Consider a classical macroscopic system in thermal contact with a heath bath at temperature $T$. 
Let $\lambda$ denote a degree of freedom of the system which can be controlled 
from outside; for example, the 
system can be a gas contained in a cylinder and the degree of freedom $\lambda$ can be its
volume which can be controlled from outside by moving a piston; the system can be 
a spin lattice and the 
parameter can be an external magnetic field whose strength and direction can be changed.
To begin with, at time $t=0$,  let $\lambda = \lambda_1$ and the system is in equilibrium 
with the heat bath. 
Then switch the value of $\lambda$ from $\lambda _1$ to $\lambda_2$. 
What is the change in entropy? Without loss of generality let us assume that the 
switching of $\lambda$ from $\lambda_1 $ to $\lambda_2$ is carried out over a time 
duration $t_s$ at a constant rate $(\lambda_2 -\lambda_1)/t_s$. 

Consider first an ideal scenario in which the switching is carried out infinitely 
slowly; in other words $t_s=\infty$; the system is sort of dragged through 
a continuous succession of equilibrium states; during the entire process of switching, 
the system is continuously in equilibrium  
 with the heat bath; such a process is often called a quasi-static equilibrium process;
the process is  reversible. Under such 
an ideal thermodynamic process, Clausius defines 
free energy (change)
$\Delta F = F(\lambda_2) -F(\lambda_1)$ as equal to the reversible work done on the system.  
\footnote{Reversible processes are an idealization; however, in experiments 
or in simulations, we can  be as close to the ideal reversible process as possible;
change $\lambda$ extremely slowly through  a succession of infinitesimal 
increments; estimate the   work done  during each increment; sum these up and 
equate the result to the free energy change.
There are slow growth algorithms,  
staged thermodynamic integration 
methods and thermodynamic perturbation 
methods to calculate/measure  free energies.
I do not intend to describe these techniques here and those interested can consult 
for example \cite{thermodynamics_methods}.}

What happens when the switching process takes place over a finite 
time duration, {\it i.e.} $t_s\ <\ \infty $ ? The system would all the time be lagging behind 
the corresponding equilibrium states; the work done  will 
depend on the particular microstate at which   the system and the heat bath
happen to be present at the precise moment the switching process starts. Hence 
the work done on the system would differ from one switching experiment to the other.
\footnote{All the switching experiments are carried out with the same protocol:  
set $\lambda =\lambda_1$; let the system 
attain thermal equilibrium with the heat bath; start the clock: $t=0$; switch 
$\lambda$  from $\lambda_1$ to $\lambda_2$ uniformly until the clock reads $t=t_s$; 
measure the work done on the system during the switching process; reset $\lambda$ to $\lambda_1$.}
In other words for $t_s < \infty$,
 the work done on the system
$W$ is a random variable; carrying out a switching experiment and 
measuring the work done is equivalent to sampling $W$ independently and 
randomly from its distribution 
$\rho (W; t_s)$. In other words $\rho (W, t_s)dW$ 
is the  probability that
the work done on the system during the switching process lies between
$W$ and $W+dW$. Let $\langle  W\rangle $ denote the average work done on the system, 
\begin{eqnarray}
\langle  W\rangle  &\equiv & \int dW\ W\ \rho (W,t_s) .
\end{eqnarray}
In the ideal quasi-static equilibrium limit of $t_s\to\infty$, we have
$\rho(W, t_s \to \infty) = \delta (W-W_R)$;  $W$ does not
change from one experiment to the other and it equals $W_R$. The reversible
work $W_R$ equals $\Delta F$, which is the definition of free energy (change)
given by Clausius. 

However such reversible thermodynamic processes, very useful though, 
 are idealizations and are not  strictly
realized in experiments or simulations. In general we have 
$\langle  W\rangle \ge  \Delta F$, with equality holding good for 
reversible processes. The difference $\langle  W\rangle -\Delta F$ is called 
the (irreversible) dissipative work and is denoted by $W_d$. Thus, to calculate 
$\Delta F$ from experiments/simulation, we need a good estimate of 
dissipation, in the absence of which,  we can,  at best provide only an upper limit:
$\Delta F \le \langle  W\rangle$. 

In the year 1951, H. B. Callen and T. A. Welton \cite{Callen} 
\index{Callen H. B.}\index{Welton T. A.}
showed that if the 
switching process keeps the system all the time very close to equilibrium,
then the dissipation $W_d$ is proportional to fluctuations:
\begin{equation}
W_d \ \equiv\  \langle  W\rangle -W_R \  \approx \ \beta\ \sigma_W ^2 /2
\end{equation}
where the fluctuations,  $\sigma^2 _W $ is given by,
\begin{equation}
\sigma^2 _W = \int dW\ W^2 \rho (W; t_s) - \langle  W\rangle ^2
\end{equation}

Recently, in the year 1997, C. Jarzynski, \cite{Jarzynski_Hamiltonian}\index{Jarzynski C.}
discovered a remarkable 
and exact identity, valid even when the switching process drives the system 
far from equilibrium.
Jarzynski's identity relating nonequilibrium work to equilibrium
free energy differences is given by,
\begin{eqnarray}
\langle  \exp\  (-\beta W) \rangle \equiv \int dW\ \exp\  (-\beta W)\ \rho (W,t_s) =
\exp\  (-\beta\  \Delta F) \ .
\end{eqnarray}
Jarzynski's equality 
\footnote{It may be noticed that since the exponential function is 
convex,
\begin{eqnarray}
\left\langle \exp (-\beta W  )\right\rangle \ge \exp (-\beta\left\langle W\right\rangle)\nonumber
\end{eqnarray} 
which in conjunction with Jarzynski's identity implies that,
\begin{eqnarray}
\exp (-\beta\Delta F ) & \ge & \exp (-\beta\left\langle W\right\rangle)\nonumber\\
-\beta\Delta F   &    \ge  &  -\beta\langle W\rangle\nonumber\\
\Delta F & \le & \left\langle W\right\rangle \nonumber
\end{eqnarray}
which is a statement of the second law, if we modify the 
switching protocol and let the system equilibrate at $\lambda = \lambda_2$ at the end of the 
switching process. In this sense, proof of Jarzynski's identity is a proof of the 
second law}
has been rigorously proved for 
Hamiltonian evolution, \cite{Jarzynski_Hamiltonian}
as well as stochastic evolution \cite{Jarzynski_Stochastic}; 
its validity has been established in computer simulation 
\cite{Jarzynski_Stochastic} and 
more importantly in experiments \cite{Jarzynski_Expt}. 

Let us express Jarzynski's equality as a cumulant expansion \cite{cumulant_expansion},
\index{cumulant expansion}
\begin{eqnarray}\label{cumulant_expansion}
\langle \exp (-\beta W)\rangle  &\equiv & 
\exp\left[ \sum_{n=1}^{\infty} {{ (-\beta)^n C_n}\over{n!}}\right]
=\exp (-\beta \Delta F)
\end{eqnarray}
where $C_n$ denotes the $n-$th cumulant of $W$. The cumulants and  the moments are 
related to each other. The $n$-th cumulant can be expressed in terms of the 
moments of order $n$ and less.  Given below are first four cumulants expressed in terms 
of moments.
\begin{eqnarray}
C_1 &=& \langle W\rangle \ ,\nonumber\\ 
C_2 &=&
\langle W^2\rangle -\langle W\rangle ^2 = \sigma^2 _W\ ,\nonumber\\
C_3 &= &\langle W^3 \rangle - 
3\langle W^2\rangle\langle W\rangle + 2\langle W \rangle^3 \ ,\nonumber\\
C_4 &= &\langle W^4\rangle -
4\langle W^3\rangle\langle W\rangle - 2\langle W^2 \rangle^2 +
12 \langle W^2\rangle\langle W\rangle^2-6\langle W\rangle ^4 \ .
\end{eqnarray}
From the cumulant expansion, see Eq. (\ref{cumulant_expansion}),  we get,
\begin{eqnarray}
\Delta F &=& \left\langle W\right\rangle -{{\beta\sigma_W ^2}\over{2}} + \sum_{n=3}^{\infty} {{(-\beta)^{n-1} C_n}\over{n!}} 
\end{eqnarray}
Consider a quasi-static equilibrium switching process;   
$\rho~(W)=\delta~(W-W_R)$, by definition. Then, in the above, only the first term (of the cumulant 
expansion) is non-zero. We get the definition of free energy as given by
Clausius:  $\left\langle W\right\rangle = W_R =  \Delta F$.

Now consider a switching process, during which the system remains very close to 
equilibrium; it is reasonable to expect the statistics of $W$ to obey 
the Central Limit Theorem \index{Central Limit Theorem}; hence $\rho (W)$ shall be a Gaussian;
for a Gaussian, all the cumulants from the third upwards are identically zero; hence in the 
above expansion only the first two terms survive and we get 
$\Delta F =  \langle W\rangle -\beta C_2 /2 = \langle W\rangle -\beta\sigma_W ^2 /2 $.
This result is identical to the fluctuation dissipation relation of Callen and Welton \cite{Callen}.
However, if  the switching process 
drives the system far from equilibrium, the work distribution would  no longer 
be  Gaussian and we need to include contributions from higher order cumulants to
calculate the dissipation $W_d$  and hence free energy: $\Delta F = \langle W\rangle - W_d$
 
Jarzynski's equality \index{Jarzynski's equality} is helpful to experimenters for measuring
free energies; the experimenters do not any more need to keep the system
at or close to equilibrium during the switching process. Jarzynski's identity 
is a powerful tool for calculating free energy,  when suitably incorporated
in a Monte Carlo or Molecular dynamics programs. A dynamical ensemble of 
nonequilibrium trajectories all starting off from an equilibrium ensemble is all 
that is required for free energy computations. 

Jarzynski's equality \index{Jarzynski's equality} 
forms a part of recent exciting and important developments in 
nonequilibrium statistical mechanics. These new developments, related to each other and 
derivable one from the other,  constitute what has come to be known as 
 {\it Fluctuation 
Theorems} \cite{Crooks,Jarzynski_Hamiltonian,Jarzynski_Stochastic,Jarzynski_Expt,Evans} 
and have led to fresh insights into  nonequilibrium
statistical mechanics, reversible and irreversible processes and the second law of 
thermodynamics. 
\section{Epilogue}
\label{epilogue}

All things, good  and  bad, must come to an end; so must, our Monte 
Carlo journey, good or bad.  
I have taken you, randomly though, through the 
lanes and by-lanes of the Monte Carlo metropolis, confining 
 to those regions  
relevant for statistical physics applications.
Table~(6)  lists what I consider as 
milestones in  Monte Carlo statistical physics. 
\begin{table}[thp]\label{milestones} 
\caption{\protect\small  Milestones in Monte Carlo Statistical Physics} 
\bigskip
\begin{center}
\begin{tabular}{|p{1.2cm}|p{5.5cm}|p{6.5cm}|}
\hline
      &                                                    & \\
When? &  What happened? & Who did it and where do I find it?  \\
      &                 &                                     \\
\hline         
      &                                                    & \\
1951  & Linear congruential (pseudo  random number) generator  & D. H. Lehmer\cite{rngenlit}\\[2mm]
1953  & Metropolis algorithm                               & N. Metropolis\index{Metropolis}, A. W. Rosenbluth, 
                                                             M. N. Rosenbluth,\index{Rosenbluth} A. H. Teller 
                                                             and E. Teller\index{Teller}\cite{metropolis}\\
      &                                                    & \\ 
1963  & Glauber dynamics\index{Glauber dynamics}                                   & R. J. Glauber\cite{Glauber}\index{Glauber}\\[2mm]
1964  & Hand Book on Monte Carlo  Methods                  & J. M. Hammersley\index{Hammersley} and 
                                                             D. C. Handscomb\index{Handscomb}\cite{mcref2}\\[2mm]
1969  & Random cluster model                               & P. W. Kasteleyn\index{Kasteleyn}, C. M. Fortuin\cite{f                                                                                       k}\index{Fortuin}\\
      &                                                    & \\
1975  & n-fold way                                         & A. B. Bortz\index{Bortz}, M. H. Kalos\index{Kalos}
                                                              and J. L. Lebowitz\index{Lebowitz}\cite{bortz}\\[2mm]
1976  & Cluster counting algorithm                         & J. Hoshen and R. Kopelman\index{Hoshen-Kopelman}                                                                                              \cite{hoshen_kopelman}\\[2mm]
1977  & Umbrella Sampling                                  &  G. M. Torrie\index{Torrie} and J. P. Valleau
                                                              \index{Valleau}\cite{TV}\\
      &                                                    & \\
1980  & Ising critical droplets                            &  A. Coniglio\index{Coniglio} and W. Klein
                                                              \index{Klein}\cite{ck}\\[2mm]
1987  & Swendsen-Wang cluster algorithm                        &  R. H. Swendsen\index{Swendsen} and 
                                                              J.-S. Wang\index{Wang}\cite{sw}\\[2mm] 
1988  & Histogram reweighting                              & A. M. Ferrenberg and R. H. Swendsen\cite{fs}\\[2mm]
1989  & Wolff cluster algorithm                   &  U. Wolff\index{Wolff}\cite{wolff}\\
      &                                                    & \\
1991  & Multicanonical Monte Carlo                         & B. A. Berg\index{Berg} and T. Neuhaus\index{Neuhaus}
                                                             \cite{multicanonicalsampling}\\[2mm]
1993  & Entropic sampling                                  & J. Lee\index{Lee}\cite{entropicsampling} \\[2mm]
1995  & Absorbing  Markov Chain                            & M. A. Novotny\index{Novotny}
                                                             \cite{novotny}\\
      &                                                    & \\
2000  & A Guide to Monte Carlo Simulations in 
        Statistical Physics                                & D. P. Landau\index{Landau} and K. Binder\index{Binder}
                                                             \cite{DPLKB}\\[2mm] 
2001  & Wang-Landau algorithm                              & F. Wang and D. P. Landau\cite{FWDPL}\\
      &                                                    &   \\
\hline
\end{tabular}
\end{center}
\end{table}
\index{Lehmer}
\index{Metropolis}
\index{Rosenbluth}
\index{Teller}\index{Glauber}\index{Hammersley}\index{Handscomb}
      \index{Kasteleyn}\index{Fortuin}\index{Lebowitz}\index{Bortz}\index{Kalos}
      \index{Hoshen}\index{Kopelman}\index{Torrie}
       \index{Valleau}\index{Coniglio}\index{Klein}\index{Swendsen}
       \index{Wang}\index{Ferrenberg}\index{Wolff}
       \index{Berg}\index{Neuhaus}\index{Lee}\index{Novotny}\index{Landau}
       \index{Binder}
       \index{Metropolis!algorithm}\index{Glauber dynamics}
       \index{cluster algorithm}\index{n-fold way}
      \index{cluster counting}\index{histogram}\index{reweighting}
       \index{Wolff algorithm}
       \index{Monte Carlo!multicanonical}
       \index{Monte Carlo!entropic sampling}
       \index{Monte Carlo!absorbing Markov chain}
       \index{Wang-Landau method}

In the first leg of our journey we looked at ensembles in general and Gibbs 
ensembles\index{ensemble!Gibbs} in particular. 
We saw details of microcanonical ensemble\index{ensemble!microcanonical} that 
describe isolated system\index{isolated system}; of canonical ensemble
\index{ensemble!canonical} that models 
closed system\index{closed system}; and of grand canonical ensemble,  
\index{ensemble!grand canonical} useful in the study of 
open systems.\index{open system}
 Then we took up the issue of random numbers that fuel the Monte Carlo 
machine. 
To our great discomfort 
we discovered that even today we do not  know when can we call a sequence of numbers 
random except when we know of its origin. 
This 
of course did not deter us. 
We took a pragmatic
attitude  and settled for pseudo random numbers\index{pseudo random numbers}. 
We saw how to generate them
and how to test them. We learnt how to use them to generate  ensemble
of random numbers representing the desired distribution - the so called 
random sampling techniques. We saw explicitly some techniques that help 
sample from the exponential and Gaussian distributions.
For purpose of illustration we saw how 
to calculate an 
integral by the method of Monte Carlo. We learnt of Monte Carlo 
error bars and 
how to estimate them. 

Then we took up the issue of importance sampling\index{importance sampling}
 and dealt with it in  
somewhat greater details mainly for two reasons.  
Importance sampling  is an important topic in
Monte Carlo theory and practice. Almost all Monte Carlo programmes  employ importance 
sampling in some form or the other and with varying degrees of sophistication. 
The second reason is that often
I find students have difficulties in comprehending 
what and how does importance sampling  
achieve what it purports to achieve. 
One way  is to emphasize that importance 
sampling reduces the variance of the desired statistical variable whose 
average is what you want to calculate by Monte Carlo simulation. 
Needless to say that the technique  should leave the average
unchanged. I have taken this view and elucidated it through a simple 
example for which all the quantities can be calculated analytically. 

Often we need to resort to importance sampling \index{importance sampling} 
as a matter of necessity;
for otherwise, any finite Monte Carlo sample generated should 
rarely contain microstates of interest to the phenomenon under investigation.  
 Metropolis importance sampling 
is an example. A randomly chosen microstate belonging to a
 closed system is most likely
to be of high energy
\footnote{the microcanonical entropy increases rather rapidly with increase of 
energy; more the energy the more is the number of ways of distributing it. 
} and hence contributes negligibly to the canonical partition sum. Hence 
we have to resort to importance sampling.  The 
Metropolis algorithm\index{Metropolis!algorithm}
 generates a Markov chain\index{Markov!chain}
 whose (asymptotic) distribution 
is the equilibrium canonical distribution at the desired temperature. 
Indeed it is the  
Metropolis algorithm that launched the Monte-Carlo-for-statistical-physics 
business; it remains
to date  the best algorithm in this field. We took Ising model\index{Ising}
 as an example to illustrate
the Metropolis algorithm, continuous phase 
transition\index{phase transition!second order}, critical exponents,
\index{exponent!critical} finite size effects and finite size scaling.
\index{finite size effect}\index{finite size scaling}  

When the actual dynamics slows down, the corresponding time-driven Monte Carlo 
simulation also becomes slow.\index{Monte Carlo!time driven}  
Bortz, Kalos and Lebowitz
\index{Bortz}\index{Kalos}\index{Lebowitz}
 suggested shifting of the (slow) dynamical information
to the time axis and derived the so-called n-fold way; 
\index{n-fold way}
this event-driven algorithm  
\index{Monte Carlo!event driven} is rejection free and
helps speed up the computations. Novotny\index{Novotny}  
generalized this technique
to Monte Carlo with Absorbing Markov Chain (MCAMC). 
\index{Monte Carlo!absorbing Markov chain}
However  critical slowing down in continuous phase transition 
\index{critical slowing down}
\index{phase transition!second order}
calls for special techniques.
We discussed the cluster algorithm of Swendsen and Wang, of Wolff, 
and of the one based on invasion percolation
{\it etc.}  Wolff algorithm became famous for quite another reason also. It was while studying 
its  performance that Ferrenberg, Landau and Wong discovered how dramatically 
the \lq\ pseudoness~\rq\ of  pseudo random numbers can affect your Monte Carlo results and we had a 
glimpse of the 
well documented R250 story.\index{R250 generator} 

While on clusters we learnt of the celebrated 
Hoshen-Kopelman cluster counting algorithm; we also discussed blocking technique that 
helps estimate statistical error bars from correlated data and   the 
histogram technique often employed in the study of systems very close to 
critical points. The histogram technique helps you calculate 
macroscopic properties of a system at temperatures
different but reasonably close to the simulation temperature by suitable reweighting. 

A first order phase transition
is associated with the so called supercritical slowing down; we found how 
non-Boltzmann sampling techniques like entropic sampling/multicanonical Monte Carlo would help.
\index{Monte Carlo!entropic sampling}\index{Monte Carlo!multicanonical}
Then we saw of the Wang-Landau algorithm which improves the performance of the 
multicanonical Monte Carlo methods. 
We ended our journey with a discussion on the recently proposed  
Jarzynski's equality \index{Jarzynski's equality} and  the promise this technique holds for computation of
equilibrium free energies from nonequilibrium simulation/experiment.

As you would have rightly noticed, the journey has been somewhat random and 
in places rather rickety. I made no serious efforts to be otherwise; it had 
never been my intention to take you through all the topics in this vast and 
 growing field. Nor did I try to place  uniform emphasis on the 
various topics covered in these notes.  
I just picked up topics that  caught my fancy and on which 
I have some hands-on experience and discussed them in a pedagogic style. 
Nevertheless I believe these notes cover quite a bit of ground and 
would serve as an introduction to this fascinating field.

The  topics we have not visited  include
 simulation of microcanonical\index{ensemble!microcanonical} 
and other ensembles, 
quantum Monte Carlo\index{Monte Carlo},
renormalization employing Monte Carlo\index{Monte Carlo}, 
replica Monte Carlo\index{Monte Carlo},
Monte Carlo for vector and parallel computers, 
off-lattice models, 
flat/broad histogram sampling,
transition matrix Monte Carlo,
exchange Monte Carlo,
simulated tempering,
cluster hybrid algorithms
and 
the whole field of simulation of nonequilibrium phenomena. 
Perhaps we can visit these  topics during another journey on another occasion.
Until then bye. 

The  recent comprehensive book by 
Landau\index{Landau} and Binder\index{Binder}~\cite{DPLKB} 
should  provide an excellent guide.
\newpage
\addcontentsline{toc}{section}{\numberline{}References}

\addcontentsline{toc}{section}{\numberline{}Index}
\printindex
\end{document}